# From Coated to Uncoated: Scanning Electron Microscopy Corrections to Estimate True Surface Pore Size in Nanoporous Membranes

Sima Zeinali Danalou[1], Dian Yu[2], Niher R. Sarker[1], Hooman Chamani[1], Jane Y. Howe[2], Patrick C. Lee[3], and Jay R. Werber[1*]

[1]*Department of Chemical Engineering and Applied Chemistry, University of Toronto, 200 College St, Toronto, ON M5S 3E5, Canada*

[2]*Department of Materials Science and Engineering, University of Toronto, 184 College St, Toronto, ON M5S 3E4, Canada*

[3]*Department of Mechanical & Industrial Engineering, University of Toronto, 5 King's College Rd, Toronto, ON M5S 3G8, Canada*

*\*Corresponding author: Jay Werber (email: jay.werber@utoronto.ca)*

# Abstract

Scanning electron microscopy (SEM) is the premier method for characterizing the nanoscale surface pores in ultrafiltration (UF) membranes and the support layers of reverse osmosis (RO) membranes. Based on SEM, the conventional understanding is that membranes typically have low surface porosities of <10%. We hypothesized that high acceleration voltage during SEM imaging and sputter metal coatings required for SEM have led to systematic underestimations of surface porosity and pore sizes. We showed that imaging a commercial UF membrane at 1, 5, and 10 kV reduced the measured surface porosity from $10.3 \pm 0.3\%$ (1 kV) to $6.3 \pm 0.4\%$ (10 kV), while increasing Pt coating thickness from 1.5 to 5 nm lowered porosity by 54% for the UF membrane ($12.9 \pm 0.9\%$ to $5.8 \pm 0.6\%$) and 46% for an RO support ($13.1 \pm 0.6\%$ to $7.0 \pm 0.2\%$). To account for the coating thickness, we then developed a digital correction method that simulates pore dilation, enabling the surface pore structure to be estimated for uncoated membranes. Pore dilation yielded uncoated surface porosity values of 23% for the UF membrane and 20% for the RO support, which are approximately 3-fold greater than the directly observed values for a typical coating thickness of 4 nm. Similarly, mean pore diameters for uncoated membranes were 2-fold greater for the UF membrane and 1.5-fold greater for the RO support than directly observed. Critically, the dilation-derived pore-size distributions agreed with low-flux dextran-retention measurements fitted with the Bungay–Brenner model. Our results suggest that surface porosities and pore sizes of nanoporous membranes are much larger than previously understood, which has major implications for structure/transport relationships. For future nanoscale pore analysis of membranes (and other nanoporous materials), we recommend low acceleration voltage (1 kV), minimal coatings (1–2 nm), and digital dilation to account for coating-induced artifacts.

## TOC Graphic

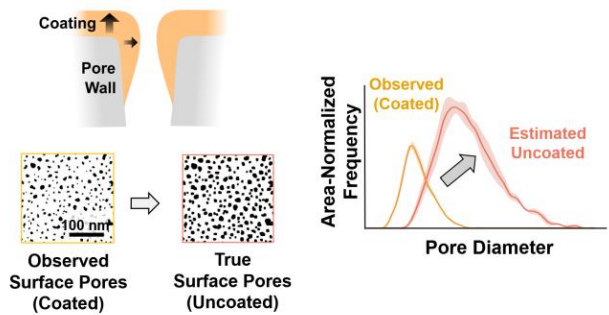

# Introduction

Modern water treatment processes make extensive use of a diverse array of membrane filtration technologies. Three of the most important categories are ultrafiltration (UF), nanofiltration (NF), and reverse osmosis (RO), all of which rely on polymeric membranes. UF membranes are characterized by their nano-sized surface pores (1–100 nm diameters), and are used extensively in membrane bioreactors, as tertiary wastewater treatment processes, and as pre-treatment for RO.[1] UF operates as a screen filtration process, where particles such as proteins, polysaccharides, pathogens, and natural organic matter are retained by the surface nanopores. In NF and RO, membranes have a thin-film composite (TFC) structure consisting of an ultra-thin (10–100 nm thick) polyamide active layer that selectively allows water permeation and an underlying nanoporous support layer that contributes to mechanical stability while minimizing resistance to flow.[2] In other words, nanoporous materials are critical components in all membranes for UF, NF, and RO.

The surface pore structure—namely the pore size distribution (PSD) and surface porosity—of UF membranes and RO/NF support layers largely defines their impact on fluid flow and particle retention. For UF membranes, a brief review of UF transport theory highlights the importance of surface pore structure. Water permeability through individual pores is typically modeled using a simple Hagen-Poiseulle relation for laminar flow in a pipe, which is dependent on pore length ($\delta_m$) and pore diameter ($d_p$). Summing up individual pores using a pore size distribution enables estimation of the overall water permeability. For a uniform pore size, water permeability ($A$) is:[3]

$$A = \frac{\varepsilon d_p{}^2}{32\tau^2 \mu \delta_m}$$ (1)

In Eqn. 1, $\varepsilon$ is surface porosity and $\mu$ is fluid viscosity. The tortuosity ($\tau$) is often assumed to be 1 to reflect a straight pore near the surface. Pore diameter is hugely important. For a fixed surface porosity, $A \propto d_p{}^2$, but surface porosity is also a function of pore diameter:

$$\varepsilon = \frac{n_{pores}\pi d_p{}^2}{4}$$ (2)

where $n_{pores}$ is the number of pores per area. In other words, for a given pore or a membrane with a fixed number of pores, $A \propto d_p{}^4$. For particle retention, a variety of relations have been derived for pore flow, all of which are heavily dependent on pore diameter.[4] For example, the Ferry equation estimates solute retention ($R$) for hydraulic flow through a single pore based solely on $\lambda = d_s/d_p$, where $d_s$ is the diameter of the solute:[5]

$$R = 1 - [2(1-\lambda)^2 - (1-\lambda)^4]$$ (3)

Another commonly used formulation from Zeman and Wales[6–8] makes minor modifications to Eqn. 3. All of the above highlight the critical dependence of UF performance on pore diameter and, secondarily, on surface porosity.

In TFC membranes for RO and NF, the impact of the porous support layer on performance is less straightforward, but still a function of pore diameter and porosity. The main impact of the porous support is on the path length of permeating species (i.e., water) through the active layer: molecules must diffuse a greater distance through the active layer if not directly above an underlying pore. Modeling has suggested that this effect can lead to sharp reductions in overall water permeance.[9] Computational fluid dynamics simulations of a simple composite structure—a smooth active layer on a support layer with repeating, circular pores—found a strong dependence of the total permeability reduction on the support porosity and the ratio of active layer thickness to support pore diameter.[9] The effect can be dramatic: for an active layer thickness equal to the pore diameter and a surface porosity of 5%, the modeled TFC water permeance is 84% lower than the "true" permeance of an unsupported active layer. Increasing the support layer surface porosity to 20% decreases the drop in overall TFC permeance to just 45%. Furthermore, experimental data on nanofilm composite membranes showed that the measured permeance of nanofilm TFC membranes could exceed values predicted by the resistance-in-series model by more than a factor of two, clearly demonstrating the non-negligible role of the support layer[10]. These results highlight that support layer structure, including pore size and surface porosity, can substantially affect water transport in TFC membranes.

Surface pore structure (i.e., pore size distribution and porosity) is therefore critical in defining membrane performance in UF, RO, and NF membranes. Unfortunately, existing pore characterization methods have significant flaws. Several physical techniques have been developed for estimating pore size distribution in porous membranes. These include gas-liquid porometry[11], liquid-liquid porometry,[12,13] liquid-vapor equilibrium method,[14] gas-liquid equilibrium method (permporometry),[15] liquid-solid equilibrium method (thermoporometry),[16] and mercury porosimetry.[17] Among these, gas-liquid and liquid-liquid porometry techniques are the most frequently used methods.[18] Critically, these methods must assume cylindrical and separated pores, which limits accuracy when dealing with the complex and non-idealized pore structures of real membranes.[19–21] They also require estimation of the pore length, which is not constant and difficult to estimate in UF membranes and RO supports formed from non-solvent-induced phase separation (NIPS). Molecular weight cut-off (MWCO) experiments with neutral particles (i.e., dextrans) are a classical and valuable tool for indirectly measuring the pore size of UF membranes. However, one must typically assume a log-normal distribution, from which just the mean and standard deviation can be estimated from the filtration data. This method can also underestimate pore size due to surface interactions of particles and concentration polarization.[22–25] Atomic force microscopy (AFM) provides direct visualization of membrane pores, yet it struggles with resolution constraints to characterize nano-sized pores and artifacts from tip-sample interactions.[26]

In contrast to the above techniques, scanning electron microscopy (SEM) provides the ability to directly visualize the surface pore structure, without the need for any assumptions about pore size distribution type, pore shape, or pore length. As such, SEM has long been the premier method to characterize membrane surface properties both qualitatively (e.g., visual observation) and quantitatively (e.g., pore size, shape, and porosity), with quantitative details from SEM analyses arguably being the most trusted values for nanoporous membrane pore structure.[9, 26, 28] Notably, SEM analyses have consistently produced low surface porosities ($\varepsilon < 10\%$) for NIPS-based membranes. For example, commercial UF membranes and RO supports made from polyethersulfone (PES),[29] polyvinylidene fluoride (PVDF),[30] and polysulfone (PSF)[31] were all reported as having surface porosities ranging from 3–10%. Similarly, hand-cast PSF [32–34] and PES[35] UF membranes and RO supports have reported porosities between 1–15%. As such, it is commonly understood in the membrane science field that UF membranes and RO supports typically have surface porosities around or below 10%.

While SEM image analysis can accommodate irregular pores, there are experimental factors that can significantly affect the visibility of the pores. SEM reveals the surface morphology of samples by detecting secondary electrons (SEs), which are low-energy electrons between 0 and 50 eV generated from the inelastic scattering of the incident and backscattered electrons (BSEs). Only SEs close to the sample surface would have sufficient kinetic energy to escape from the surface, limiting the escape depth to a few nanometers.[36] This property of SEs enables the high surface specificity of SEM. However, stable imaging conditions can only be achieved when the number of incident electrons equals the sum of the electrons exiting the sample. For conductive samples such as metals, electrical grounding can facilitate an overall balance of charge, but non-conductive polymeric membranes need sputter coating with a thin layer (typically 3–5 nm) of a conductive metal, such as Au, Pd, Ir, or Pt, on the surface to avoid charging. Coating thickness has been qualitatively shown to impact pore analysis,[27,37] but a comprehensive analysis of coating thickness on pore structure has not been performed systematically. In fact, it is the general practice in membrane science for coatings to be *ignored* during pore analysis, with coating thickness and coating angle often not reported (Supporting Table S1); the reported pore size distributions and porosities then correspond to the coated sample, not the actual, uncoated membrane. We hypothesized that sputter coating thickness could be having a large impact on reported pore properties, leading to a widespread and systematic underestimation of surface porosity. A simple thought exercise suggests that thin coating layers could have a substantial impact. For example, if a conformal coating layer of just 2 nm is deposited on a pore with a 10-nm diameter, the visualized pore will have a diameter of 6 nm. This corresponds to a 64% drop in perceived pore area (i.e., a 64% drop in observed surface porosity) compared with the actual pore area of the uncoated pore.

In this study, we demonstrate unequivocally that both Pt sputter-coating thickness and high SEM acceleration voltage lead to dramatic drops in observed surface porosity. To do so, we varied Pt coating thickness from 1.5 nm to 5 nm on a commercial UF membrane and RO support and acquired images at low (1 kV) and higher (5–10 kV) voltages. To obtain true values for surface porosity and pore size, we developed a simple digital correction approach that dilates observed pores to account for the coating thickness, effectively undoing the impact of coating artifacts. The method is robust, with consistent "uncoated" pore structures approximated from varying coating levels, especially for the RO support that had larger pores. To provide further validation, we show that the SEM-derived uncoated PSDs have good agreement with PSDs derived from low-flux dextran-retention MWCO experiment (Bungay–Brenner model fit). Altogether, this study provides a new fundamental understanding that surface porosities and surface pore sizes of nanoporous membranes are greater than previously understood, while providing a simple correction method for researchers to conduct similar analyses of any nano-scale porous materials to estimate uncoated porosity and pore-size distributions.

## Materials and Methods

### Preparation of membranes for SEM imaging

Commercial AD-90 seawater reverse osmosis (RO) membrane and PW ultrafiltration (UF) membrane were kindly provided by Veolia Water Technologies. To assess the surface pores of the RO support, the top selective polyamide (PA) layer was selectively degraded. Dry membranes were soaked in 10,000 ppm sodium hypochlorite (NaOCl) under basic conditions (pH 12)[38] for three days on the orbital shaker at room temperature. Following this treatment, the membranes were sonicated for 1 minute to ensure complete detachment of residual PA from the support. Subsequently, the membranes were thoroughly washed with DI water to remove any residual NaOCl. The removal of the PA layer was confirmed using attenuated total reflectance-Fourier transform infrared (ATR-FTIR) spectroscopy using a Thermo Scientific iS50: Spectral range 4600 -50 cm$^{-1}$ (see Supporting Fig. S1.1). To confirm that this treatment did not alter the support pore structure, we compared SEM images of supports fabricated in-lab from Udel® polysulfone (PSF) solution in N-methyl-2-pyrrolidone (NMP) before and after bleach exposure under identical conditions (see Supporting Fig. S1.2). No morphological differences were observed. Prior to sputter coating and SEM imaging, membranes were pre-washed in sequential steps to eliminate manufacturing agents inside the membranes (e.g., glycerol protective layer).[39] The pre-wash procedure included immersing in deionized (DI) water for 10 minutes on an orbital shaker, followed by 10 minutes in 1:3 v/v isopropanol-DI water, 5 minutes pure isopropanol, and 20 minutes hexane. Membranes were then removed and left to air dry, typically overnight.

**SEM Characterization**

Dried membranes were Pt sputter-coated with a Leica ACE600 sputter coater at thickness-control mode. Among common metal targets, we selected Pt due to its fine grain size and ability to form uniform coatings at low thicknesses (sub-5 nm). The coating thickness was estimated by the instrument using internal calibration with a quartz crystal microbalance (QCM) and the coating thickness was displayed on the screen after each run. Due to residual particles in the sputter chamber, the actual deposited thickness may slightly exceed the target set value. Therefore, we recorded and used the system-displayed thickness after each coating as the system-measured Pt thickness throughout our study. For a 90° coating angle, the stage was tilted at 25°, while for a 65° coating angle, the stage was kept horizontal. The surface morphology of the membranes was visualized using a Hitachi SU-7000 Schottky Field Emission SEM. To obtain satisfactory images, a low acceleration voltage of 1 kV,[36,40] a low working distance of < 4 mm and high magnification (100,000× and 150,000×) were used.

**Image processing**

SEM images were processed in Fiji/ImageJ (version 1.54f) software. For more accurate pore segmentation, the gray-value SEM images were contrast-adjusted and de-noised. Brightness and contrast for each image were optimized in Fiji, considering the gray-value histogram of the images. Subsequently, excess noise was eliminated using the "Despeckle" function. For pore segmentation, the Trainable Weka Segmentation (TWS) plugin (version 3.3.4)[41] was used. During model training, features (i.e., pores and polymer walls) were labeled. Then, to separate the adjacent pores that are in contact with each other and accurate pore size measurement, the watershed algorithm[42] was applied on the binary image. The surface porosity was measured as the total area of pores divided by the total area of the binary image. The mean pore diameter was measured as the circular-equivalent diameter—i.e., the diameter of a circle having the same area as the given pore. Additionally, we report the maximum inscribed circle (MIC) diameter, defined as the largest circle fully contained within each pore opening. Three SEM images were analyzed for each data point.

**Artificial pore alteration**

The developed Python code facilitates the process of pore dilation towards zero coating thickness and pore constriction towards higher coating thickness. Initially, the binary image is read, where pores are represented by black pixels and the polymer solid walls by white pixels. The code identifies the pore regions, and we digitally 'remove' or 'apply' the coating strip rim by shifting the pore boundary outward or inward by a chosen thickness using a distance-transform on a binary mask; this preserves pore shape and makes the added/removed strip rim width equal to our chosen thickness. To achieve a specific coating thickness, the image was resized accordingly based on the desired thickness. The resizing process adjusted the image dimensions to ensure that the number of iterations accurately represented the coating thickness. Full details and the Python script are provided in the Supporting Information Section S4.1-2.

## Membrane performance: molecular weight cut-off (MWCO) retention test

Membranes were tested with dextrans of varying molecular weight (MW), and experimental dextran rejection curves were determined. Dextrans derived from *Leuconostoc* were dissolved in a buffer solution at the following concentrations: 0.4 g/L ($M$ ~6,000 g/mol), 0.4 g/L ($M$ ~40,000 g/mol), 0.89 g/L ($M$ ~100,000 g/mol), and 1.2 g/L ($M$ ~450,000–650,000 g/mol). The buffer was prepared by dissolving 0.823 g/L sodium phosphate dibasic anhydride, 0.504 g/L sodium phosphate monobasic monohydrate, and 16.99 g/L sodium nitrate in DI water, with 0.02 w% sodium azide to prevent algae growth. Sodium hydroxide was added to adjust the pH to 7. All materials were obtained from Sigma-Aldrich.

Membranes were pretreated by soaking in DI water for 10 minutes, followed by immersion in 25 vol % isopropanol for 10 minutes, and then 5 minutes in pure isopropanol, while being placed on a shaker. After pretreatment, membranes were thoroughly rinsed with DI water to remove residual isopropanol before further use. A custom cross-flow filtration system was operated for 30 minutes with DI water, followed by filtration of dextran solution for 1 h at 0.5 psi, a cross-flow rate of 4.5 L/h (0.08 m/s average crossflow velocity) and a flux of 1.3 L m$^{-2}$ h$^{-1}$. The custom cross-flow cell has feed channel dimension of 0.1 cm height, 1.6 cm width, and 2.2 cm length, giving an effective membrane area of 3.52 cm$^2$ (Supporting Fig. S2.1).

Feed and permeate samples (1 mL) were collected and pre-filtered through 0.2 µm polyvinylidene fluoride (PVDF) membranes to remove particulates before analysis. The samples were then analyzed using an Agilent Technologies Model 1260 Infinity LC system equipped with a RID detector. The columns were two PL aquagel-OH Mixed H 8 µm 300 x 7.5 mm linked in series with a PL aquagel-OH Guard 8 µm 50 x 7.5 mm guard column (Agilent Technologies). The column temperature was maintained at 30 °C, with a flow rate of 1 mL/min. Calibration was performed using solutions of individual, narrow molecular weight dextran standards at a concentration of 1 mg/mL, providing an equation for converting chromatogram signals to corresponding molecular weights. Subsequently, dextrans' radius were calculated using Equations (4) and (5).

The dextran radius was determined using the Stokes-Einstein equation:[23]

$$a = \frac{k_B T}{6\pi\mu D} \qquad (4)$$

Where $k_B$ is Boltzmann's constant ($1.380649 \times 10^{-23}$ J/K), T is the absolute temperature, $\mu$ is the dynamic viscosity of the solution ($0.893 \times 10^{-9}$ m$^2$/s for water), and D is the dextran diffusion coefficient, calculated as:[23]

$$\log_{10}(D) = -8.1154 - 0.47752 \log_{10} MW \tag{5}$$

Where D is in m$^2$/s and MW is in Da.

The observed retention ($R_o$) was calculated using the following formula:

$$R_o = 1 - \left(\frac{C_p}{C_f}\right) \tag{6}$$

Where $C_p$ and $C_f$ represent the solute concentrations in the permeate and bulk feed, respectively.

A major challenge in using dextran retention tests for membranes is concentration polarization. When the feed solution contains solutes with different molecular weights, some solutes may accumulate at the membrane surface, creating a boundary layer where the solute concentration at the membrane surface ($C_m$) is higher than in the bulk feed ($C_f$)[43] (Supporting Fig. S2.2). This effect decreases the observed retention values compared to true retention values, the latter of which are relevant to transport models.

The actual retention ($R_a$) was calculated using:

$$R_a = 1 - \left(\frac{C_p}{C_m}\right) \tag{7}$$

To assess the concentration polarization tests were performed at three relatively low permeate fluxes: 1.3 L m$^{-2}$ h$^{-1}$, 2 L m$^{-2}$ h$^{-1}$, and 5 L m$^{-2}$ h$^{-1}$. The extent of concentration polarization was assessed using a concentration polarization factor (see Supporting Section S3). Differences between actual and observed retention were minimal at lower fluxes, and with 1.3 L m$^{-2}$ h$^{-1}$ providing lower retention for lower dextran molecular weights (see Supporting Fig. S2.3). Consequently, this flux was selected for our analysis to achieve more reliable retention values.

**Pore size distribution from retention test**

We inferred the surface pore-size distribution by fitting the actual sieving data $S_a$ with the hydrodynamic model of Bungay and Brenner.[4,44,45] A log-normal pore size distribution is commonly employed for UF membranes:[46]

$$n(r) = \frac{n_0}{r} \frac{1}{\sqrt{2\pi \ln{(1+(\sigma/\bar{r})^2)}}}$$
$$\exp\left\{-[\ln{(r/\bar{r})}]^2 \left[\frac{1+(\sigma/\bar{r})^2}{2\ln{[1+(\sigma/\bar{r})^2]}}\right]\right\} \tag{8}$$

Where $n(r)$ is the number of pores of radius r per unit area, $n_o$ is the total number of pores per unit area, $\bar{r}$ is the mean pore size, and $\sigma$ is the standard deviation of the log-normal distribution. Mean pore radius ($\bar{r}$) and standard deviation ($\sigma$) were determined by fitting models for the actual experimental sieving coefficients ($S_a$) values to the values for range of dextran radii.

Actual $S_a$ is measured as follows:

$$S_a = 1 - R_a \tag{9}$$

$R_a$ is the actual experimental retention value, after accounting for concentration polarization.

In the Bungay–Brenner model, the actual sieving coefficient of a membrane of pore radius r and a solute of radius a (measured using Equation (4)), $S_a(r,a)$, was determined using:

$$S_a(r,a) = \frac{S_\infty(r,a)\exp[Pe_m(r,a)]}{S_\infty(r,a) + \exp[Pe_m(r,a)] - 1} \tag{10}$$

Where $S_\infty(r,a)$ is referred to as the asymptotic sieving coefficient:

$$S_\infty(r,a) = \phi(r,a)K_c(r,a) \tag{11}$$

and $Pe_m(r,a)$, the Peclet number in a pore of radius $r$, is:

$$Pe_m(r,a) = \left(\frac{S_\infty(r,a)}{\phi(r,a)K_d(r,a)}\right)\left(\frac{J_v\delta_m}{D_\infty\varepsilon}\right) \tag{12}$$

Where $J_v$ is volumetric water flux (1.3 L m$^{-2}$ h$^{-1}$), $D_\infty$ is solute diffusivity (measured by Equation (5)), and $\phi(r,a)$ is the equilibrium partition coefficient of the solute between the fluid inside the membrane pore and the fluid adjacent to the membrane, measured as follow:

$$\phi(r,a) = \left[1 - \left(\frac{a}{r}\right)^2\right] \tag{13}$$

Surface porosity, $\varepsilon$ was estimated from uncoated samples. The membrane thickness, $\delta_m$ was estimated from cross-sectional SEM images (see Supporting Fig. S2.4). Expressions for $K_c(r,a)$ and $K_d(r,a)$ are listed at the end of Supporting Section S2.

To fit the experimental and theoretical sieving coefficients derived from the Bungay and Brenner model, $\sigma$ was expressed as $y \times \bar{r}$. Adjusting $y$ between 0.1 and 0.3 yielded relatively minor changes in the error between theoretical and experimental retention values, as shown in Supporting Fig. S2.5. Consequently, y was set to 0.2, consistent with values used in previous studies.[7]

## Results and Discussion

### Optimizing Image Segmentation and SEM Operation for Pore Analysis

The first step in quantitative analysis of pore structure is image segmentation, also known as binarization, which involves partitioning the gray-scale image into distinct segments that separate the pores from the surrounding polymer matrix, resulting in a binary image. Accurate image segmentation is essential, yet challenging[47]. Precise binarization ensures reliable pore size measurements, preventing underestimation or overestimation and thus enhancing the accuracy of further analysis.[48]

Fig. 1a illustrates the steps involved in precise pore segmentation, which include three main stages: pre-processing, binarization, and post-processing. In the pre-processing phase, the original gray-scale SEM image is refined to eliminate factors in pore segmentation such as blurring effects and artifacts. This involves enhancing contrast and reducing noise, resulting in a more uniform threshold value that improves pore segmentation accuracy. As shown in Fig. 1b, the gray values across a specific image distance are smoother after pre-processing. For image binarization, a machine learning algorithm—Trainable Weka Segmentation (TWS)[41]—was employed to classify the polymer matrix and pore regions, effectively segmenting the pores. A step-by-step example of the full workflow, including de-noise, contrast tuning, supervised TWS training with added labels, removal of tiny specks (<3–4 px) as noise, is provided in Supporting Fig. S1.3.The post-processing was followed by the watershed algorithm to minimize over-segmentation and prevent the misclassification of adjacent pores as a single entity, leading to an overestimated equivalent diameter. We applied the watershed step only when computing the pore size distribution (PSD), not during digital dilation or constriction, to preserve the original pore geometry. Splitting touching pores with watershed gives a PSD that better represents the pores controlling size-exclusion during filtration and therefore better reflects the membrane's rejection behavior. Although no standardized and fixed method exists for the binarization of gray-scale SEM images, we applied a consistent approach to pore segmentation across all SEM images in this study. Despite being largely user-independent, minor variations in thresholding decisions among different users can still slightly influence the segmentation outcomes.

To minimize artifacts associated with electron beam penetration depth at high acceleration voltages, we used a relatively low voltage of 1 kV for SEM imaging in this study. Our goal was to maximize surface sensitivity and the topographical contrast between pores and the polymer matrix for accurate segmentation. Monte Carlo simulation of electron trajectories shows that, at 1 kV, the BSEs are generated within 20 nm from the point of incidence (see Supporting Fig. S1. 3. a), so the SEs generated by BSEs (SE2s) contribute to the nanoscale topographical contrast. At 5 and 10 kV, the BSEs can travel a lateral distance of over 500 nm before escaping the sample (see Supporting Fig. S1. 3. b and c). In this case, SE2s do not contribute to the nanoscale topographical contrast in the pores, instead forming a diffuse background that reduces the overall contrast.[36] For most materials, the SE yield at 1 kV would be a few times higher than that of 10 kV, enabling a lower probe current for lower beam-induced contamination and radiation damage in the bulk polymer without increasing noise.[49] Modern SEM instruments can maintain high spatial resolution of 2 nm at 1 kV,[50] which would be less than the expected mean pore size of the polymer membranes.

To directly assess the impact of acceleration voltage on surface pore appearance and segmentation, a UF membrane (PW, 20 kDa MWCO, Veolia) was imaged and analyzed at 1, 5, and 10 kV acceleration voltages (Fig. 1 c). At lower acceleration voltages (1 kV), surface pores appear more pronounced, with less noise caused by coating particles, while at higher voltages, the contrast between pores and the surrounding matrix is reduced, and image enhancement (i.e., contrast adjustment and de-noising) plays a more critical role, causing extra challenges in pore segmentation and affecting segmentation accuracy. For each acceleration voltage, three SEM images were analyzed using image segmentation (Supporting Fig. S1. 5-7). Fig. 1d quantifies the observed surface porosity as a function of acceleration voltage, showing a marked decrease in porosity from 10.3% ± 0.3% at 1 kV to 6.9% ± 0.2% at 5 kV and 6.3% ± 0.4% at 10 kV. The decreased observed surface porosity at higher accelerations voltages is likely an artifact of the deeper penetration of the electron beam, which captures subsurface features and reduces contrast at the surface, leading to inaccurate pore segmentation. We note that many SEM analyses of porous membranes use acceleration voltages of 5–10 kV, which has likely contributed to underestimation of porosity (Supporting Table S1). Acceleration voltage also impacts the estimated pore size distribution (Fig. 1 e), with lower acceleration voltages yielding a broader distribution of pore sizes, reflecting a higher level of surface sensitivity, while higher voltages result in a narrower distribution, potentially underestimating pore dimensions due to deeper electron interaction volumes. We also performed simulations to illustrate the reduction of visibility of surface topography with acceleration voltage, considering a polymer sample with continuous Pt coating on the flat surface and hemispherical pores in the polymer without coating (Fig. 1f). The lower visibility at 5 and 10 kV for pores with diameters less than 5 nm can be attributed to higher random noise in the matrix region due to a decrease in the SE yield.

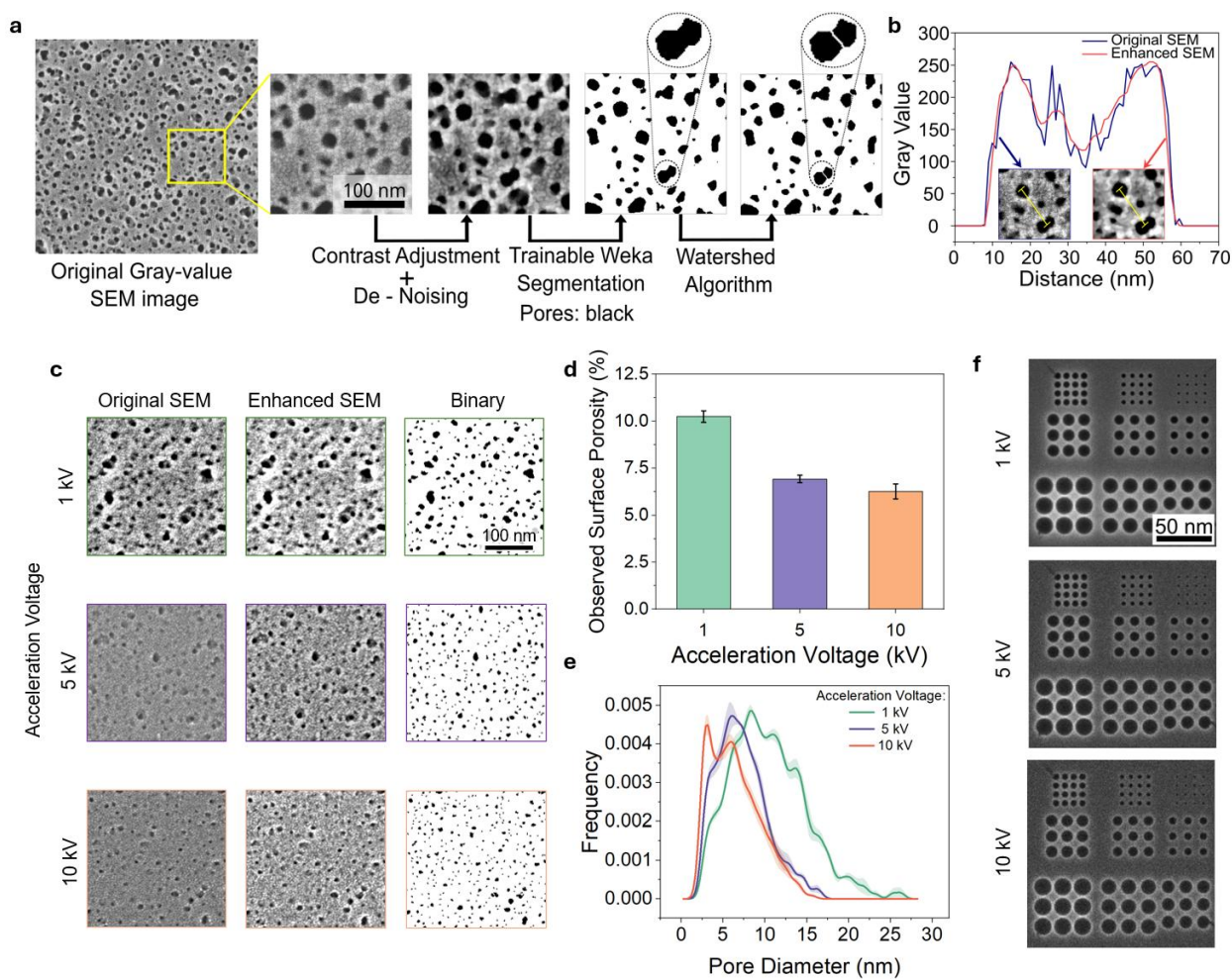

**Fig. 1. Image enhancement, segmentation, and impact of acceleration voltage. a,** Image segmentation procedure of a gray-value SEM image. The diagram describes a series of steps for optimal pore segmentation: pre-thresholding steps including contrast adjustment and de-noising, followed by pore segmentation using the trainable Weka segmentation plugin in Fiji (ImageJ). After binarization, a watershed algorithm is applied to separate connected pores. **b,** Gray value distribution before and after contrast adjustment and de-noising for the profile shown by the yellow line in the insets. **c,** Gray-value, enhanced and binary SEM images of the UF membrane at 1.7 nm system-measured Pt coating thickness at various acceleration voltages. **d,** Effect of acceleration voltage on the observed surface porosity measured from binary images from part (c). **e,** Effect of acceleration voltage on the area-normalized pore size distribution (bin width = 0.25 nm; Gaussian smoothing σ = 1). Frequency is the total area of pores of a given pore size divided by the total image area. **f,** Simulated SE images of a porous polymer (polymethylmethacrylate) with 1.7 nm Pt coating on the top surface at 1, 5, and 10 kV.

## Sputter Coating Reduces the Observed Surface Porosity and Pore Size

Given the low conductivity of most polymers, sputter coating with a conductive material like Pt is essential to obtain high-quality SEM images and to avoid charging during imaging.[51] We hypothesized that the deposition of metal particles during the coating process would partially cover the nano-sized pores, potentially leading to an underestimation of pore sizes (see Fig. 2a). To validate this hypothesis, the UF membrane and a PSF seawater RO support were sputter coated in a Leica ACE600 system with Pt at varying thicknesses, ranging from approximately 1.5 to 5 nm (Fig. 2b). The RO support layer was isolated from an AD-90 seawater RO membrane (Veolia) by selectively degrading the polyamide selective layer using sodium hypochlorite (see methods). The Leica ACE600 uses a quartz crystal microbalance to quantify the applied coating thickness; we call this thickness the system-measured thickness. We verified these coating thicknesses off-line using spectral reflectometry (Supporting Fig. S1. 8). SEM imaging was then performed to visualize the top surfaces of the membranes, and pores were segmented from gray-scale SEM images following the procedure described in Fig. 1. Visual observations clearly indicated that the pores become more constricted as the Pt coating thickness increases (Fig. 2b). For each coating thickness, three SEM images were analyzed using image segmentation (Supporting Fig. S1. 9-16).

To quantitatively analyze the impact of coating on the pore structure, the surface porosity was measured. Reductions of 54% in surface porosity for the UF membrane and 46% for the RO support were observed when we increased the Pt coating thickness from ~1.5 nm to ~5 nm (Fig. 2c). At a 90° coating angle (i.e., the stage tilted so that samples are parallel to the sputter target), the observed surface porosity for the UF membrane decreased from $12.9\% \pm 0.9\%$ to $5.8\% \pm 0.6\%$. For the RO support, the observed surface porosity decreased from $13.1\% \pm 0.6\%$ to $7.0\% \pm 0.2\%$. To investigate the effect of coating angle, the RO support was coated at two angles: 90° (tilted stage) and 65° (horizontal stage). At 65°, the observed surface porosity was slightly lower than at 90° (Supporting Fig. S1. 17). This difference can be attributed to increased particle self-shadowing at lower angles, where Pt particles are more likely to adhere to each other rather than conformally coat the surface.[52–54] The variation, however, remains relatively small, suggesting that while coating angle impacts porosity measurements, its influence is less pronounced compared to coating thickness.

To compare changes in the surface pore size distribution, the normalized pore area frequency at different coating thicknesses is presented in Fig. 2d. As the coating thickness increases, the area under the curve, representing surface porosity, decreases, and the curve shifts toward smaller pore diameters. This shift is more pronounced for the UF membrane, likely due to its smaller initial pore sizes, as shown in Fig. 2b. Qualitatively, the smaller pores appeared to promote sputter coating particle adhesion and the formation of bridges as the coating thickness increased. This process may have caused individual pores to split into two or more smaller pores after coating.

To further illustrate the impact of Pt coating on membrane surface pore size, we measured the water flux at 30 psi of the UF membrane and RO support before and after sputter coating. The UF membrane exhibited a flux of $712 \pm 31$ L m$^{-2}$ h$^{-1}$ in the uncoated state, which dropped to $108 \pm 21$ L m$^{-2}$ h$^{-1}$ after a system-measured 3.9 nm Pt coating. Similarly, the RO support showed a decrease from $288 \pm 16$ L m$^{-2}$ h$^{-1}$ to $107 \pm 14$ L m$^{-2}$ h$^{-1}$ after a system-measured 3.7 nm of Pt. This reduction in water flux is attributed to pore narrowing caused by Pt deposition, which partially blocks or constricts the membrane surface pores.

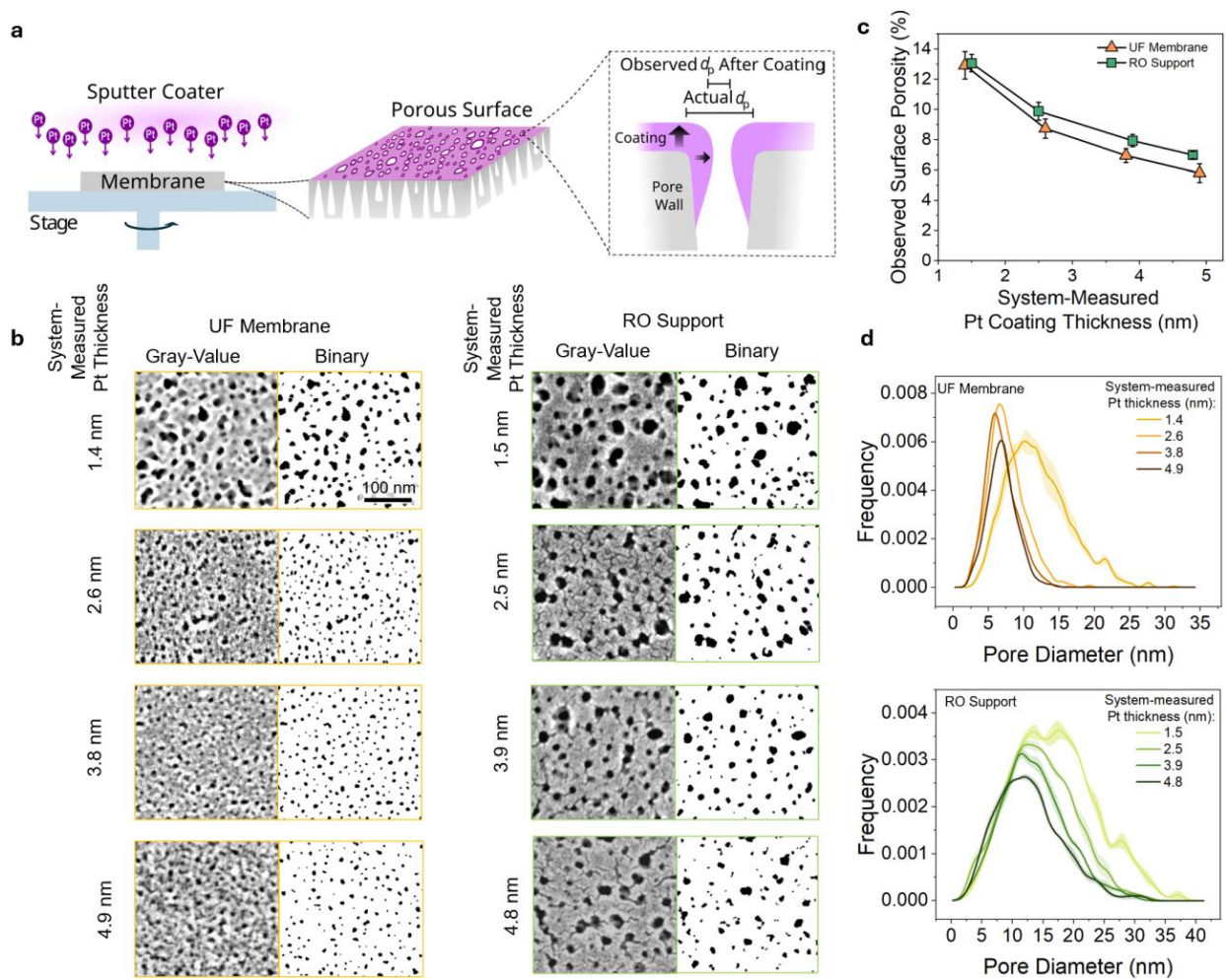

**Fig. 2. Influence of sputter coating thickness on surface porosity and pore size. a,** Schematic of the sputter coating process on a membrane (left), illustrating how the coating layer surrounds the pores and reduces the observed pore diameter (right). **b,** Gray-value and binary SEM images of the UF membrane and RO support at various system-measured coating thicknesses at 90° coating angle. **c,** Observed surface porosity of the UF membrane and RO support measured at various system-measured coating thicknesses. Error bars represent s.d. (n=3). **d,** Area-normalized observed pore size distributions of the UF membrane and RO support at different system-measured coating thicknesses

(bin width = 0.25 nm; Gaussian smoothing σ = 1). Frequency is the total area of pores of a given pore size divided by the total image area. Shaded areas represent s.d. (n=3).

**Digital Pore Constriction Reflects Effective Coating Thickness**

The above data clearly shows that metal coating layers lead to dramatic errors in surface porosity measurements. The ideal way to circumvent this would be conduct SEM directly on uncoated membranes; however, this is generally not possible due to charging effects during the imaging process.[51] Consequently, we developed a Python program to estimate the pore properties of the uncoated membrane. Starting with binary images at specific coating thicknesses, the program was used to digitally add or remove the coating layer around the segmented pores. In the first path, pores are constricted conformally in size, which simulates the addition of extra coating layer thickness. The second path digitally dilates pores conformally to approximate the surface pore structures of the uncoated membrane, from which PSDs and surface porosity can be quantified. This process is illustrated in Fig. 3a. Similar "erosion" functions are available in some pore analysis software, such as Fiji/ImageJ and Dragonfly (Objects Research Systems (ORS) Inc., Canada).

For the initial pore constriction, binary images with the thinnest coating thickness (1.4 nm for UF membrane and 1.5 nm for RO support) were digitally constricted in 1 nm intervals to generate binary SEM images representing greater coating thicknesses. A comparison between surface porosity measurements from binary images derived from experimentally coated SEM images and those digitally constricted is presented in Fig. 3b. For the initial step from ~1.5 nm coating to ~2.5 nm coating, the digitally constricted samples and experimental samples provide similar porosity values with a difference of ~0.8% for the UF membrane and ~1.1% for the RO support, suggesting that the conformal-deposition assumption is reasonable at low thickness. However, beyond ~2 nm, the conformal-deposition assumption appears to no longer be valid, as the digitally constricted SEM images showed a sharper reduction in surface porosity compared to the experimental samples. As coating thickness increases, particles tend to adhere to each other more, and due to the self-shadowing effect, vertical growth of particles dominates over radial growth inside the pores.[52–54] This effect would diminish the impact of extra coating on perceived pore size at higher coating thicknesses, as observed in our data.

To estimate porosity and pore size of membranes with zero coating thickness, the dilation of pores was guided by an "adjusted coating thickness." We obtain this adjusted coating thickness by matching each image's measured surface porosity to that of a digitally constricted image generated from the lowest coating baseline (1–2 nm); this adjusted coating thickness is then used for dilation to approximate the uncoated surface. For example, the system-measured coating thicknesses of 2.6 nm and 3.8 nm for UF membranes were adjusted in the model to 2.2 nm and 2.6 nm, respectively, while for the RO support, 2.5 nm and 3.9 nm were adjusted to 2.3 nm and 2.8 nm, respectively. At low thickness (1–2 nm), the conformal-deposition assumption is reasonable, and the system-measured thickness can be used for digital dilation without adjustment. Given that deviations escalate above ~3 nm, we restrict dilation inputs to the first three coating thicknesses.

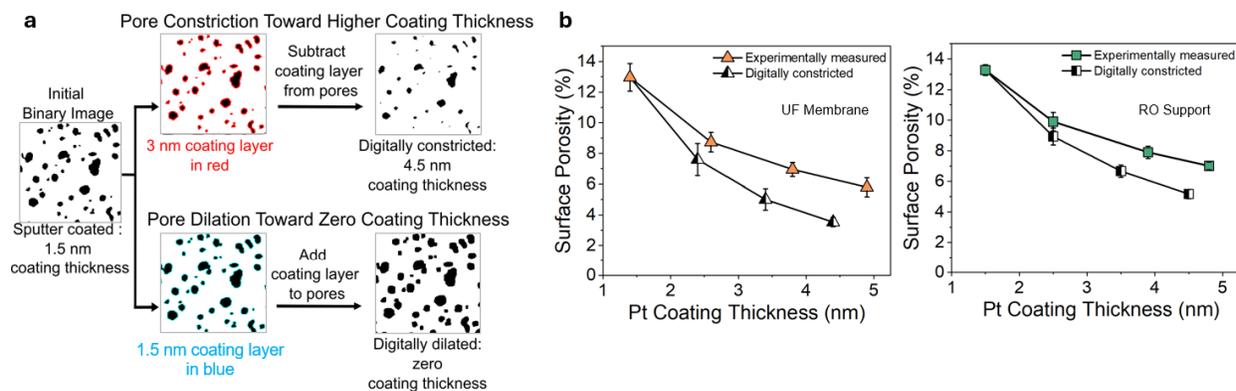

**Fig. 3. Digital pore constriction and pore dilation by varying coating layer thickness. a,** Schematic representation of pore constriction and dilation in a binary SEM image, achieved through the *in silico* addition or removal of a layer of defined width around the pores. **b,** Comparison of experimental and digitally constricted surface porosities at increasing coating thicknesses. Colored symbols represent the system-measured coating thickness, while the black symbols indicate the porosity values that resulted from sequentially increasing digital pore constriction, starting from samples with the thinnest Pt coating of ~1.5 nm. Errors represent s.d. (n=3)

## True Surface Porosity and Pore Size are Sharply Greater Than Directly Observed

To generate binary SEM images representing zero coating thickness, we digitally dilated the binarized pores by the adjusted coating thickness. The dilation was calibrated by strip width: we set it so that the mean coating thickness added to each pore matched the target coating thickness. This keeps the intended offset while preserving pore shape; implementation details and code are provided in Methods and Supporting Information. Fig. 4a and Fig. 4b illustrate the images of pore dilation for the UF membrane and RO support, respectively. We conducted pore dilation from multiple coating thicknesses to validate the approach, as the estimated uncoated pore structure should be nearly equivalent even if different starting thicknesses are used. Visually, the pore structures at zero coating thickness obtained from the three adjusted coating thicknesses appear similar, especially for the RO support. For quantitative comparison, pore size distributions and porosity were then measured (Fig. 4c and Fig. 4d). The porosity values at zero coating thickness derived from the three adjusted coating thicknesses fell within a reasonably close range, 22.7%–23.9% for the UF membrane and 19.7%–21.0% for the RO support.

Fig. 4c and Fig. 4d show the pore size distribution of the UF membrane and RO support at zero coating thickness. For the RO support, the three pore size distributions are similar, suggesting that for this sample, the methodology reliably produced statistically equivalent uncoated pore structures from different starting points. For the UF membrane, the pore size distribution from dilation from the lowest coating thickness deviated substantially from the other two distributions. There are two likely reasons for this deviation: first, UF membrane pores are small and close together, so at thicker Pt some small features drop below visibility or appear partly merged (see Fig. 2), and dilation cannot fully recover them; second, UF pore sizes approach the Pt grain size (2–3 nm) which complicates the assumption of conformal coating that is required for pore dilation. Accordingly, to better estimate the uncoated sample, we rely on the thinnest available start (1-2 nm at 1 kV), which is necessary for the UF membrane and would be neutral or beneficial for all membranes generally.

To further validate the pore dilation method on a membrane with well-defined and uniform pore structure, we tested a commercial polycarbonate track-etched (PCTE) membrane from Sterlitech (0.01 μm pore size). After Pt coating and digital pore dilation, the corrected mean pore diameter (10.5 ± 0.5 nm) and porosity closely matched the nominal specifications, further confirming the reliability of our correction approach. Full results are provided in Supporting Fig. S1.17.

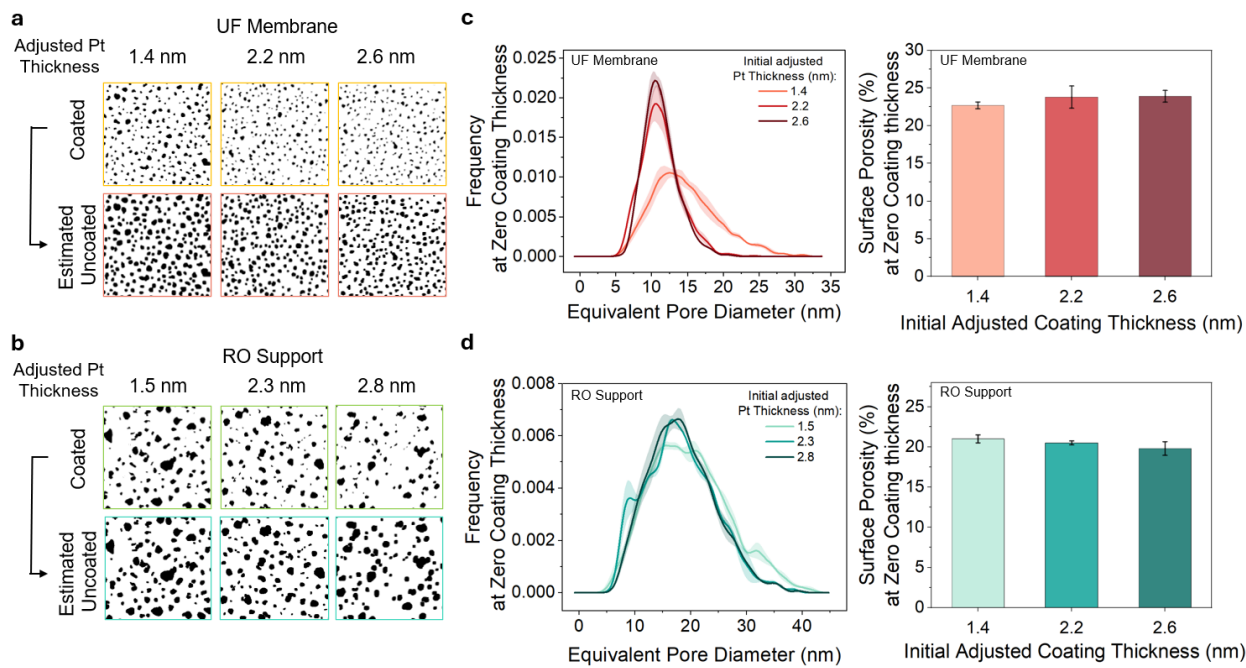

**Fig. 4. Estimating porosity and pore size distributions of uncoated membranes via digital pore dilation. a-b,** Visualization of pore dilation to approximate the true pore structure of the (a) uncoated UF membrane and (b) uncoated RO support. **c-d,** Pore size distributions and surface porosity at zero coating thickness, with different colors representing the initial coating thickness for the (c) UF membrane and (d) RO support (bin width = 0.25 nm; Gaussian smoothing σ = 1). Frequency is the total area of pores of a given pore size divided by the total image area. Shaded areas represent s.d. (n=3).

Fig. 5 compares the observed porosity and pore size distributions of samples with a ~4 nm system-measured coating thickness to their uncoated counterparts approximated through digital pore dilation using the lowest coating thickness images (1-2 nm), highlighting the dramatic impact of coating on pore size analysis. For a system-measured coating thickness of approximately 4 nm (corresponding to a ~3 nm adjusted thickness) in both the UF membrane and RO support, the true porosity values are 3.4-fold and 2.7-fold greater than the experimentally observed porosities, respectively (see Fig. 5a). UF membrane shows a slightly larger underestimation between the coated and estimated-uncoated values, consistent with its smaller, closely spaced pores being more prone to pore narrowing at higher Pt thickness (see Fig. 2). These findings suggest that nearly all porosity values reported in the literature for nanoporous membranes are likely substantially underestimated, as the effects of sputter coating on membrane characterization have thus far been neglected. Additionally, a noticeable shift in pore size distribution is observed when comparing the coated samples with a 4 nm system-measured coating thickness and uncoated samples. The uncoated samples display a significantly larger area under the curve, indicating a higher frequency of larger pores, and the distribution curve shifts to the right, reflecting the presence of these larger pores. This shift results in mean pore diameters of 14.7 nm and 6.8 nm for the uncoated and coated UF membrane, respectively, and 19.8 nm and 13.6 nm for the uncoated and coated RO support, respectively. This noticeable shift in the area under the curve and the position of the pore size distribution highlights the significant effect of coating-induced artifacts on the analysis of nano-sized pores.

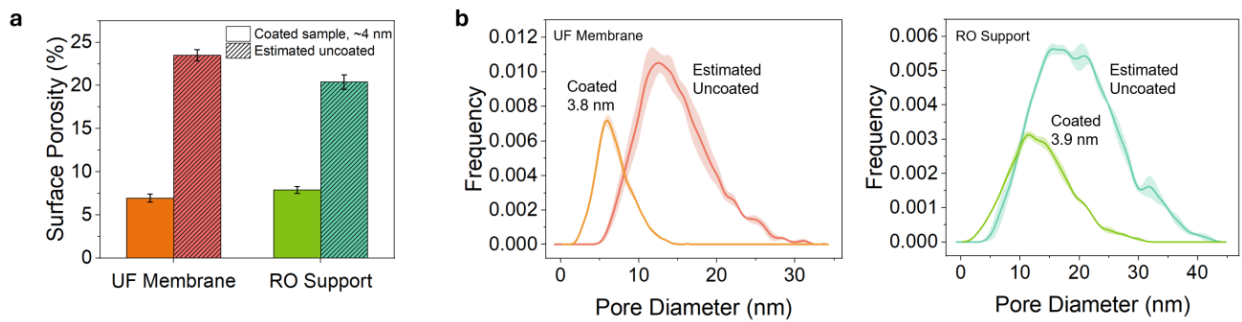

**Fig.5. Quantitative comparison of observed pore structure of coated samples and estimated pore structure of uncoated samples. a,** Surface porosity of coated samples with a 4 nm system-measured coating thickness and the estimated uncoated samples generated through digital pore dilation using the lowest coating thickness images (1-2 nm). **b,** Pore size distribution of coated samples with a 4 nm system-measured and estimated uncoated samples for UF membrane and RO support (bin width = 0.25 nm; Gaussian smoothing σ = 1). Frequency is the total area of pores of a given pore size divided by the total image area. Shaded areas represent s.d. (n=3).

## Estimated Uncoated Pore Size Distributions Are Similar to Those Modeled from Filtration

As discussed earlier, membrane pore size distributions are frequently estimated through modeling retention data of macromolecules. One challenge with modeling retention data is the existence of multiple models, all of which require simplifying assumptions that lead to varying results[4]. For example, the commonly used Zeman-Wales model assume that pores are cylindrical and convection dominates, allowing the surface pore size distribution to be the only input [8,55,56]. In contrast, the Bungay and Brenner model incorporates diffusive contributions but requires additional parameters such as surface porosity and pore length[4,44]. In NIPS-based membranes, pore length must be crudely estimated, as the fine, tortuous pores in the selective layer gradually transition into larger pores in the support structure. Other challenges and limitations include accurately accounting for concentration polarization, which is mitigated by operating at low water flux [57], and the assumption that linear polysaccharides behave as uncharged, diffusing spheres.

With these limitations in mind, we conducted filtration experiments using linear dextran polysaccharides on UF membranes and RO supports. The Zeman-Wales model's validity is determined by a wall Peclet number ($Pe_m > 1$), which compares convective and diffusive transport. Using Equation (12), we found that for dextran molecules with retention below 99% (Fig. 6 a and b), the Peclet number remained below 1 (see Supporting Fig. S2.6), indicating diffusion-dominated transport. The Python code used to compute and plot $Pe_m$ is provided in Supporting Information Section S4.3.

Our retention test was intentionally conducted at low flux to mitigate concentration polarization (CP). While this approach helps reduce CP, it also further lowers the $Pe_m$. Consequently, the Zeman-Wales model was unsuitable for modeling retention data in this case. To capture solute transport behavior across the full range of relative dextran-to-pore size ratios ($\lambda$ = a/r, from 0 to 1), we employed the Bungay and Brenner model, which provides an appropriate hydrodynamic formulation for the entire range of solute-to-pore size ratios. [4,57]

To compare SEM-derived pore-size distributions with filtration consistently, we report pore size as the maximum inscribed circle (MIC), the largest circle entirely inside each pore opening (see Supporting Fig. S2.7). MIC captures the narrowest aperture that governs solute exclusion and maps directly to the size ratio $a/r$ in the Bungay–Brenner model; in contrast, the circular-equivalent diameter (EqD) is an area-based estimate that can overstate size for irregular or throat-constricted pores.

Observed sieving data was first corrected for concentration polarization to obtain actual sieving data. Then, we fitted the Bungay–Brenner model to the actual sieving data from low-flux dextran-retention experiments. The fits were constrained by surface porosity ($\varepsilon$) and selective-layer thickness ($\delta_m$) obtained from SEM. The Python code was used to fit model to retention data is provided in Supporting Information Section S4.4. The model reproduces the actual retention curves well (Fig. 6 a,b; dashed red). The modeled log-normal PSDs from retention data also matched the MIC-based SEM PSDs in both mean and distribution shape (Fig. 6 c,d). The fitted model mean PSDs were 9.0 nm (UF membrane) and 12.7 nm (RO support), close to the SEM MIC means of 10.3 nm and 13.0 nm, respectively. Repeating the comparison with circular-equivalent diameters (EqD) shifts the SEM-derived PSDs to larger apparent diameters, as expected for an area-based metric (Supporting Fig. S2.8), with less agreement with PSDs estimated from retention data.

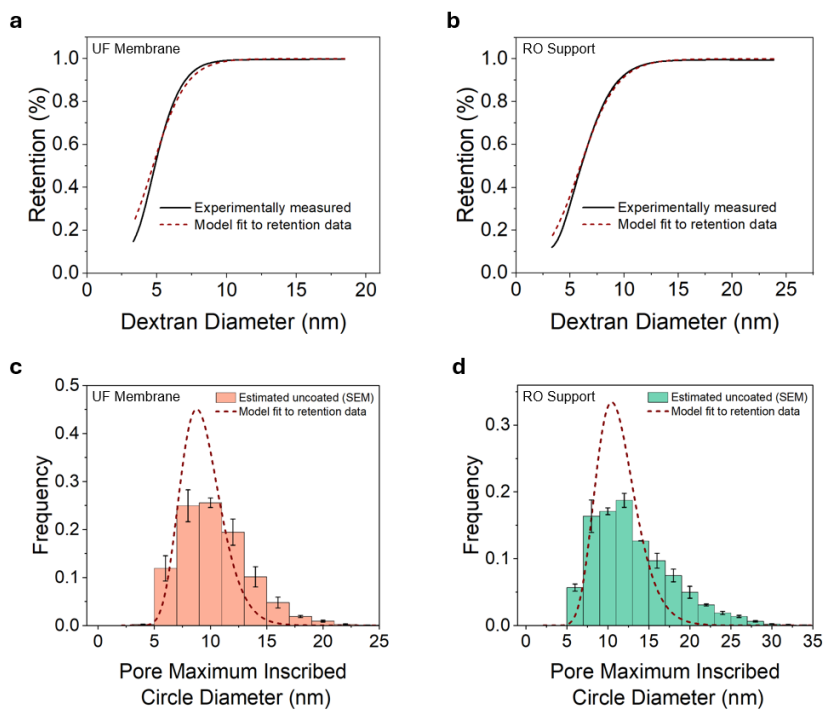

**Fig. 6. Comparison between SEM pore analysis and modeled pore sizes from fitting to filtration. a** and **b**, Actual retention values of dextran as a function of dextran diameter. The solid black line represents experimentally determined values, while the dashed red line denotes the Bungay and Brenner model fit for the UF membrane and RO support. **c** and **d**, Pore-size distributions expressed as maximum inscribed circle (MIC) diameter from SEM after digital dilation to estimate the uncoated surface, starting from the lowest coating thickness images (1-2nm), overlaid with the model-inferred PSDs from the fits in a,b (red dashed) for UF membrane and RO support. The frequency represents the number of pores at each pore size divided by the total number of pores. The model-fitted pore size distributions are normalized so that the area beneath the curves matches that of the estimated uncoated pore size distribution.

## Implications and Outlook for Structure/Transport Relationships in Nanoporous Membranes

Our findings have important implications for researchers and engineers relying on SEM imaging for pore analysis. As the membrane science field typically uses high acceleration voltages and has not accounted for metal coating thicknesses in the past, surface porosities and surface pore sizes have likely been sharply underestimated for nearly all nanoporous membranes. In other words, it is likely that UF membranes and RO supports generally have surface porosities substantially greater than 10%. To further test this hypothesis, we hope that other researchers will adopt the method in this paper to approximate the surface pore structures of other uncoated nanoporous membranes. Our results suggest that the assumption of conformal coating is most appropriate for membranes with larger pores—evidenced by the overlapping uncoated pore size distributions for the RO support—and for the thinnest coating thicknesses (1–2 nm). For such low coating thicknesses, researchers could simply dilate segmented pores by the measured coating thickness. However, a similar analysis as done in this study with multiple coating thicknesses may be needed for other systems (e.g., use of non-Pt targets).

In terms of structure/transport relationships, the results in this study provide a revised perspective on the surface pore structures of nanoporous membranes—namely that surface pores make up a sharply greater area of membrane surfaces and that these pores are larger than previously understood (Fig. 5). These findings open many research questions about transport behavior through these materials. As described earlier, UF membrane transport is typically modeled with major simplifications, such as each pore being circular and having a uniform pore diameter throughout its depth. These and other assumptions are also built into many of the pore characterization methods described earlier, such as the fitting of particle retention experiments to estimate a log-normal pore size distribution.[25,55]

Advancements in pore characterization by electron microscopy—i.e., the present study and recent 3-D tomographic studies[58–61]—should enable even more comprehensive analyses of the applicability of classical models to the tortuous pore structures of the NIPS-based membranes used in the real world. With accurate 2-D and 3-D pore structures in hand, future studies could compare classical models using simplified inputs from 2-D analyses (i.e., pore radius and length), computational fluid dynamics (CFD) simulations using 3-D structures, and experimental filtration data. We note that the coating layer impact described in the present study has also likely lead to underestimations of 3-D porosity and pore sizes in state-of-the-art tomographic 3-D structure analyses that rely on SEM, such as focused ion beam SEM (FIB-SEM) and serial block face SEM (SBF-SEM). Conductive coatings are needed for each 2-D slice in these methods, and as with top-view analyses, coating layer thickness has similarly not been accounted for in these studies.[58,60] Coating layer artifacts may explain why hydraulic permeabilities of a commercial UF membrane were recently underestimated by CFD simulations using a 3-D structure from FIB-SEM.[58,61]

For transport in TFC membranes, much remains unknown about the impact of support layer pore structure on performance. The surface pore structures estimated using our method may be directly relevant. A modeling study by Ramon and Hoek, for example, used a simplified structure where a smooth active layer rested on top of a circular pore, wherein flow resistance through the pore was assumed to be negligible.[9] NF membranes typically have a smooth active layer reminiscent of this model[62]. If a smooth active layer of a TFC membrane similarly rests above the surface pore bodies of the support layer, then the pore sizes estimated through SEM should be applicable for transport analyses using such simplified models. In contrast to NF membranes, the active layers of RO membranes have a "crumpled nanofilm" morphology in which the 3-D structure of the active layer is highly heterogeneous.[63,64] Combinations of our method to analyze support layer pore structure and recent advanced tomographic reconstructions of active layers[65–68] could yield the ability to comprehensively analyze transport through TFC membranes used in RO.

## Acknowledgements

Funding for this work came from the University of Toronto WaterSeed Program and from the Natural Sciences and Engineering Research Council of Canada (NSERC) through the Alliance Missions Program (ALLRP 570714-2021). We thank Salvatore Boccia from the OCCAM facility for assistance with SEM imaging and the Abdul Wasay from the CRAFT facility for their assistance with sputter coating. We also acknowledge Prof. Molly Shoichet and Timothy Cheung for their help with GPC testing. Additionally, we appreciate the support and feedback from Andrew Back and Steve Harrold from Veolia Water Technology & Solutions.

**Declaration of generative AI in scientific writing**

During the preparation of this work, the authors used ChatGPT (OpenAI) to assist with language refinement (e.g., rewording, and improving readability) and to suggest improvements to analysis scripts (e.g., commenting, error handling and robustness checks). All AI-assisted text and code suggestions were reviewed, tested, and edited by the authors, who take full responsibility for the final manuscript and results.

**Supporting Information:**

**Section S1. SEM Imaging**

Table S1. Summary of membrane support materials, fabrication methods, sputter coating conditions, SEM acceleration voltages, and surface porosity measured via top surface SEM.

| Membrane Support Material | Fabrication Method | Coating Material/ Thickness (nm) | Acceleration Voltage (kV) | Surface Porosity (%) | Reference |
|---|---|---|---|---|---|
| **Polysulfone (PSF)** | Hand-cast | Pt/3 | 10 | 1.2 -13.3 | [55] |
| **Polyethersulfone (PES)** | Commercial | Pt/ NA | 5 | 10 | [29] |
| **Polyvinilidenefluoride (PVDF)** | Commercial | Pt-Pd/ NA | 15 | 2.8 | [30] |
| **Polycarbonate (PC)** | Commercial | NA | NA | 10.5 | [69] |
| PVDF + PMMA | Hand-cast | Pt/NA | 10 | 0.8-14.7 | [70] |
| **Polysulfone (PSF)** | Hand-cast | Au/2 | 5 | 0.99-1.38 | [32] |
| **Polysulfone (PSF)** | Hand-cast | Au/NA | NA | 0.33-1.33 | [34] |
| **Polysulfone (PSF)** | Hand-cast | Au/NA | NA | 0.24-4.3 | [33] |
| **PAN** | Hand-cast | Au-Pd/NA | 10 | 4-10 | [71] |
| PES/SPSf + nano-CaCO$_3$ | Hand-cast | Au/NA | 5 | 5-15.4 | [35] |
| **PVC** | Hand-cast | Au/2 | 5 | 1.01-11.5 | [72] |

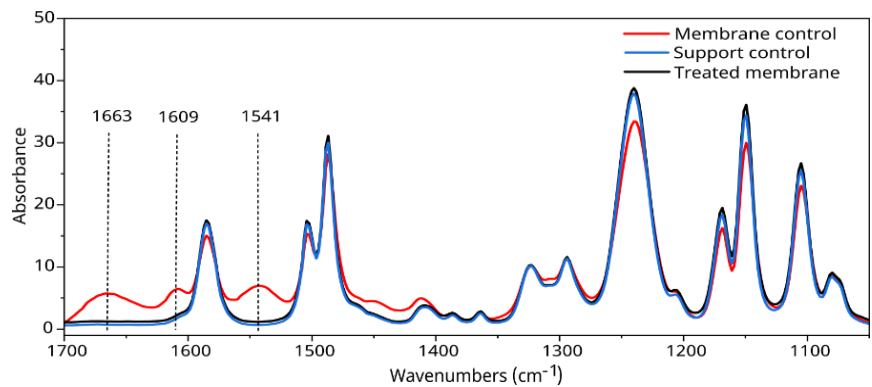

**Supporting Fig. S1.1. ATR-FTIR spectra of the treated membrane, control membrane, and control support layer.** The NaOCl treatment effectively removed the PA layer. The wavenumbers at 1663 cm⁻¹, 1609 cm⁻¹, and 1541 cm⁻¹ correspond to characteristic vibrations of the polyamide (PA) layer: 1663 cm⁻¹ is attributed to C=O stretching (amide I), 1609 cm⁻¹ to aromatic ring C=C stretching, and 1541 cm⁻¹ to N-H bending and C-N stretching (amide II). The control membrane refers to the untreated AD-90 RO membrane with a PA selective layer. The support control layer refers to a porous polysulfone ultrafiltration membrane.

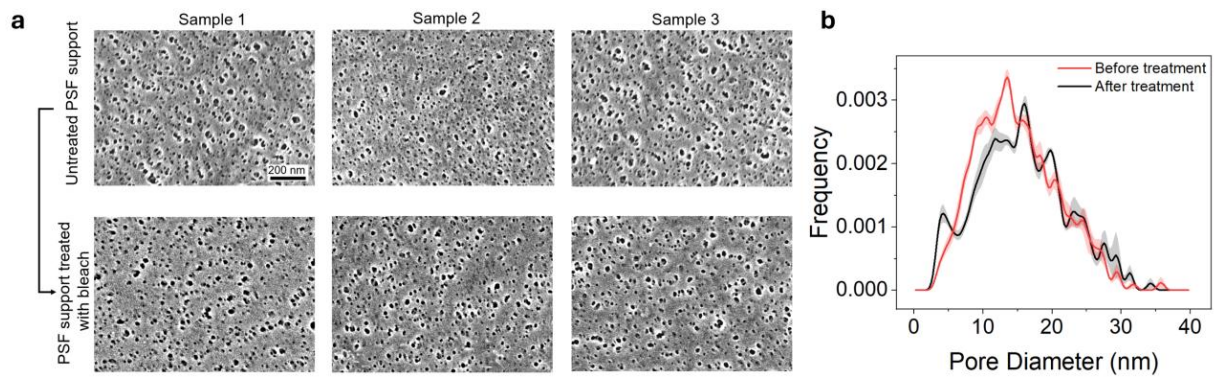

**Supporting Fig. S1. 2. Effect of Bleach Treatment on PSF Support Surface Morphology and Pore Size Distribution.** (a) Surface SEM images of polysulfone (PSF) supports fabricated in-lab using the nonsolvent-induced phase separation (NIPS) method with 17 wt% Udel® polysulfone (PSF) in N-methyl-2-pyrrolidone (NMP), cast at a thickness of 6 mil onto a commercial nonwoven backing fabric and immersed in a water coagulation bath. The top row shows untreated supports; the bottom row shows supports treated with sodium hypochlorite under the same conditions used for active layer removal. All samples were sputter-coated with 1.6 nm of Pt at a 90° coating angle, and imaging magnification was 100k. (b) Area-normalized pore-diameter distributions from n = 3 images per condition (bin width = 0.25 nm; Gaussian smoothing σ = 1). No significant changes in pore morphology or diameter were observed, confirming that the bleach treatment does not alter the support structure.

**Step A:**
**De-noise**

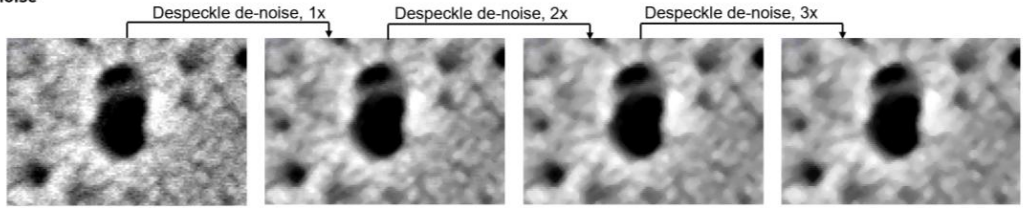

Despeckle de-noise, 1x    Despeckle de-noise, 2x    Despeckle de-noise, 3x

**Step B:**
**Contrast enhancement (Window/level in Fiji)**

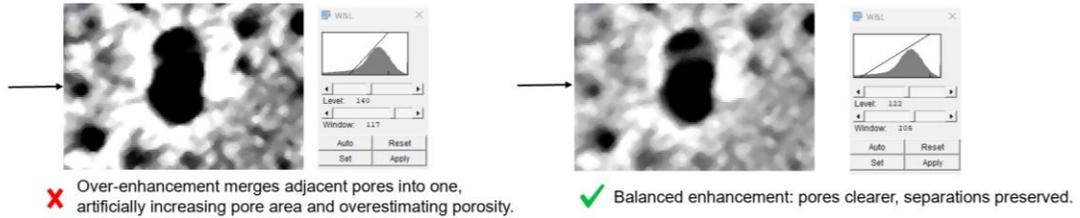

✗ Over-enhancement merges adjacent pores into one, artificially increasing pore area and overestimating porosity.

✓ Balanced enhancement: pores clearer, separations preserved.

**Step C:**
**Supervised binarization in Trainable Weka Segmentation (TWS) in Fiji**

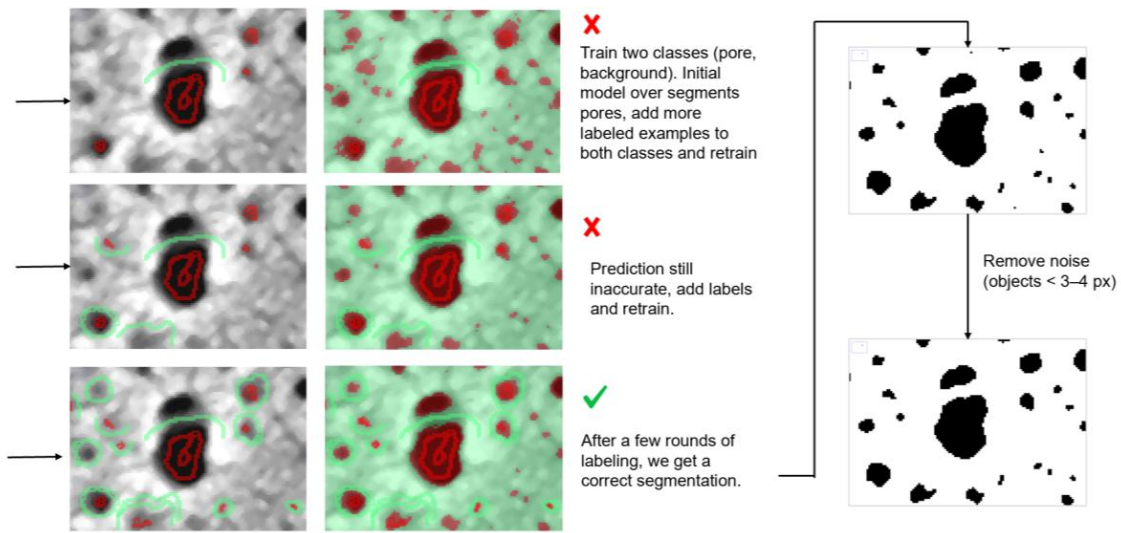

✗ Train two classes (pore, background). Initial model over segments pores, add more labeled examples to both classes and retrain

✗ Prediction still inaccurate, add labels and retrain.

✓ After a few rounds of labeling, we get a correct segmentation.

Remove noise (objects < 3–4 px)

**Supporting Fig. S1. 3. SEM pore-segmentation workflow (de-noise → contrast → supervised segmentation → clean-up).** Step A, De-noise. The raw SEM image is lightly filtered in Fiji (Despeckle) two to three times to suppress salt-and-pepper noise. Step B, Contrast adjustment. Window/Level in Fiji is used to improve pore–wall contrast. Left: settings that are too strong merge close pores and overestimate porosity. Right: balanced settings make pores clearer while keeping them separated. Step C, Supervised segmentation in Fiji (Trainable Weka). Two classes are trained (pore, wall). Top-to-bottom: initial model over-calls pores; more labeled examples are added, and the model is retrained; after a few rounds the segmentation is correct and repeatable. Then a binary mask is obtained. Small specks are removed as noise (objects < 3–4 px).

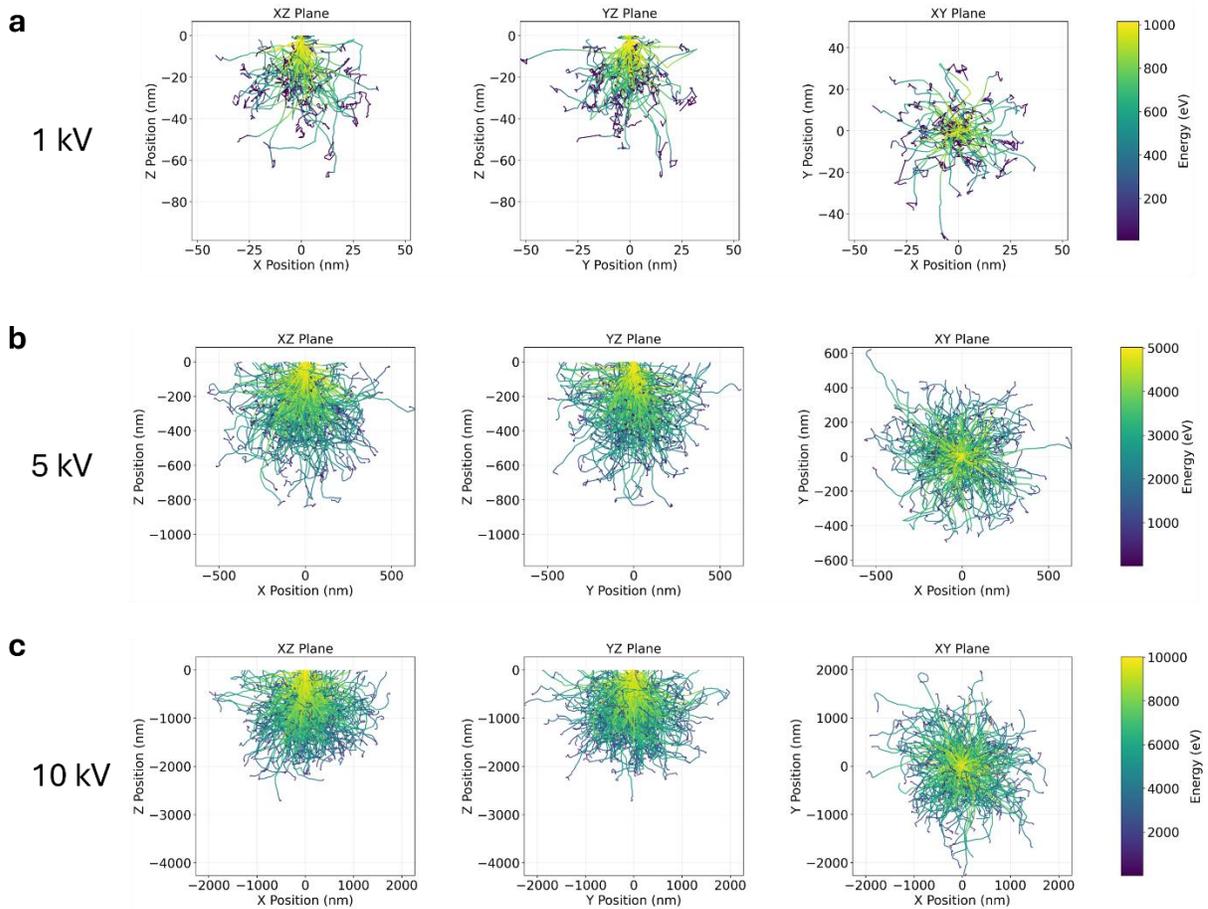

**Supporting Fig. S1. 4. Electron Interaction Volume in Poly(methyl methacrylate) (PMMA) with Pt Coating at Varying Acceleration Voltages.** 2D projections of the interaction volume of the incident electrons at (a) 1 kV, (b) 5 kV, and (c) 10 kV on PMMA with 1.7 nm Pt coating. The simulations were performed using an open-source Monte Carlo software package called Nebula, with 500 incident electrons travelling in -z direction towards the sample for each acceleration voltage. The z-direction refers to depth of penetration into the sample, with z = 0 referring to the top surface. The x-y plane is the lateral dimensions, with (0,0) referring to the central focus of the beam (i.e., point of incidence). At 1 kV, the BSEs can escape from the surface of the sample up to 25 nm from the point of incidence. This distance increases to 400 nm at 5 kV and 1.8 μm at 10 kV. Since the $SE_2$s generated by BSEs carry the same topographic information, a lower acceleration voltage would allow more localized generation of $SE_2$s for nanoscale topography.

UF membrane, imaged at acceleration voltage of 1 kV
System measured coating thickness : 1.7 nm

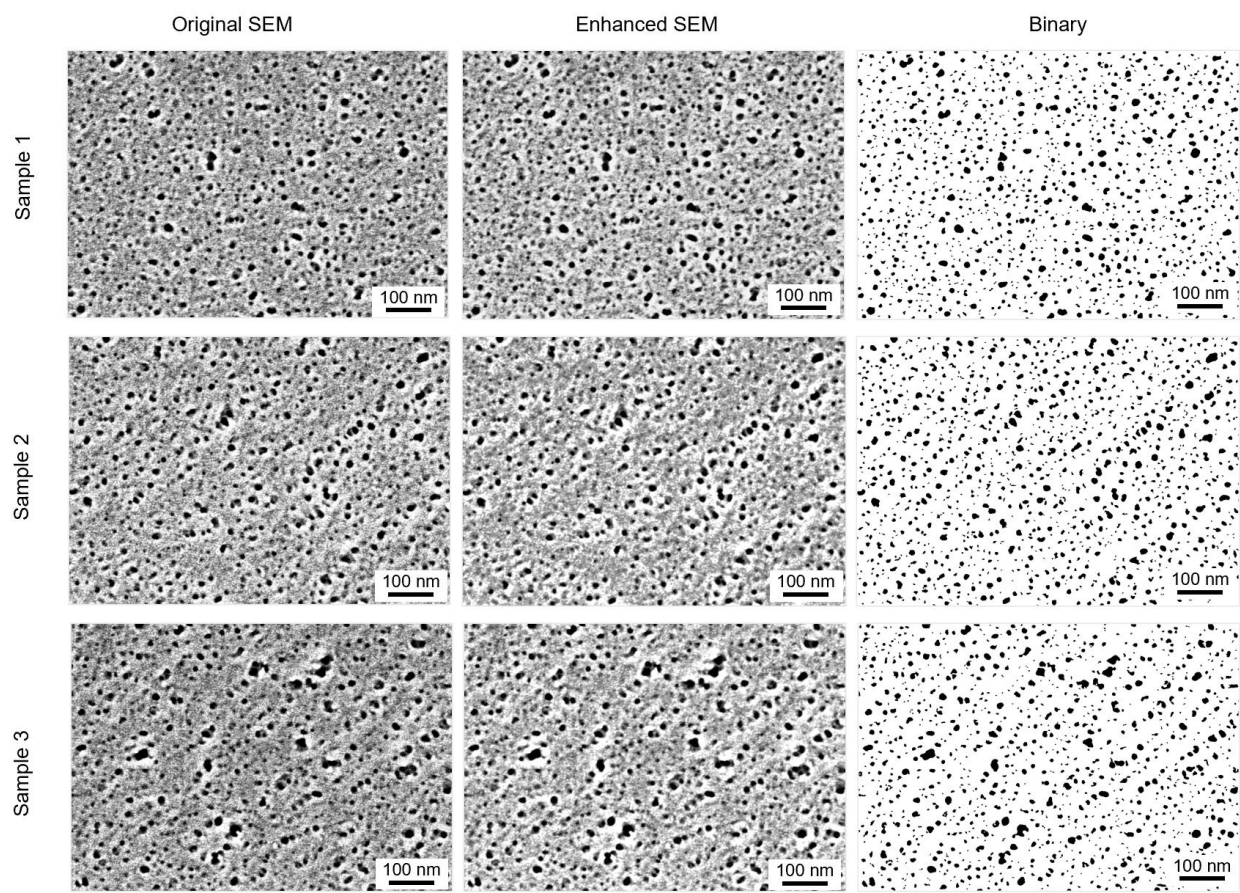

**Supporting Fig. S1. 5. Raw SEM images, enhanced SEM images, and corresponding binary images of the UF membrane at 1 kV acceleration voltage.** System-measured coating thicknesses: 1.7 nm. Imaging condition: 150k magnification.

UF membrane, imaged at acceleration voltage of 5 kV
System measured coating thickness : 1.7 nm

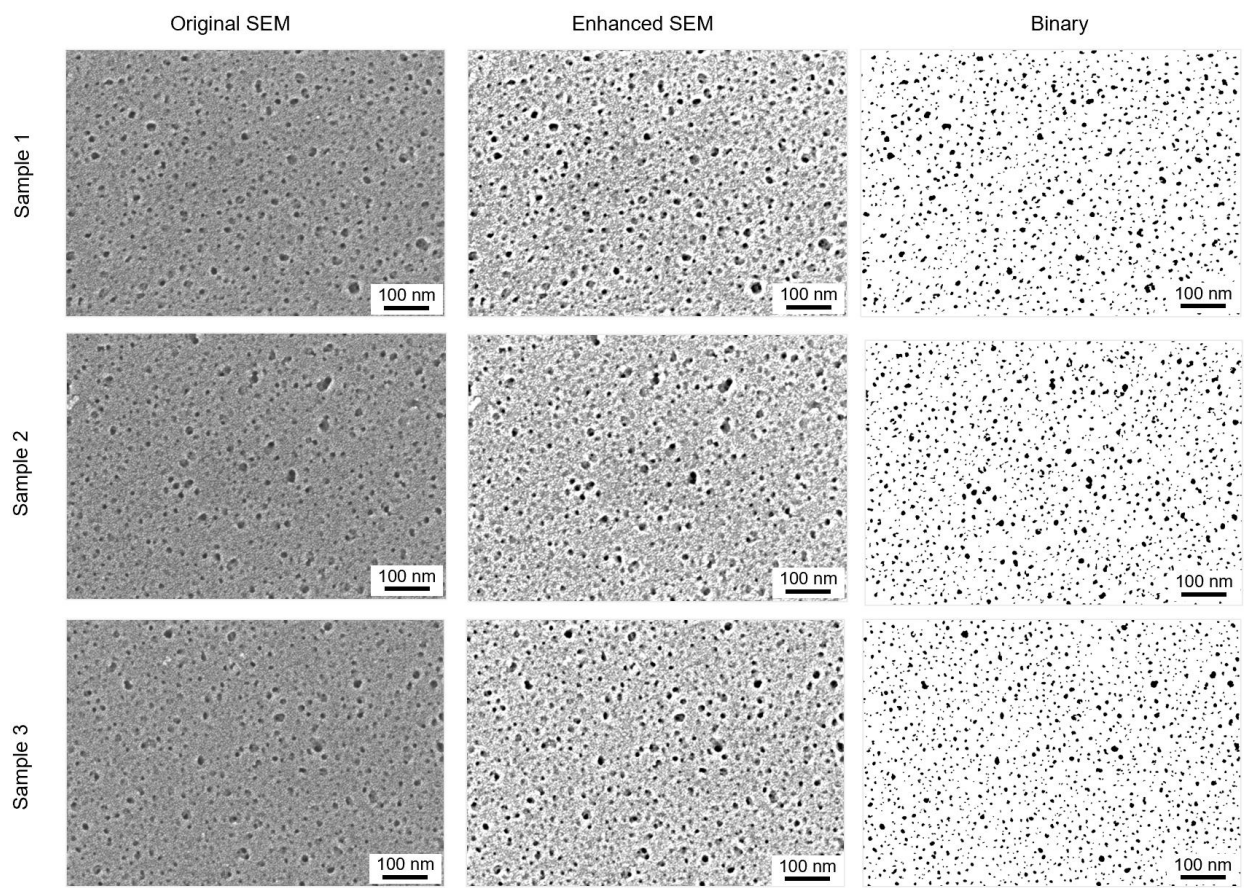

**Supporting Fig. S1. 6. Raw SEM images, enhanced SEM images, and corresponding binary images of the UF membrane at 5 kV acceleration voltage.** System-measured coating thicknesses: 1.7 nm. Imaging condition: 150k magnification.

UF membrane, imaged at acceleration voltage of 10 kV
System measured coating thickness : 1.7 nm

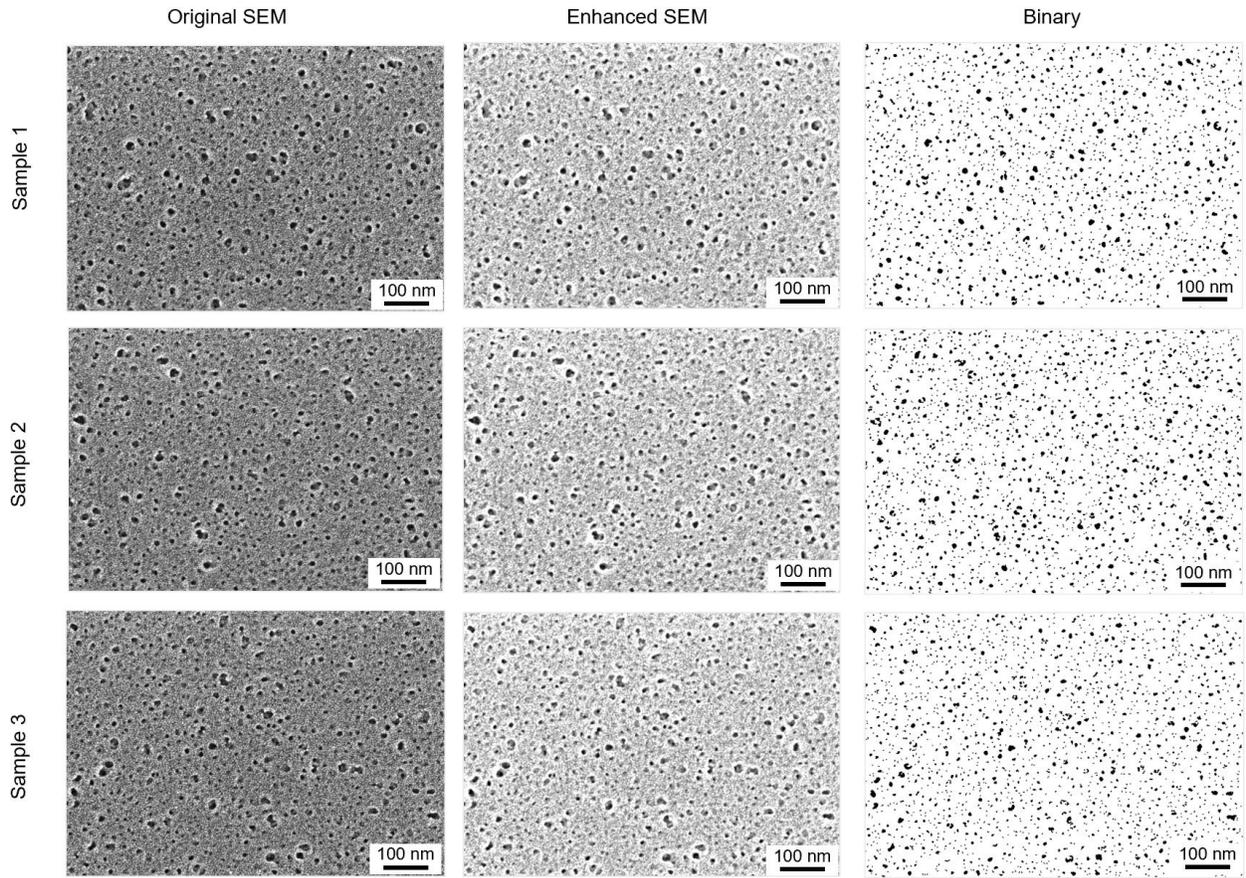

**Supporting Fig. S1. 7. Raw SEM images, enhanced SEM images, and corresponding binary images of the UF membrane at 10 kV acceleration voltage.** System-measured coating thicknesses: 1.7 nm. Imaging condition: 150k magnification.

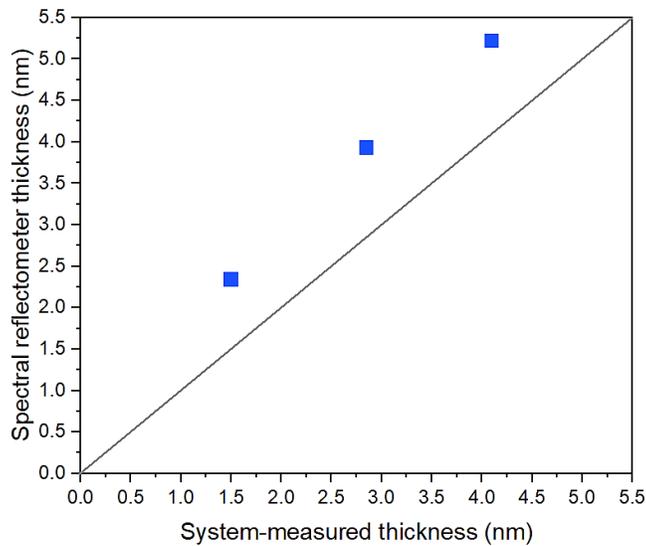

**Supporting Fig. S1. 8. Comparison of platinum (Pt) coating thickness measured using the sputter coating system (system-measured) and spectral reflectometer (F20-UV, Filmetrics, KLA Corporation).** The solid line represents a 1:1 slope. The measured values between the two systems show reasonable agreement. The system-measured thickness from the microbalance within the Leica ACE600 sputter coater was used in the analysis.

UF Membrane, imaged at acceleration voltage of 1 kV
System measured coating thickness : 1.4 nm

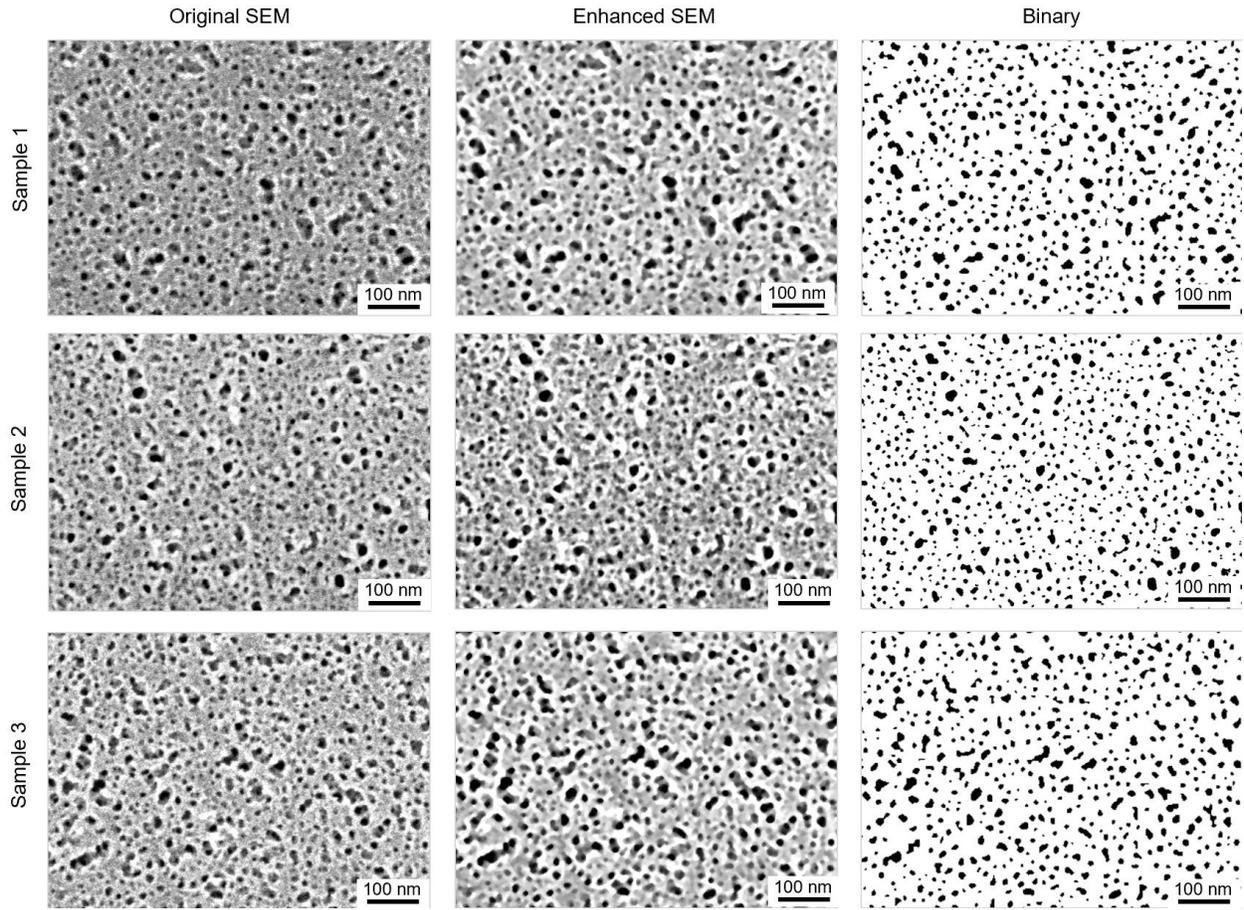

**Supporting Fig. S1. 9. Raw SEM images, enhanced SEM images, and corresponding binary images of the UF membrane at 1.4 nm system-measured coating thicknesses.** Imaging conditions: 1 kV acceleration voltage and 150k magnification.

UF Membrane, imaged at acceleration voltage of 1 kV
System measured coating thickness : 2.6 nm

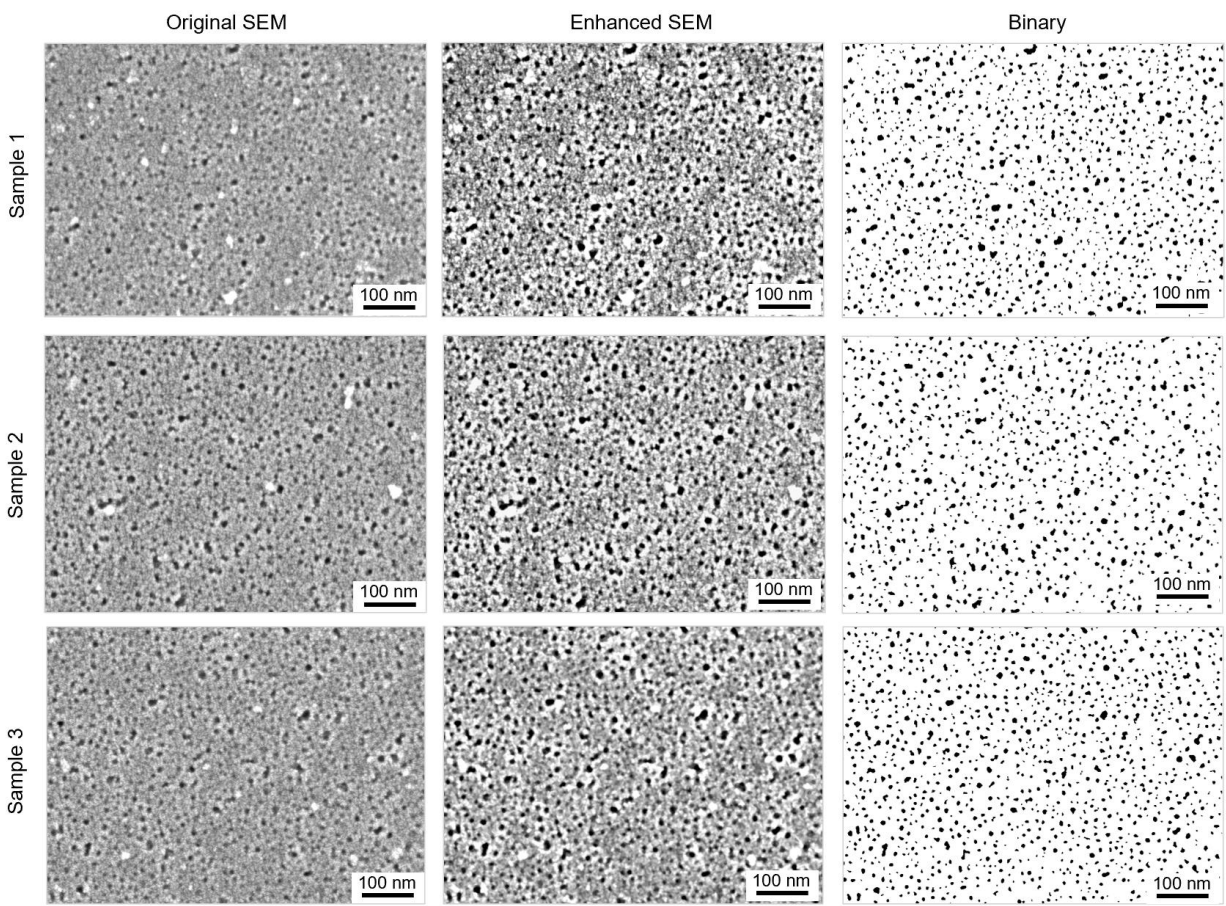

**Supporting Fig. S1. 10. Raw SEM images, enhanced SEM images, and corresponding binary images of the UF membrane at 2.6 nm system-measured coating thicknesses.** Imaging conditions: 1 kV acceleration voltage and 150k magnification.

UF Membrane, imaged at acceleration voltage of 1 kV
System measured coating thickness : 3.8 nm

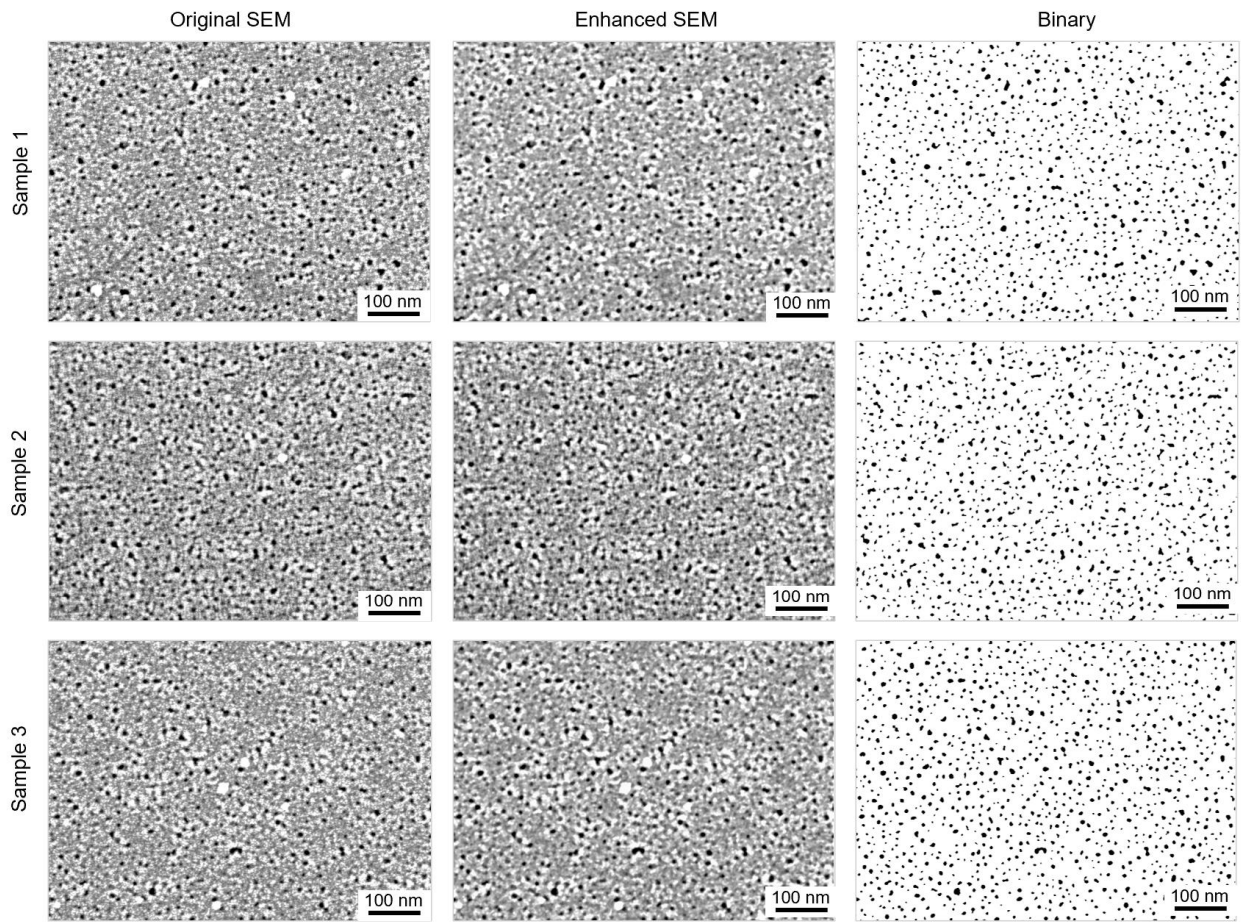

**Supporting Fig. S1. 11.  Raw SEM images, enhanced SEM images, and corresponding binary images of the UF membrane at 3.8 nm system-measured coating thicknesses.** Imaging conditions: 1 kV acceleration voltage and 150k magnification.

UF Membrane, imaged at acceleration voltage of 1 kV
System measured coating thickness : 4.9 nm

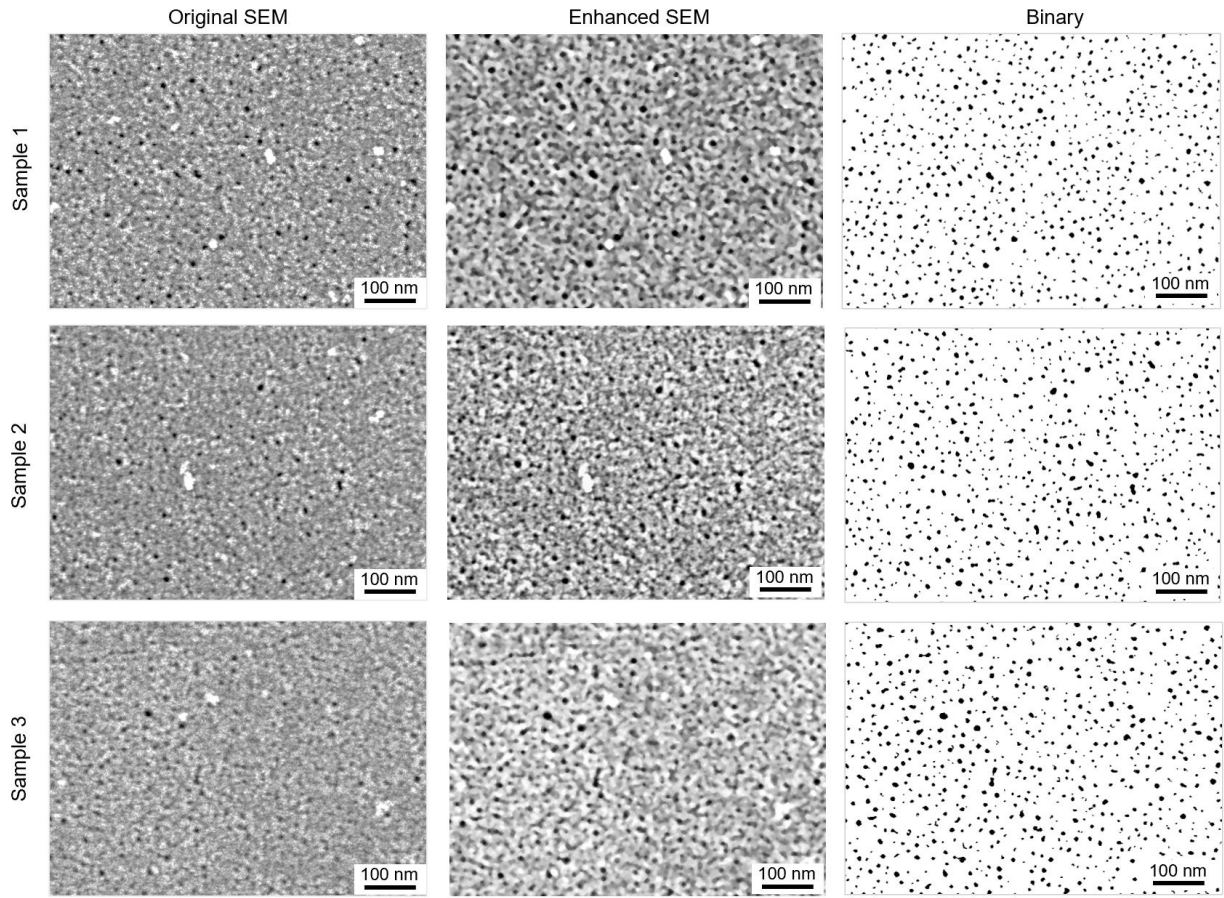

**Supporting Fig. S1. 12.  Raw SEM images, enhanced SEM images, and corresponding binary images of the UF membrane at 4.9 nm system-measured coating thicknesses.** Imaging conditions: 1 kV acceleration voltage and 150k magnification.

RO Support, imaged at acceleration voltage of 1 kV
System measured coating thickness : 1.5 nm

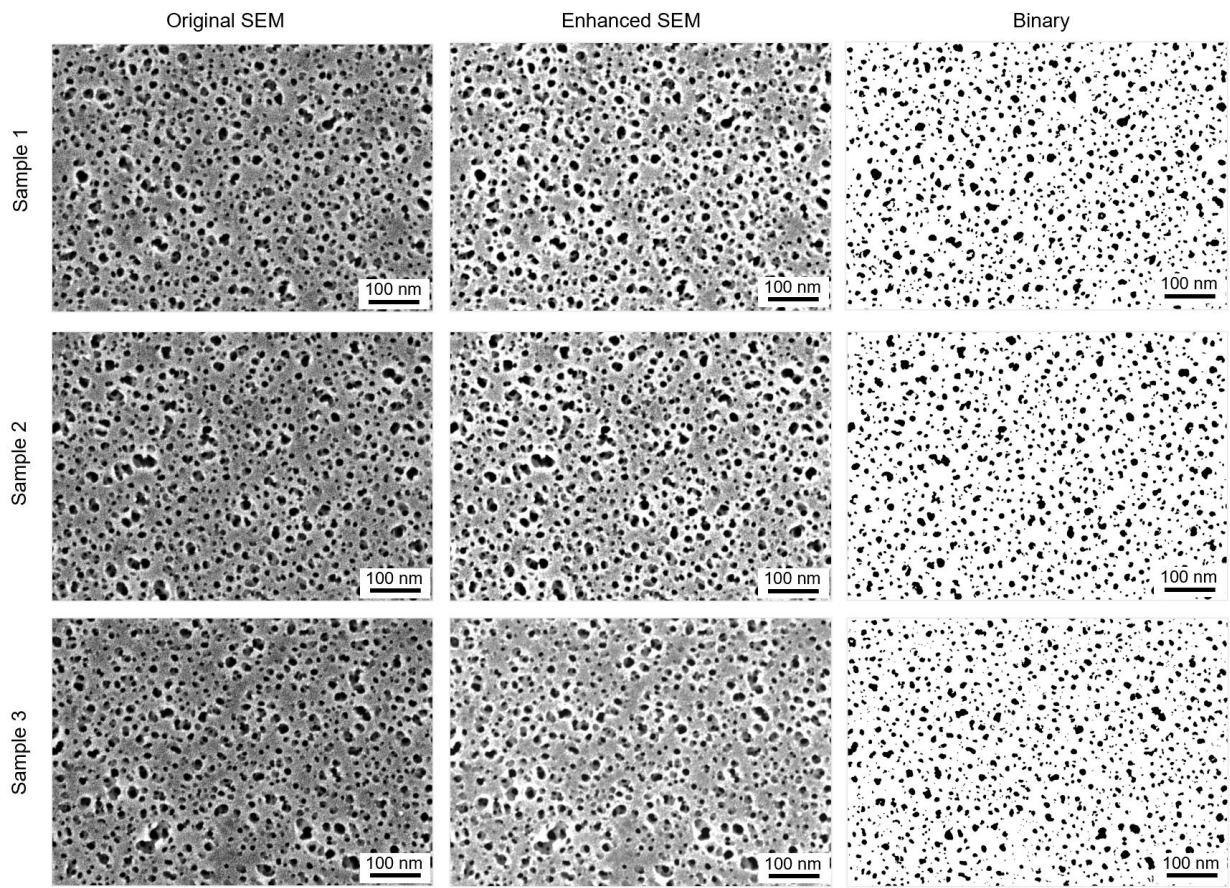

**Supporting Fig. S1. 13.  Raw SEM images, enhanced SEM images, and corresponding binary images of the AD support at 1.5 nm system-measured coating thicknesses.** Imaging conditions: 1 kV acceleration voltage and 100k magnification.

RO Support, imaged at acceleration voltage of 1 kV
System measured coating thickness : 2.5 nm

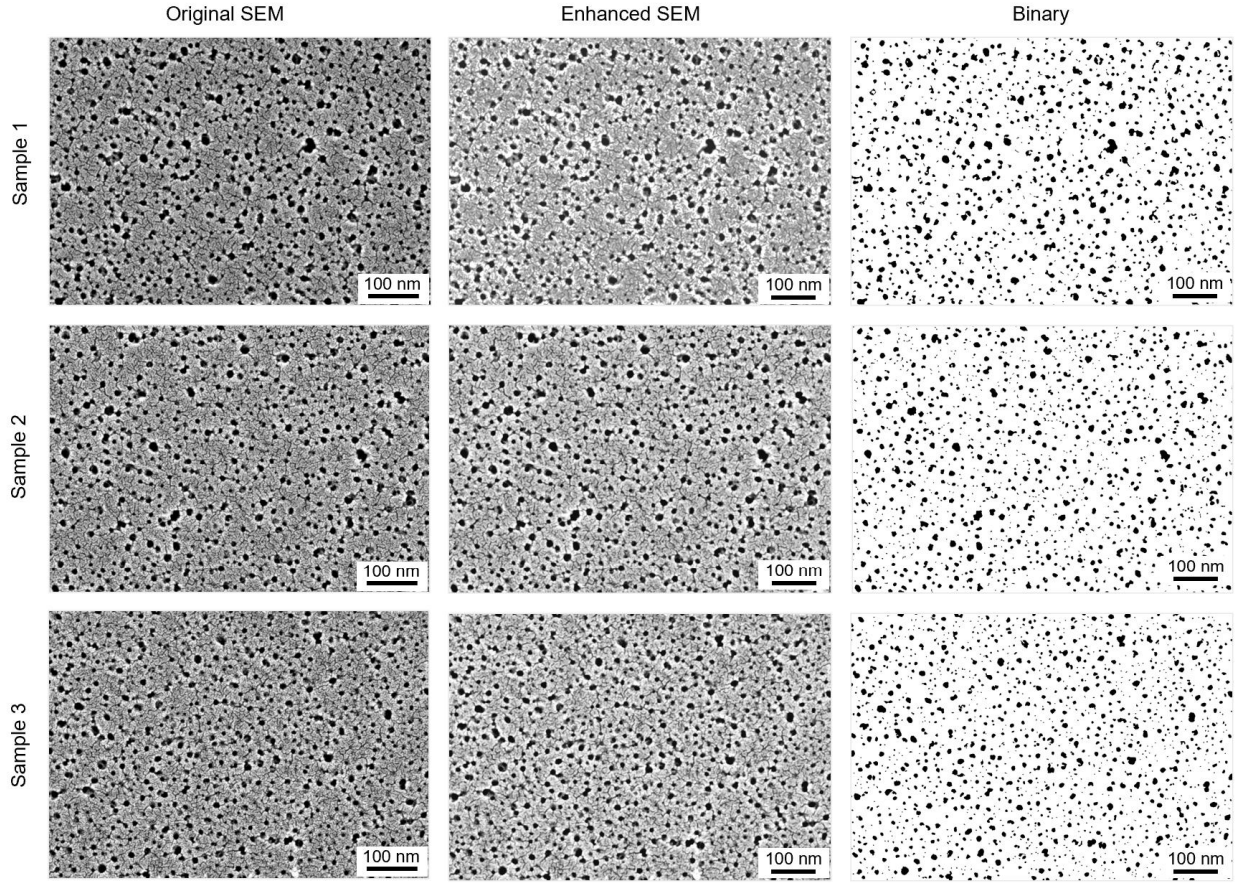

**Supporting Fig. S1. 14. Raw SEM images, enhanced SEM images, and corresponding binary images of the AD support at 2.5 nm system-measured coating thicknesses.** Imaging conditions: 1 kV acceleration voltage and 100k magnification.

RO Support, imaged at acceleration voltage of 1 kV
System measured coating thickness : 3.9 nm

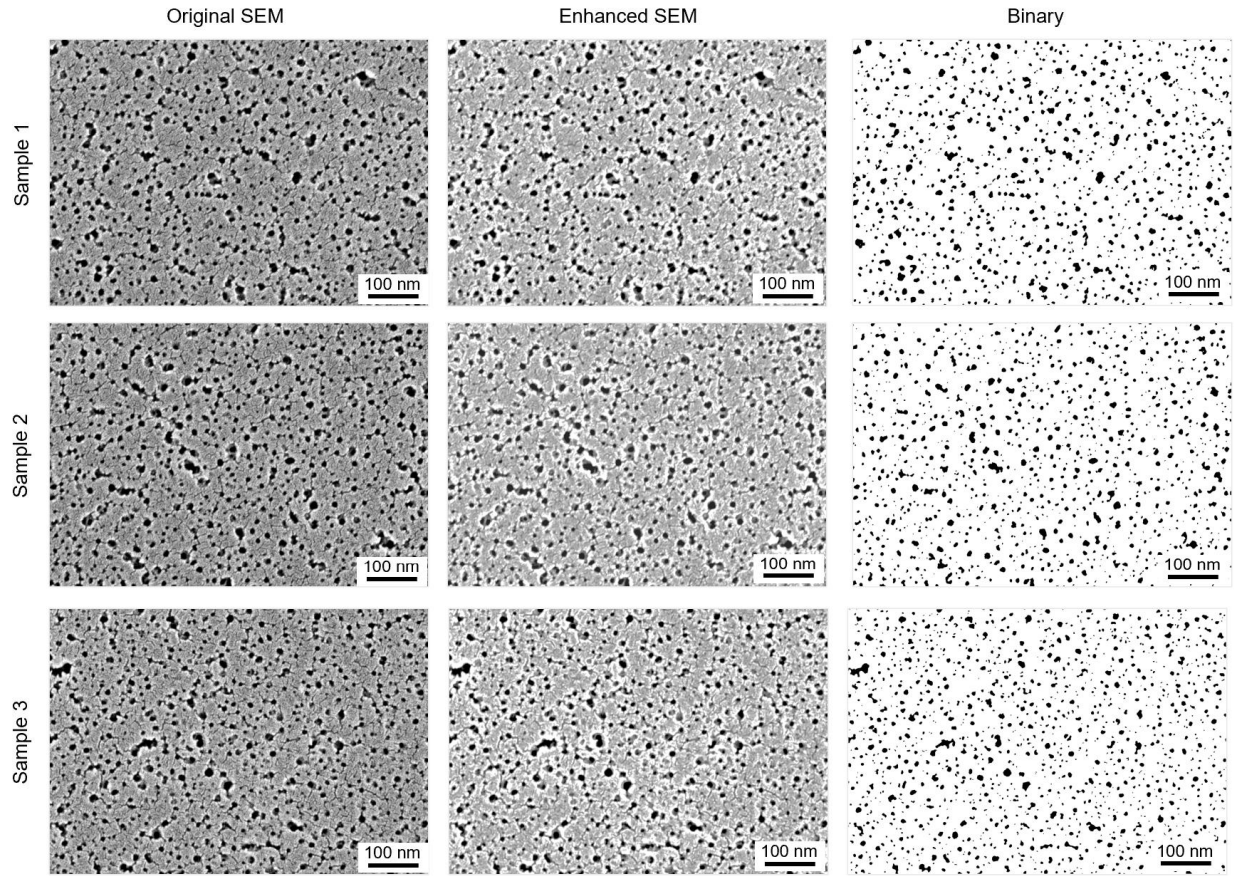

**Supporting Fig. S1. 15. Raw SEM images, enhanced SEM images, and corresponding binary images of the AD support at 3.9 nm system-measured coating thicknesses.** Imaging conditions: 1 kV acceleration voltage and 100k magnification.

RO Support, imaged at acceleration voltage of 1 kV
System measured coating thickness : 4.8 nm

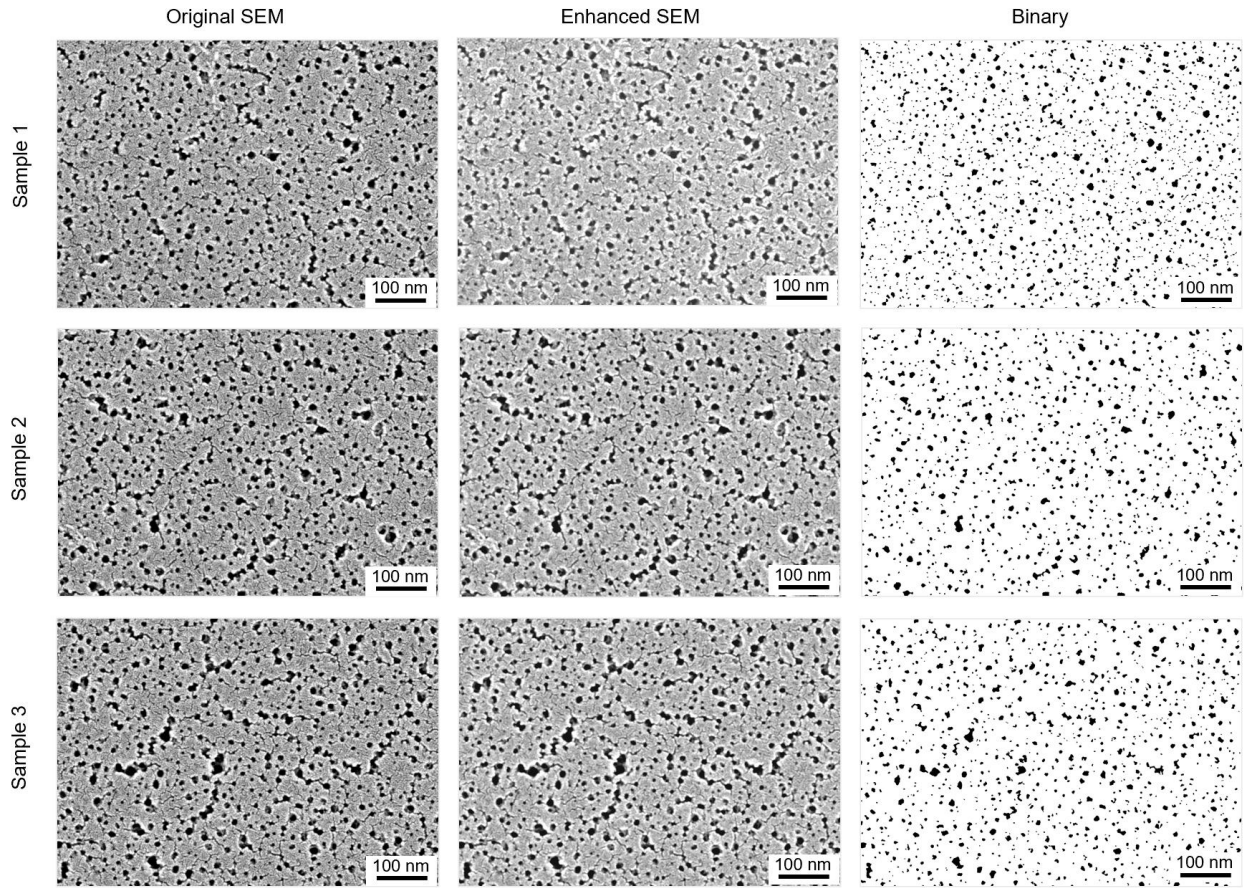

**Supporting Fig. S1. 16.   Raw SEM images, enhanced SEM images, and corresponding binary images of the AD support at 4.8 nm system-measured coating thicknesses.** Imaging conditions: 1 kV acceleration voltage and 100k magnification.

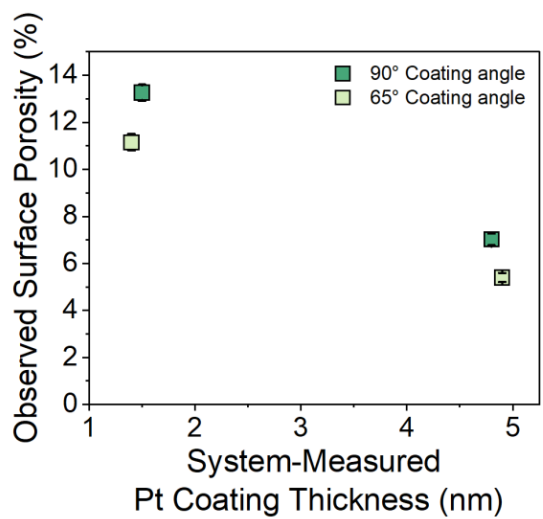

**Supporting Fig. S1. 17. Effect of coating angle on the measured surface porosity at different sputter coating thicknesses for UF membranes and RO supports.** In 90° the stage and target are parallel and in 65° the stage is horizontal.

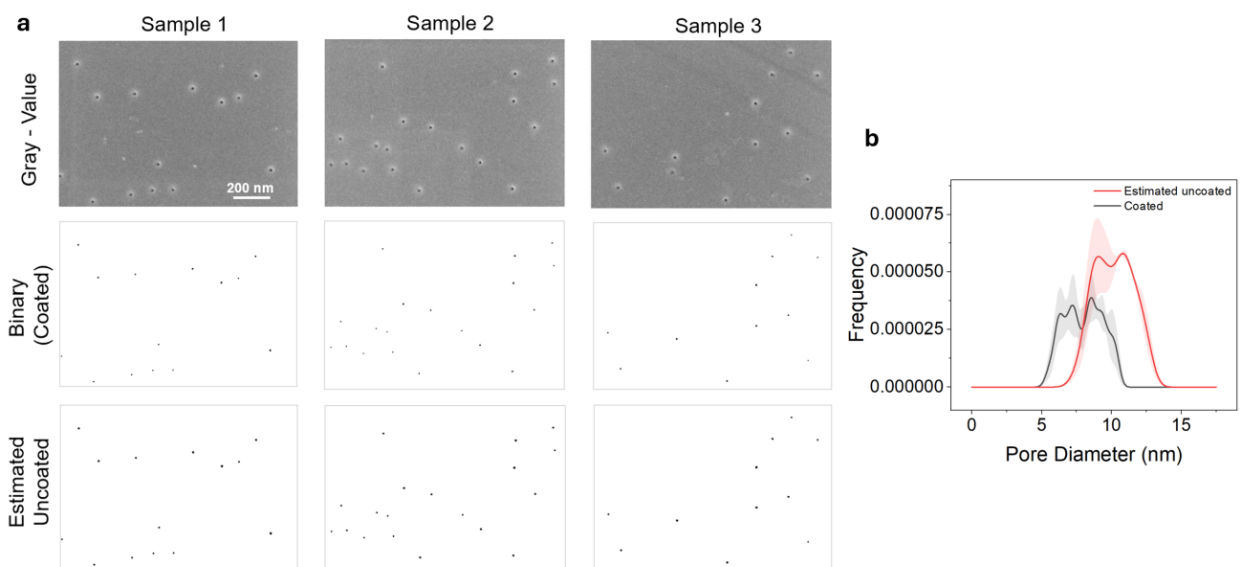

**Supporting Fig. S1. 18. Validation of pore correction method using a commercial PCTE membrane.** (a) Grayscale SEM images of a track-etched polycarbonate membrane (Sterlitech, nominal 10 nm pore size) coated with 1.2 nm of Pt at a 90° deposition angle, along with binarized pore images before and after digital pore dilation. (b) Area-normalized pore-diameter distributions from n = 3 images per condition (bin width = 0.25 nm; Gaussian smoothing σ = 1). Prior coating, the membrane was rinsed with DI water followed by 1:3 v/v isopropanol-DI water. For the coated samples the area-weighted mean pore diameter was 7.6± 0.2 nm, and the surface porosity was 5.9% ±1.0%. Following digital pore dilation, the corrected area-weighted mean pore diameter was 10.4± 0.2 nm, and the surface porosity was 10.4% ±2.0%, closely matching the membrane's nominal specifications.

## Section S2. Filtration Experiments and modeling

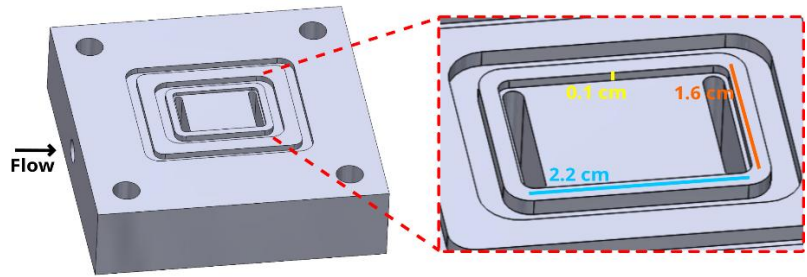

**Supporting Fig. S2.1. Cross-flow cell dimensions.**

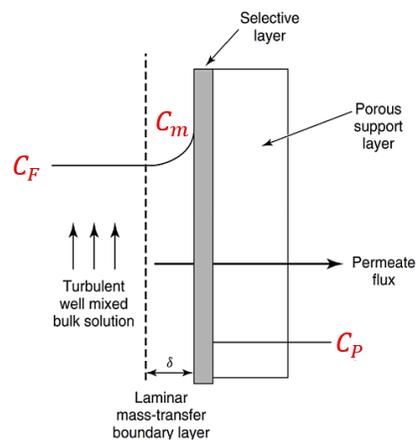

**Supporting Fig. S2.2. Concentration gradients of solutes adjacent to membrane** [73]. $C_F$, $C_m$, and $C_p$ represent solute concentrations in the feed bulk, at the membrane surface, and in the permeate, respectively.

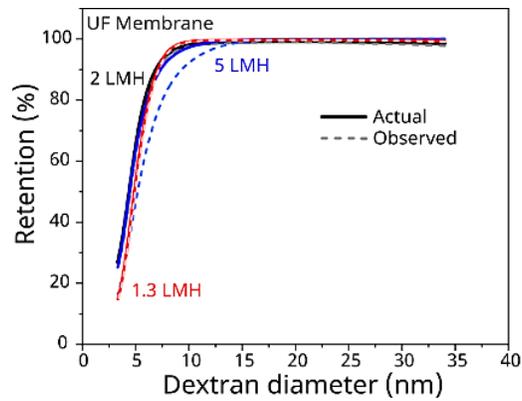

**Supporting Fig. S2.3. Actual (considering CP) and observed retention values of dextran as a function of dextran diameter at different flux values for the UF membrane.**

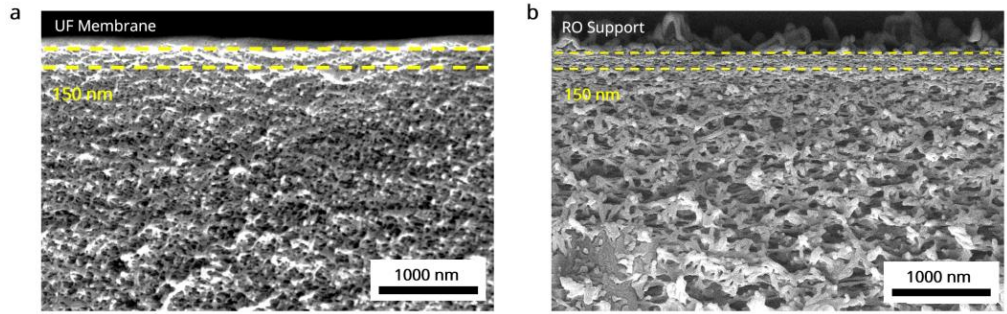

**Supporting Fig. S2.4. Selective Layer Thickness of UF Membrane and RO Support.** Cross-sectional SEM images of (a) the UF membrane and (b) the RO support, highlighting the selective layer thickness ($\delta_m$) estimated as 500 nm for the UF membrane and 300 nm for the RO support (indicated by yellow dashed lines).

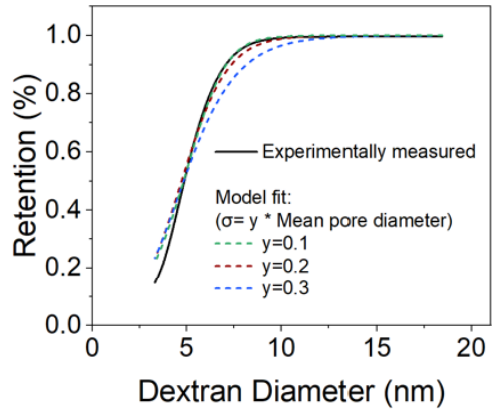

**Supporting Fig. S2.5. Experimental and Bungay and Brenner model fitted retention values of dextran as a function of dextran diameter for the UF membrane.**

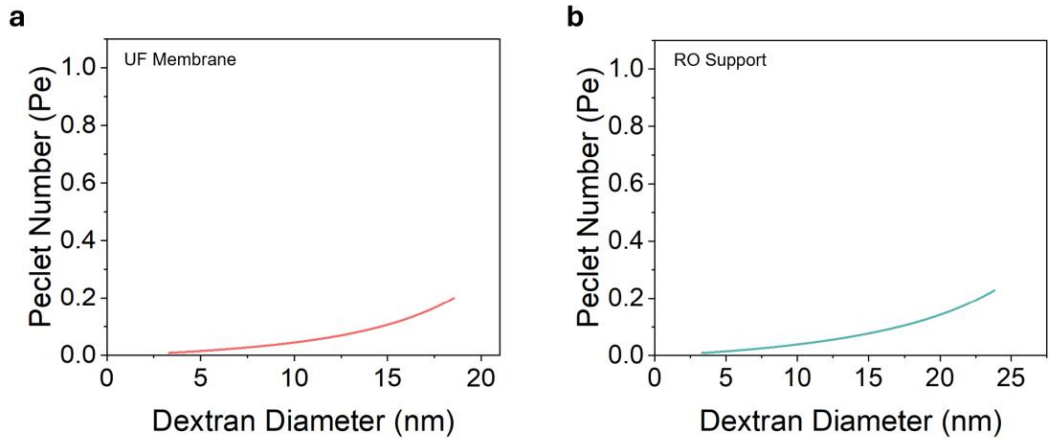

**Supporting Fig. S2.6. Peclet Number Trends in UF Membrane and RO Support.** Peclet number (Pe) as a function of dextran diameter for (a) the UF membrane and (b) the RO support. A Peclet number around 1 indicates diffusion-dominated transport for smaller dextran molecules.

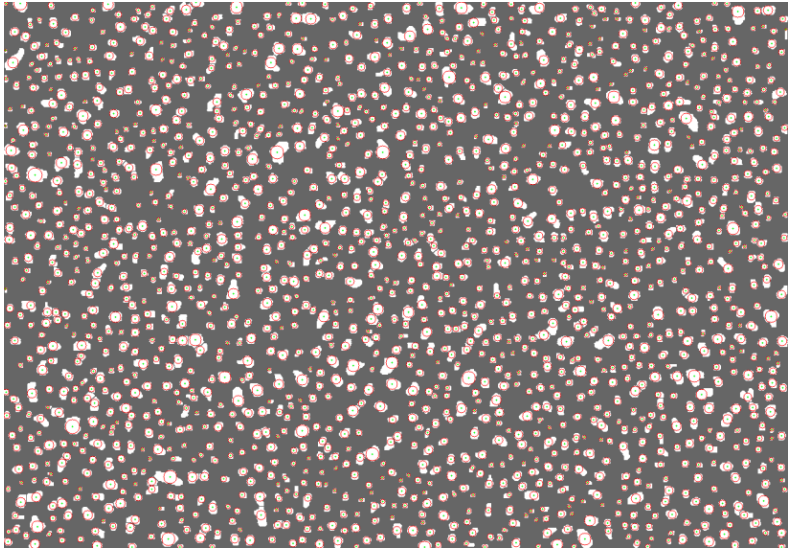

**Supporting Fig. S2.7. Extraction of maximum inscribed circle (MIC from SEM.** Representative membrane surface showing the segmented pore mask (white) with the maximum inscribed circle (MIC) for each pore opening overlaid (red); gray denotes solid polymer.

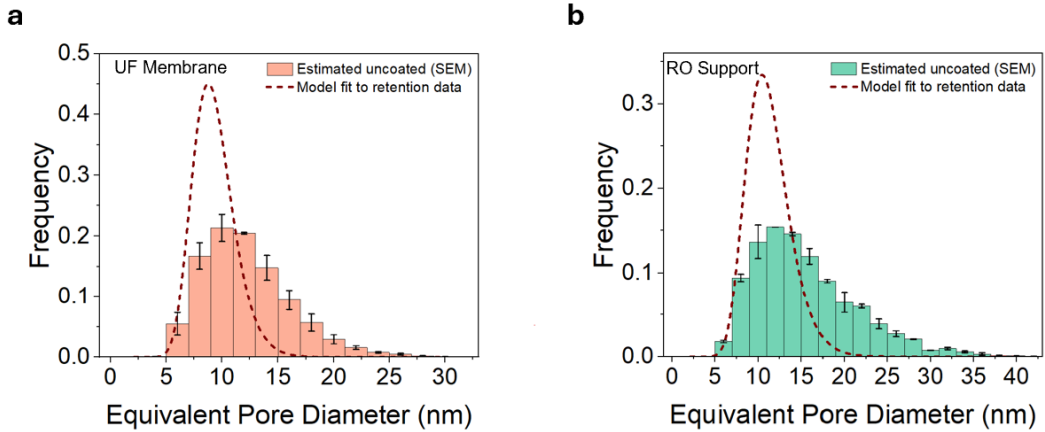

**Supporting Fig. S2.8. Comparison between SEM pore analysis and modeled pore sizes from fitting to filtration. a** and **b**, Pore-size distributions expressed as circular-equivalent diameters (EqD) from SEM after digital dilation to estimate the uncoated surface, overlaid with the model-inferred PSDs from the fits for UF membrane and RO support. The frequency represents the number of pores at each pore size divided by the total number of pores. The model-fitted pore size distributions are normalized so that the area beneath the curves matches that of the estimated uncoated pore size distribution.

**Expressions for $K_c(r, a)$ and $K_d(r, a)$ are as follows:**

$$K_c(r, a) = \frac{\big(2 - \phi(r, a)\big) K_s(r, a)}{2 K_t(r, a)} \tag{1}$$

$$K_d(r,a) = \frac{6\pi}{K_t(r,a)} \tag{2}$$

$$K_s(r,a) = \frac{9}{4}\pi^2\sqrt{2}\left(1-\left(\frac{a}{r}\right)\right)^{-\frac{5}{2}}\left[1+\frac{7}{60}\left(1-\left(\frac{a}{r}\right)\right)-\frac{2227}{50400}\left(1-\left(\frac{a}{r}\right)\right)^2\right]+ \\ 4.0180 - 3.9788\left(\frac{a}{r}\right) - 1.9215\left(\frac{a}{r}\right)^2 + 4.392\left(\frac{a}{r}\right)^3 + 5.006\left(\frac{a}{r}\right)^4 \tag{3}$$

$$K_t(r,a) = \frac{9}{4}\pi^2\sqrt{2}\left(1-\left(\frac{a}{r}\right)\right)^{-\frac{5}{2}}\left[1-\frac{73}{60}\left(1-\left(\frac{a}{r}\right)\right)+\frac{77293}{50400}\left(1-\left(\frac{a}{r}\right)\right)^2\right]- \\ 22.5083 - 5.6117\left(\frac{a}{r}\right) - 0.3363\left(\frac{a}{r}\right)^2 - 1.216\left(\frac{a}{r}\right)^3 + 1.647\left(\frac{a}{r}\right)^4 \tag{4}$$

**Section S3. Theoretical background for measuring concentration polarization factor**

Solute transport through porous membranes is commonly characterized by an observed retention coefficient, $R_o$. $R_o$ is measured as follows:

$$R_o = 1 - \frac{C_p}{C_F} \tag{5}$$

Where $C_p$ and $C_f$ are the solute concentration in permeate and bulk feed, respectively. Due to concentration polarization, where the solute concentration at the membrane surface is higher than in the bulk feed, the actual retention coefficient ($R_a$) is greater than $R_o$. $R_a$ is measured as follows:

$$R_a = 1 - \frac{C_p}{C_m} \tag{6}$$

Where $C_m$ is the concentration of solute on the membrane surface.

To measure the $C_m$, the concentration polarization factor must be measured. Based on convective and diffusive solute transport in the concentration boundary layer, the mass transfer coefficient, k, relates permeate flux, $J_w$, and solute concentration driving force as follows[46]:

$$\frac{C_m - C_p}{C_F - C_P} = \exp\left(\frac{J_w}{k}\right) \tag{7}$$

Here, mass transfer coefficient is calculated using Sherwood−Reynolds− Schmidt correlations.

The relationship between k and the Sherwood number (Sh) is given b:

$$Sh = \frac{k \times d_H}{D} \tag{8}$$

Where D is species diffusivity, which can be measured by Equation (5) in the main text. And $d_H$ is hydraulic diameter and can be measured suing the channel dimensions as follow:

$$d_H = \frac{4 \times (\ cross\ sectional\ area\ for\ flow\ )}{2 \times channel\ perimeter} \tag{9}$$

The dimensions of the cell used for this experiment are illustrated in Supporting Fig. S2.1.

When the velocity profile is fully developed Sherwood number can be estimated as Equation (10) as well [74], and can be used in Equation (8) to estimate $k$.

$$Sh = 1.62 \left(\frac{d_H}{L}\right)^{0.33} (Re)^{0.33} (Sc)^{0.33} \tag{10}$$

The Reynolds (Re) and Schmidt (Sc) numbers can be measured as follow:

$$Re = \frac{u_0 \times d_H}{\mu} \tag{11}$$

$$Sc = \frac{\mu}{D} \tag{12}$$

$u_0$ is superficial velocity and can be measured as follows:

$$u_0 = \frac{Q}{cross-sectional\ area\ for\ flow} \tag{13}$$

Where Q is feed flow in mL/min.

## Section S4. Python Code

These codes were developed by Sima Zeinali Danalou in August 2025 for: (i) pore dilation to predict uncoated SEM images, (ii) pore erosion to apply additional coating thickness, (iii) plotting the Peclet number, and (iv) fitting the Bungay–Brenner model to retention data to estimate the pore-size distribution (PSD).

### Digital pore alteration (overview)

We start from a binary pore mask where pores are white (255) and polymer is black (0). The image is up-scaled by an integer factor (e.g., 10×) so that small thickness steps can be applied cleanly. Using a distance transform, we then shift the pore boundary by a uniform amount: outward to "remove" coating (dilate pores) or inward to "add" coating (constrict pores). The shift is set in pixels on the up-scaled grid and maps to a target coating thickness. Because the boundary is offset everywhere by the same amount, the pore shape is preserved (no arbitrary smoothing or corner clipping beyond the chosen metric.

After the offset on the fine grid, we save a "parity" binary that matches the up-scaled porosity exactly by selecting the top-K blocks during down-sampling (this avoids small porosity drift). For that, we also save a coating rim overlay that highlights only the added band relative to the input mask for quick visual checks. The script prints: porosity on the up-scaled grid before→after, porosity of the saved parity, and the strip width for dilation defined as removed area divided by the original perimeter ($\Delta A$ / perimeter, using 4-connected pixel edges), and for erosion as added area divided by the eroded perimeter ($\Delta A$ / perimeter, using 4-connected pixel edges). If pores are black in your masks, invert them before processing.

## S4.1. Pore Dilation:

```python
# -*- coding: utf-8 -*-
"""
Created on Tue Aug 26 11:41:06 2025

@author: szein
"""

# -*- coding: utf-8 -*-
"""
Minimal pipeline for A1/A2/A3:
- Upscale -> shape-preserving dilation (distance transform offset)
- Save ORIGINAL-size 'parity' binary (porosity exactly matches upscaled)
- Save rim overlays for both (what was added)
- Print: Porosity @ upscaled grid (raw), and strip width (ΔA/P4) on original grid
"""

import os
import cv2
import numpy as np

# -------------------------- INPUTS (A1/A2/A3) --------------------------
A1 = r"C:/Users/…/….tif"
A2 = r"C:/Users/…/….tif"
A3 = r"C:/Users/…/….tif"

# Choose ONE way to set the offset thickness:
UPSCALE  = 10              # fine grid multiplier (integer), pixel size originally was 1 nm, now after
resizing the image pixel sized is 0.1 nm.
METRIC   = 'Linf'          # 'L2' (Euclidean), 'L1' (city-block), 'Linf' (chessboard)

# Option A (pixel-based on upscaled grid): set this directly if you already know it.
PX_DILATE_UP = 14          # <-- upscaled-pixel offset used in your earlier runs, Important: you may
change this number so the Strip width (this will be printed) would match your target coating thickness
(here the target is 1.5 nm); Important: this is the only number you would change based on your pixel size
and coating thickness.

# Option B (physical thickness): if you prefer nm, set these instead of PX_DILATE_UP
# NM_PER_PX    = 1.00      # image resolution (nm per original pixel)
# THICKNESS_NM = 1.5       # coating thickness to remove/apply (nm)
# PX_DILATE_UP = int(round(THICKNESS_NM * UPSCALE / NM_PER_PX))

# -------------------------- UTILS --------------------------
def metric_to_dist(metric: str):
    m = metric.lower()
    if m == 'l2': return cv2.DIST_L2
    if m == 'l1': return cv2.DIST_L1
    return cv2.DIST_C   # Linf / chessboard

def porosity(mask_255: np.ndarray) -> float:
```

```python
        """Fraction of white pixels (pores) in [0,255] mask."""
        return float(np.count_nonzero(mask_255 == 255)) / mask_255.size

def binarize_gray(img_gray: np.ndarray) -> np.ndarray:
        """Go to 0/255 binary. Pores must be white (255)."""
        _, b = cv2.threshold(img_gray, 127, 255, cv2.THRESH_BINARY)
        return b

def offset_mask(mask255: np.ndarray, t_px: float, metric: str) -> np.ndarray:
        """
        !!!! Shape-preserving outward offset by distance transform:
        turn to white any background pixel within <= t_px of a pore pixel.
        This adds a uniform 'rim' of width ≈ t_px (in the chosen metric).
        """
        if t_px <= 0:
            return mask255.copy()
        dist_type = metric_to_dist(metric)
        bg01 = (mask255 == 0).astype(np.uint8)            # distance computed in background
        dt   = cv2.distanceTransform(bg01, dist_type, 5)  # distance to nearest pore pixel
        out  = mask255.copy()
        out[(mask255 == 0) & (dt <= t_px + 1e-6)] = 255    # add rim: background → pore
        return out

def block_sum_counts_01(up_mask255: np.ndarray, factor: int) -> np.ndarray:
        """
        Count white pixels in each factor×factor upscaled block that maps to ONE original pixel.
        Used for 'parity' downsample: we will pick the top-K blocks to match target porosity exactly.
        """
        up01 = (up_mask255 == 255).astype(np.uint8)
        Hup, Wup = up01.shape
        H0, W0 = Hup // factor, Wup // factor
        up01 = up01[:H0*factor, :W0*factor]                # crop to multiple of 'factor'
        return up01.reshape(H0, factor, W0, factor).sum(axis=(1,3)).astype(np.int32)  # 0..factor^2

def parity_binary_topk_from_counts(counts_int: np.ndarray, target_por: float):
        """
        Make an ORIGINAL-size 0/255 binary with porosity == round(target_por*N)/N exactly
        by selecting the top-K blocks with largest white counts.
        """
        N = counts_int.size
        K = int(round(target_por * N))
        K = max(0, min(N, K))
        flat = counts_int.ravel()
        if K == 0:
            return np.zeros_like(counts_int, dtype=np.uint8), 0.0
        if K == N:
            return np.full_like(counts_int, 255, dtype=np.uint8), 1.0
        kth = N - K
        idx_top = np.argpartition(flat, kth)[kth:]          # indices of the K largest scores
        out = np.zeros(N, dtype=np.uint8)
        out[idx_top] = 255
        return out.reshape(counts_int.shape), K / float(N)

def overlay_magenta(base_gray_255: np.ndarray, added_gray_255: np.ndarray) -> np.ndarray:
        """
        Make a color overlay where only the ADDED rim (difference) is shown in magenta.
        Base is grayscale; magenta = [255,0,255] (BGR).
        """
        ov = cv2.cvtColor(base_gray_255, cv2.COLOR_GRAY2BGR)
        ov[added_gray_255 > 0] = [255, 0, 255]
        return ov

def perimeter_4(mask01: np.ndarray) -> int:
        """
        4-connected perimeter (pixel-edge count) on ORIGINAL grid.
        mask01 is 0/1 where 1 = pores (white).
        """
        vert_edges  = np.sum(mask01[:, 1:] != mask01[:, :-1])
        horiz_edges = np.sum(mask01[1:, :] != mask01[:-1, :])
```

```python
    return int(vert_edges + horiz_edges)

# ------------------------- CORE (one image) -------------------------
def process_one(path: str, upscale: int, px_dilate_up: int, metric: str):
    print(f"\n[Processing] {os.path.basename(path)}  |  "
upscale={upscale}x  px_dilate_up={px_dilate_up}  metric={metric}")
    g = cv2.imread(path, cv2.IMREAD_GRAYSCALE)
    if g is None:
        print(f"  ! Could not read file.")
        return
    # If your pores are black and walls white, invert here:
    # g = 255 - g

    # 1) Ensure binary (pores=255, polymer=0)
    img0 = binarize_gray(g)

    # 2) Upscale for fine-grid offset (nearest keeps edges crisp)
    H0, W0 = img0.shape
    img_up = cv2.resize(img0, (W0*upscale, H0*upscale), interpolation=cv2.INTER_NEAREST)
    init_por_up = porosity(img_up)

    # 3) DILATE on the upscaled grid (shape-preserving offset of boundary by px_dilate_up)
    dil_up = offset_mask(img_up, float(px_dilate_up), metric)
    final_por_up = porosity(dil_up)

    # ---- print requested upscaled porosity line ----
    print(f"  Porosity @ upscaled grid:           {init_por_up:.6f} -> {final_por_up:.6f}
(raw)")

    # 4) Build ORIGINAL-size 'parity' binary with EXACT porosity match
    counts = block_sum_counts_01(dil_up, upscale)         # white counts per original pixel
    bin_parity, por_parity = parity_binary_topk_from_counts(counts, final_por_up)

    # 5) Also compute DIRECT original-grid dilation with same physical thickness
    t_orig = float(px_dilate_up) / max(1, upscale)        # convert upscaled-px to original-px
    dil_orig_direct = offset_mask(img0, t_orig, metric)

    # 6) Build RIM overlays (what was added) for both results
    added_parity = cv2.subtract(bin_parity, img0)         # parity rim
    added_direct = cv2.subtract(dil_orig_direct, img0)    # direct rim
    ov_parity = overlay_magenta(bin_parity, added_parity)
    ov_direct = overlay_magenta(dil_orig_direct, added_direct)

    # 7) Save outputs (ORIGINAL size only)
    root, _ = os.path.splitext(path)
    tiff_params = []
    if hasattr(cv2, 'IMWRITE_TIFF_COMPRESSION'):
        tiff_params += [cv2.IMWRITE_TIFF_COMPRESSION, 5]  # LZW lossless

    out_parity     = f"{root}-dilated_orig_por_parity_{px_dilate_up}px_{metric}.tif"
    out_parity_ov  = f"{root}-added_rim_parity_{px_dilate_up}px_{metric}.tif"

    cv2.imwrite(out_parity,    bin_parity)
    cv2.imwrite(out_parity_ov, ov_parity, tiff_params)

    # 8) Log porosities at ORIGINAL size
    p0 = porosity(img0)
    pp = porosity(bin_parity)
    pd = porosity(dil_orig_direct)

    # ---- strip width (ΔA/P4) on ORIGINAL grid, using ORIGINAL P4 ----
    P4_orig = perimeter_4((img0 == 255).astype(np.uint8))
    area_added_parity = int((added_parity > 0).sum())
    area_added_direct = int((added_direct > 0).sum())
    strip_parity = area_added_parity / max(1, P4_orig)
    strip_direct = area_added_direct / max(1, P4_orig)

    print(f"  Porosity (parity, saved) : {pp:.6f}  (matches upscaled target)")
```

```
    print(f"  Strip width (px @ original, ΔA/P4 using ORIGINAL P4, target is 1.5
nm): nm={strip_parity:.4f} ")
    # If you want in nm, uncomment and set NM_PER_PX above:
    # print(f"  Strip width (nm)          :  parity={strip_parity*NM_PER_PX:.2f}   ")

    print("  Saved:")
    print(f"    {out_parity}")
    print(f"    {out_parity_ov}")

# ------------------------ RUN (A1/A2/A3) --------------------------------
for path in (A1, A2, A3):
    process_one(path, UPSCALE, PX_DILATE_UP, METRIC)
```

## S4.2. Pore erosion:

```
# -*- coding: utf-8 -*-
"""
Erode pore rims on upscaled grid, paint eroded rim in purple, and report:
  - Porosity (before / after) at original and upscaled grids
  - Mean rim (strip) thickness from:
      High-res "coarea-style" estimate using area/perimeter along the white-purple interface

Key points:
  • Upscale by UPSCALE_FACTOR (nearest) for clean, isotropic CDT steps.
  • Erode white (pore) pixels inward by n Chebyshev (8-connected) steps (or 4-connected if desired).
  • Map the upscaled erosion back to the original grid and save both preview and original-res images.
  • Strip width printout now uses P4 of the *eroded* pore mask

Author: szein (updated)
Created: Wed Aug 27 10:19:09 2025
"""

import os
import gc
import numpy as np
from PIL import Image
from scipy.ndimage import distance_transform_cdt, binary_dilation

# ===================== Parameters =====================
UPSCALE_FACTOR = 10      # integer; 10 is a good default, the pixel size was 1 nm, now turned into 0.1 nm
SCALE = 1.0              # units per pixel for the high-res width print (set to nm/px if desired)
INCLUDE_DIAGONALS = True # True → chessboard (8-connected) erosion; False → taxicab (4-connected)

# ===================== Core utilities =====================
def measure_porosity(img_pil):
    """Porosity of a binary (white=255) mask image."""
    arr = np.asarray(img_pil.convert('L'), dtype=np.uint8)
    por = float(np.sum(arr == 255)) / arr.size
    return por

def perimeter_4(mask01_bool):
    """4-connected perimeter (pixel-edge count)."""
    a = mask01_bool.astype(np.uint8)
    vert  = np.sum(a[:, 1:] != a[:, :-1])
    horiz = np.sum(a[1:, :] != a[:-1, :])
    return int(vert + horiz)

def measure_purple_strip_width(img_pil_rgb, include_diagonals=True, scale=1.0):
    """
    High-res "coarea-style" average strip width:
      width ≈ (purple area) / (length of the white boundary adjacent to purple)
    """
    arr = np.asarray(img_pil_rgb.convert('RGB'), dtype=np.uint8)
    WHITE  = np.array([255, 255, 255], dtype=np.uint8)
    PURPLE = np.array([128,   0, 128], dtype=np.uint8)
```

```python
    white_mask  = np.all(arr == WHITE,  axis=-1)
    purple_mask = np.all(arr == PURPLE, axis=-1)

    if include_diagonals:
        struct = np.ones((3, 3), dtype=bool)
    else:
        struct = np.array([[0,1,0],
                           [1,1,1],
                           [0,1,0]], dtype=bool)

    adj_to_purp   = binary_dilation(purple_mask, structure=struct)
    boundary_white = white_mask & adj_to_purp

    area_pixels      = int(purple_mask.sum())
    perimeter_pixels = int(boundary_white.sum())

    if perimeter_pixels == 0:
        print("No inner white boundary found; average width undefined.")
        return float('nan')

    avg_width = (area_pixels / perimeter_pixels) * scale
    print(f"Average purple strip width (high-res coarea-style): {avg_width:.6f} (units)")
    return avg_width

def pore_erosion(img_pil_rgb, n_steps, include_diagonals=True,
                 grow_from='non-white_and_purple', return_mask_only=False):
    """
    Mark the first n_steps (Chebyshev or taxicab) layers of WHITE (255,255,255) pixels
    that are adjacent to NON-WHITE (and optionally NOT purple). Returns a boolean mask
    of the WHITE pixels to convert to "eroded rim" (purple).
    """
    arr0 = np.asarray(img_pil_rgb.convert('RGB'), dtype=np.uint8)

    WHITE = np.array([255, 255, 255], dtype=np.uint8)
    PURPLE = np.array([128,   0, 128], dtype=np.uint8)

    white_mask = np.all(arr0 == WHITE,  axis=-1)
    purple_mask0 = np.all(arr0 == PURPLE, axis=-1)

    if grow_from == 'non-white_and_purple':
        seed_mask = ~white_mask  # any non-white pixel counts as "outside"
    elif grow_from == 'non-white_only':
        # "outside" excludes purple if present in the input
        seed_mask = (~white_mask) & (~purple_mask0)
    else:
        raise ValueError(f"Invalid grow_from: {grow_from!r}")

    metric = 'chessboard' if include_diagonals else 'taxicab'
    inv_seed = ~seed_mask
    dmap = distance_transform_cdt(inv_seed, metric=metric)

    # only white pixels can be eroded
    max_reach = int(dmap[white_mask].max()) if white_mask.any() else 0
    num_steps = min(int(n_steps), max_reach)

    rim_mask = (dmap <= num_steps) & white_mask  # boolean: WHITE pixels to become rim
    if return_mask_only:
        return rim_mask

    # Paint rim purple (preview)
    arr_p = arr0.copy()
    arr_p[rim_mask] = PURPLE
    return Image.fromarray(arr_p)

# ===================== Pipeline =====================
def process_image(name, path, n_steps):
    print(f"\nProcessing: {name}")
    img = Image.open(path)
```

```python
    # Porosity @ original (pre)
    por_pre = measure_porosity(img)

    # Upscale (nearest) for clean CDT steps
    img_up = img.resize((img.width * UPSCALE_FACTOR, img.height * UPSCALE_FACTOR), Image.NEAREST)

    # Porosity @ upscaled (pre)
    up_before_arr = np.asarray(img_up.convert('L'), dtype=np.uint8)
    por_up_before = float(np.sum(up_before_arr == 255)) / up_before_arr.size

    # --- Erode on upscaled grid ---
    rim_mask_up = pore_erosion(img_up, n_steps, include_diagonals=INCLUDE_DIAGONALS,
return_mask_only=True)

    # Create upscaled purple overlay (white→purple where rim)
    arr_up_rgb = np.asarray(img_up.convert('RGB'), dtype=np.uint8)
    PURPLE = np.array([128, 0, 128], dtype=np.uint8)
    arr_up_purple = arr_up_rgb.copy()
    arr_up_purple[rim_mask_up] = PURPLE
    purple_img_up = Image.fromarray(arr_up_purple)

    # High-res strip width (coarea-style)
    width_hi = measure_purple_strip_width(purple_img_up, include_diagonals=INCLUDE_DIAGONALS, scale=SCALE)

    # Porosity @ upscaled (post)
    up_after_arr = np.asarray(purple_img_up.convert('L'), dtype=np.uint8)
    por_up_after = float(np.sum(up_after_arr == 255)) / up_after_arr.size
    print(f"Porosity @ upscaled grid:          {por_up_before:.6f} -> {por_up_after:.6f} (raw)")

    # Downsample the boolean rim mask back to original size
    rim_mask_down = Image.fromarray(rim_mask_up.astype(np.uint8) * 255).resize(img.size, Image.NEAREST)
    rim_mask_orig = (np.asarray(rim_mask_down, dtype=np.uint8) > 0)

    # Build original-res outputs: (a) eroded (white→black), (b) purple overlay
    arr_orig_rgb = np.asarray(img.convert('RGB'), dtype=np.uint8)

    arr_eroded = arr_orig_rgb.copy()
    arr_eroded[rim_mask_orig] = [0, 0, 0]
    eroded_img = Image.fromarray(arr_eroded)

    arr_purple = arr_orig_rgb.copy()
    arr_purple[rim_mask_orig] = PURPLE
    purple_img = Image.fromarray(arr_purple)

    # Porosity @ original (post)
    por_post_original = measure_porosity(purple_img)
    por_post_highres  = measure_porosity(purple_img_up)

    # --- Mean strip thickness on ORIGINAL grid using *ERODED* perimeter ---
    # ΔA = pixels removed at original size (the rim)
    delta_A = int(rim_mask_orig.sum())

    # Build pre/eroded pore masks (white==255 is pore)
    orig_gray        = np.asarray(img.convert('L'), dtype=np.uint8)
    pore_mask_orig   = (orig_gray == 255)
    pore_mask_eroded = pore_mask_orig & (~rim_mask_orig)

    P4_eroded = perimeter_4(pore_mask_eroded)   # <-- use eroded perimeter
    strip_width_px  = delta_A / max(1, P4_eroded)       # average rim thickness in original pixels
    strip_width_div = strip_width_px / UPSCALE_FACTOR       # optional: convert to "upscaled-pixel" units

    # ---- Save outputs ----
    out_dir = os.path.dirname(path)
    eroded_path     = os.path.join(out_dir, f"eroded_no_purple_{name}.tif")
    purple_path     = os.path.join(out_dir, f"eroded_with_purple_{name}.tif")
    purple_up_path  = os.path.join(out_dir, f"highres_purple_{name}.tif")
```

```python
        eroded_img.save(eroded_path)
        purple_img.save(purple_path)
        purple_img_up.save(purple_up_path)

        print(f"Saved: {eroded_path}")
        print(f"Saved: {purple_path}")
        print(f"Saved high-res purple: {purple_up_path}")

        # Cleanup
        del img, img_up, rim_mask_up, rim_mask_down, arr_orig_rgb, arr_eroded, arr_purple, arr_up_rgb,
arr_up_purple
        gc.collect()

        return {
            "Porosity Before (orig)": por_pre,
            "Porosity After (up)":    por_post_highres,
            "Strip width hi-res (nm)": width_hi/UPSCALE_FACTOR,

        }

# ====================== Configure & Run ======================
image_configs = [
    {
        "name": "A1",
        "path": r"C:/Users/szein/Desktop/Re-visit/review/AD-1.5nm-3/Mask of Classified image-AD-1.5nm-3-
mask10.tif",
        "n_steps": 9   # tune so strip width aligns with your physical thickness (in px-units)
    },
    {
        "name": "A2",
        "path": r"C:/Users/szein/Desktop/Re-visit/review/AD-1.5nm-5/Mask of Classified image-AD-1.5nm-5-
mask10.tif",
        "n_steps": 9
    },
    {
        "name": "A3",
        "path": r"C:/Users/szein/Desktop/Re-visit/review/AD-1.5nm-9/AD after - 1.5 nm pt _0009-.tif",
        "n_steps": 8
    }
]

results = {}
for cfg in image_configs:
    stats = process_image(cfg["name"], cfg["path"], cfg["n_steps"])
    results[cfg["name"]] = stats

# ====================== Summary ======================
print("\n=== Summary ===")
for name, metrics in results.items():
    print(f"\n{name}:")
    for k, v in metrics.items():
        print(f"  {k}: {v:.6f}" if isinstance(v, float) else f"  {k}: {v}")
```

## S4.3. Plotting Pe Number:

```python
# -*- coding: utf-8 -*-
"""
Created on Tue Aug 26 17:59:32 2025

@author: szein
"""

# -*- coding: utf-8 -*-
"""
SI §S4.3 — Wall Péclet number (Pe_m) computed from a discrete pore-size distribution
```

```
--------------------------------------------------------------------------------

What this script does
---------------------
1) Loads dextran size (a) and bulk diffusion (D_inf) from an Excel sheet.
2) Uses a *discrete* pore-size distribution P_values over pore_sizes (your
   uncoated SEM-derived histogram; number-weighted) and computes weighted
   averages of K_c(r,a) and K_d(r,a) using a Poiseuille r^4 throughput weight.
3) Forms an *effective* Pe_m for each dextran size using:
       Pe_m = (J_v * δ_m / ε_s) * ( <K_c> / <K_d> ) / D_inf
   where angle brackets are r^4-weighted averages over the discrete PSD.
   (Note: this is the ratio of the averages, not the average of the ratio.)
4) Plots K_d(a), K_c(a), and Pe_m(a).

Important assumptions / checks
------------------------------
• Units: a and pore_sizes **must be in nm as *radii*** (not diameters). The
  BB formulas use λ = a / r, so if your pore_sizes are *diameters* D,
  replace r by D/2 before use. (See the comment near 'pore_sizes' below.)
• The PSD here is number-weighted and *not normalized*; that's fine because
  each weighted mean divides by the same Σ(P_values * r^4), i.e., normalization
  cancels in the mean. For Pe_m we form <K_c> / <K_d> — a common "effective
  property" approximation; if you instead want <K_c/K_d>, the integration must
  be changed (see comment near "Pe_value").
• The first several pore_sizes are ≤ 0 with P=0 to reflect empty bins; they do
  not contribute to any sums.

Caveats (leave as comments if you keep the same math)
-----------------------------------------------------
• Pe_m here ≈ (J_v δ_m / ε_s) * <K_c> / <K_d> / D_inf. If the physics you intend
  requires <K_c/K_d>, replace the two separate averages with a single average of
  the ratio using the same r^4 weighting.
• Ensure D_inf is in m^2/s, J_v in m/s, δ_m in m, ε_s dimensionless.
"""

import numpy as np
import pandas as pd
import matplotlib.pyplot as plt

# ----------------------- Load the data from Excel -------------------------
file_path = r'C:\Users\szein\Desktop\Others-membrane\mwco\20240927- 0.5 psi PW - 0.6 AD- 1.3 LMH for
both\PW-P 1.3LMH 20Sep2024.xlsx'
# file_path = r'C:\Users\...\...xlsx'  # alternate dataset
sheet_name = 'Sheet1'
df = pd.read_excel(file_path, sheet_name=sheet_name)

# Rows/columns: J = dextran size "a" [nm, hydrodynamic *radius*], K = D_inf [m^2/s]
column_J_data = df.iloc[102:536, 9].values    # a (nm) — dextran radius
column_K_data = df.iloc[102:536, 10].values   # D_inf (m^2/s)
a_values = np.array(column_J_data)            # dextran radius 'a' (nm)
D_inf    = np.array(column_K_data)            # bulk diffusion (m^2/s)

# ----------------------- Membrane / operating parameters ------------------
J_v_LMH  = 1.3                      # permeate flux in LMH (L m^-2 h^-1) +
J_v_m_s  = (J_v_LMH * 1e-3) / 3600  # convert LMH → m/s  (1 LMH = 1e-3/3600 m/s)
epsilon_s = 0.227                   # surface porosity ε_s [-] from uncoated SEM
delta_m   = 150e-9                  # skin-layer thickness δ_m [m] from ceross SEM

# ----------------------- Discrete pore-size distribution ------------------
# NOTE: The BB formulas use pore *radius* r. If your PSD is in *diameter* (D),
# you must convert to radius: r = D / 2. Keep as-is only if these are radii.
# The first six entries are ≤ 0 and P=0 (empty bins) and thus do not contribute.
# If you wish, you can assert non-negativity via: assert np.all(pore_sizes[P_values>0] > 0)
# (Kept as-is per your request.)
import numpy as np

'''
# Example PSD with extended range (commented out in your script).
# (Keep for reference; not used below)
```

```python
pore_sizes = np.array([...], dtype=float)          # centers
P_values   = np.array([...], dtype=float)          # number frequencies
'''

import numpy as np
#the following values are derived from uncotaed SEM (Number-weighted frequencies )
pore_sizes = np.array([
    -0.75, -0.25, 0.25, 0.75, 1.25, 1.75, 2.25, 2.75, 3.25, 3.75,
    4.25, 4.75, 5.25, 5.75, 6.25, 6.75, 7.25, 7.75, 8.25, 8.75,
    9.25, 9.75, 10.25, 10.75, 11.25, 11.75, 12.25, 12.75, 13.25, 13.75,
    14.25, 14.75, 15.25, 15.75, 16.25, 16.75, 17.25, 17.75, 18.25, 18.75,
    19.25, 19.75, 20.25, 20.75, 21.25, 21.75, 22.25, 22.75, 23.25, 23.75,
    24.25, 24.75, 25.25, 25.75, 26.25, 26.75, 27.25, 27.75, 28.25, 28.75,
    29.25, 29.75, 30.25, 30.75, 31.25, 31.75, 32.25, 32.75, 33.25, 33.75
], dtype=float)

P_values = np.array([
    0.0, 0.0, 0.0, 0.0, 0.0, 0.0, 1.18992e-09, 2.94003e-08, 3.81504e-07, 2.88882e-06,
    1.53658e-05, 6.68045e-05, 2.31579e-04, 5.97756e-04, 1.17e-03, 1.87e-03, 2.68e-03, 3.52e-03, 4.37e-03,
    5.25e-03,
    6.09e-03, 6.97e-03, 8.05e-03, 9.15e-03, 9.87e-03, 1.028e-02, 1.049e-02, 1.05e-02, 1.034e-02, 1.014e-
02,
    9.85e-03, 9.37e-03, 8.92e-03, 8.54e-03, 7.96e-03, 7.28e-03, 6.74e-03, 6.25e-03, 5.72e-03, 5.19e-03,
    4.67e-03, 4.26e-03, 3.82e-03, 3.25e-03, 2.84e-03, 2.70e-03, 2.54e-03, 2.17e-03, 1.78e-03, 1.55e-03,
    1.48e-03, 1.46e-03, 1.37e-03, 1.20e-03, 9.39212e-04, 6.37496e-04, 4.40341e-04, 3.84517e-04, 3.77368e-
04, 3.17222e-04,
    2.08524e-04, 1.39425e-04, 1.44224e-04, 1.96334e-04, 2.04249e-04, 1.24160e-04, 4.06933e-05, 9.07372e-
06, 1.32130e-05, 3.57975e-05
], dtype=float)

# ----------------------- Single-pore transport factors -----------------------
# Minimal scalar versions (kept exactly as in your code) with guards at λ≥1.

def lam_simp(x, a):
    """Return λ = a/x if x > a else ~1 (clamped to avoid singularities). Both x and a in nm."""
    return a / x if x > a else 0.99999999

def K_s(r, a):
    """Hindered convection factor for solute in a cylindrical pore (scalar r)."""
    lam = lam_simp(r, a)
    if lam >= 1:
        return 0
    term1 = (9/4) * (np.pi**2) * np.sqrt(2) * ((1 - lam)**(-5/2))
    term2 = 1 + (7/60) * (1 - lam) - (2227/50400) * ((1 - lam)**2)
    term3 = 4.0180 - 3.9788 * lam - 1.9215 * lam**2 + 4.392 * lam**3 + 5.006 * lam**4
    return term1 * term2 * term3

def K_t(r, a):
    """Hydrodynamic drag factor (scalar r)."""
    lam = lam_simp(r, a)
    if lam >= 1:
        return np.inf
    term1 = (9/4) * (np.pi**2) * np.sqrt(2) * ((1 - lam)**(-5/2))
    term2 = 1 - (73/60) * (1 - lam) + (77293/50400) * ((1 - lam)**2)
    term3 = 22.5083 - 5.6117 * lam - 0.3363 * lam**2 - 1.216 * lam**3 + 1.647 * lam**4
    return term1 * term2 * term3

def K_c(r, a):
    """Composite convection factor K_c = ((2 - φ) K_s) / (2 K_t)."""
    lam = lam_simp(r, a)
    if lam >= 1:
        return 0
    phi = (1 - lam) ** 2
    k_t_value = K_t(r, a)
    if np.isinf(k_t_value) or k_t_value == 0:
        return 0
    return ((2 - phi) * K_s(r, a)) / (2 * k_t_value)

def K_d(r, a):
```

```python
    """Diffusive factor K_d = 6π / K_t."""
    lam = lam_simp(r, a)
    if lam >= 1:
        return np.inf
    k_t_value = K_t(r, a)
    if np.isinf(k_t_value) or k_t_value == 0:
        return np.inf
    return 6 * np.pi / k_t_value

# ------------------------ Weighted integration over the PSD ------------------------
Pe_values = []   # Pe_m(a) per dextran size
Kc_values = []   # r^4-weighted <K_c>
Kd_values = []   # r^4-weighted <K_d>

for i, a in enumerate(a_values):
    D_current = D_inf[i]  # D∞ for this dextran size (m^2/s)

    # Evaluate K_c and K_d on each discrete pore-size bin center.
    # IMPORTANT: 'pore_sizes' are assumed to be pore *radii* in nm. If they are
    # diameters, convert first: r = pore_sizes / 2.
    Kc_discrete = np.array([K_c(r, a) for r in pore_sizes])
    Kd_discrete = np.array([K_d(r, a) for r in pore_sizes])

    # r^4 throughput weighting (Poiseuille) times discrete number frequency P_values.
    # No explicit normalization is needed; we divide by the same sum below.
    weights = P_values * (pore_sizes**4)

    numerator_Kc  = np.sum(Kc_discrete * weights)
    denominator_ps = np.sum(weights)
    Kc_integrated = numerator_Kc / denominator_ps if denominator_ps > 0 else 0

    numerator_Kd  = np.sum(Kd_discrete * weights)
    Kd_integrated = numerator_Kd / denominator_ps if denominator_ps > 0 else 0

    # Effective Pe_m using the *ratio of weighted means* <K_c>/<K_d>.
    # If you instead want the *weighted mean of the ratio* (K_c/K_d), replace the two
    # blocks above with a single sum over (Kc_discrete / Kd_discrete) * weights.
    Pe_value = (J_v_m_s * delta_m / epsilon_s) * (Kc_integrated / Kd_integrated) / D_current

    Pe_values.append(Pe_value)
    Kc_values.append(Kc_integrated)
    Kd_values.append(Kd_integrated)

# Convert results to NumPy arrays for plotting/printing
Pe_values = np.array(Pe_values)
Kc_values = np.array(Kc_values)
Kd_values = np.array(Kd_values)

# ------------------------ Plots ------------------------
# K_d vs dextran *diameter* (we multiply a by 2 only for x-axis labeling)
plt.figure(figsize=(8, 6))
plt.plot(a_values*2, Kd_values, marker='o', linestyle='-', color='r', label="K_d (r^4-weighted mean)")
plt.xlabel("Dextran Diameter, D = 2a (nm)", fontsize=14, fontweight="bold")
plt.ylabel("K_d (dimensionless)", fontsize=14, fontweight="bold")
plt.title("K_d vs. Dextran Size", fontsize=16, fontweight="bold")
plt.legend(); plt.grid(True); plt.show()

# K_c vs dextran diameter
plt.figure(figsize=(8, 6))
plt.plot(a_values*2, Kc_values, marker='o', linestyle='-', color='g', label="K_c (r^4-weighted mean)")
plt.xlabel("Dextran Diameter, D = 2a (nm)", fontsize=14, fontweight="bold")
plt.ylabel("K_c (dimensionless)", fontsize=14, fontweight="bold")
plt.title("K_c vs. Dextran Size", fontsize=16, fontweight="bold")
plt.legend(); plt.grid(True); plt.show()

# Pe_m vs dextran diameter
plt.figure(figsize=(8, 6))
plt.plot(a_values*2, Pe_values, marker='o', linestyle='-', color='b', label="Péclet Number (Pe_m)")
plt.xlabel("Dextran Diameter, D = 2a (nm)")
```

```python
plt.ylabel("Péclet Number, Pe_m")
plt.ylim(0, 5)              # y-axis limit; adjust if your Pe spans outside this range
plt.xlim(0, 20)             # dextran diameter range shown
plt.title("Wall Péclet Number vs. Dextran Size")
plt.legend(); plt.grid(True); plt.show()

# Print values (D = 2a in nm, Pe_m as a fraction)
for x, y in zip(a_values*2, Pe_values):
    print(f"{x}\t{y}")
```

## S4.3. Fitting Bungay–Brenner model To Retention Data:

```python
# -*- coding: utf-8 -*-
"""
Created on Tue Aug 26 17:46:19 2025

@author: szein
"""

# -*- coding: utf-8 -*-
SI §S2.x — Bungay-Brenner (BB) fits to dextran retention curves
----------------------------------------------------------------

Purpose
-------
Fit the apparent sieving/retention curves for neutral dextrans on UF/RO samples
using a Bungay-Brenner-type hindered transport model, constrained by SEM-derived
surface porosity (ε_s) and selective-layer thickness (δ_m). The fit parameterizes
the *pore-radius distribution* p(r) as a log-normal with mean radius r̂ and width
σ = y * r̂ (0.1 ≤ y ≤ 0.5), integrates the single-pore sieving S_a(r, a, D∞) over
p(r) with a *throughput* weight r^4 (Poiseuille), and minimizes the mean absolute
error between the experimental sieving S_exp(a) and the heterogeneous prediction
S_het(a; r̂, y).

Input
-----
Excel file (Sheet1) with:
- Column J: dextran hydrodynamic radius a [nm]                (rows 102:535)
- Column W: experimental *apparent sieving* S_exp = 1 - R (0..1)  (rows 102:535)
- Column K: bulk diffusion coefficient D∞ [m^2/s]             (rows 102:535)

Model (summary)
---------------
λ = a / r_p  (solute-to-pore radius ratio)  ;  clamp λ < 1
K_s(λ), K_t(λ)  →  K_c = ((2 - (1-λ)^2) K_s) / (2 K_t),   K_d = 6π / K_t
S_∞(r,a) = (1 - λ)^2 * K_c(r,a)
Pe_m = (J_v * δ_m / ε_s) * (K_c / K_d) / D∞
S_a(r,a,D∞) = S_∞ * exp(Pe_m) / ( S_∞ + exp(Pe_m) - 1 )     # numerically guarded

Heterogeneous averaging (throughput-weighted):
S_het(a) = ∫ S_a(r,a,D∞) p(r) r^4 dr  /  ∫ p(r) r^4 dr

Numerics / Guards
-----------------
- All lengths for λ are in *nm* consistently (r and a), which cancels in λ.
- Pe exponent is clipped to [-700, 700] to avoid overflow.
- r-grid: uniform nm grid (1..15 nm here; adjust if needed).
- Optimization: L-BFGS-B over (r̂, y) with bounds r̂∈[4,10] nm, y∈[0.1,0.5].

Outputs
-------
- Overlay plot of experimental retention R_exp = (1 - S_exp) and fitted R_calc.
```

```python
    - Text report of optimal (mean pore *diameter* 2 r̂) and relative width y.
    - Optional PSD plot (log-normal estimate), scaled to overlay with an SEM histogram.
    - Excel sheet "R_fit" with Dextran D (nm), R (exp), and R (calc).

    Reproducibility Notes
    ---------------------
    - Units: J_v in m/s (LMH→m/s), δ_m in m, D∞ in m^2/s, ε_s dimensionless.
    - This code fits only the *curve shape* through (r̂, y); ε_s and δ_m are fixed.
    - The optional SEM histogram scaling at the end is cosmetic (plot overlay only).
    """

import numpy as np
import pandas as pd
import matplotlib.pyplot as plt
from scipy.optimize import minimize

# -------------------- 1) Load data --------------------
file_path  = r'C:\Users\szein\Desktop\Others-membrane\mwco\20240927- 0.5 psi PW - 0.6 AD- 1.3 LMH for
both\PW-P 1.3LMH 20Sep2024.xlsx'
sheet_name = 'Sheet1'
df = pd.read_excel(file_path, sheet_name=sheet_name)

# Rows/columns holding the dextran hydrodynamic radius a [nm], experimental S, and D∞ [m^2/s]
ROW_SLICE    = slice(102, 536)  # Python slice is [start, stop] → rows 102..535 inclusive
COL_A_NM     = 9   # Column J (0-based index): a [nm]
COL_SIEVING  = 22  # Column W: S_exp = 1 - R  (fraction 0..1)
COL_D_INF    = 10  # Column K: D∞ [m^2/s]

a_nm  = df.iloc[ROW_SLICE, COL_A_NM].to_numpy(dtype=float)      # dextran hydrodynamic radius [nm]
S_exp = df.iloc[ROW_SLICE, COL_SIEVING].to_numpy(dtype=float)   # apparent sieving (fraction 0..1)
D_inf = df.iloc[ROW_SLICE, COL_D_INF].to_numpy(dtype=float)     # diffusion coefficient [m^2/s]

# -------------------- 2) Fixed membrane/operating parameters (SI) --------------------
J_v_LMH = 1.3                         # permeate flux [L m^-2 h^-1]
J_v     = (J_v_LMH * 1e-3) / 3600.0   # convert to m/s (1 L m^-2 h^-1 = 1e-3/3600 m/s)
epsilon = 0.227                       # surface porosity ε_s [-] from uncoated SEM
delta_m = 150e-9                      # selective-layer thickness [m] from cross SEM

# -------------------- 3) Pore-radius grid for integration --------------------
# Use nm here so λ = a/r is dimensionally consistent without conversions.
r_min_nm, r_max_nm = 1.0, 15.0
dr_nm = (r_max_nm - r_min_nm) / 500.0
r_grid_nm = np.arange(r_min_nm, r_max_nm, dr_nm)  # uniform grid in nm

# -------------------- 4) Hindered transport primitives --------------------
# Clamp λ = a/r to <1 to avoid singularities in K_s, K_t; when a≥r, set very near 1.
def _lambda(a_nm, r_nm):
    lam = a_nm / r_nm
    return np.where(r_nm > a_nm, lam, 0.99999999)

def K_s(r_nm, a_nm):
    lam = _lambda(a_nm, r_nm)
    valid = lam < 1.0
    out = np.zeros_like(r_nm, dtype=float)
    if not np.any(valid):
        return out
    one_minus = 1.0 - lam[valid]
    term1 = (9.0/4.0) * (np.pi**2) * np.sqrt(2.0) * (one_minus**(-2.5))
    term2 = 1.0 + (7.0/60.0)*one_minus - (2227./50400.0)*(one_minus**2)
    term3 = 4.0180 - 3.9788*lam[valid] - 1.9215*(lam[valid]**2) + 4.392*(lam[valid]**3) +
5.006*(lam[valid]**4)
    out[valid] = term1 * term2 * term3
    return out

def K_t(r_nm, a_nm):
    lam = _lambda(a_nm, r_nm)
    valid = lam < 1.0
    out = np.full_like(r_nm, np.inf, dtype=float)
    if not np.any(valid):
```

```python
        return out
    one_minus = 1.0 - lam[valid]
    term1 = (9.0/4.0) * (np.pi**2) * np.sqrt(2.0) * (one_minus**(-2.5))
    term2 = 1.0 - (73.0/60.0)*one_minus + (77293.0/50400.0)*(one_minus**2)
    term3 = 22.5083 - 5.6117*lam[valid] - 0.3363*(lam[valid]**2) - 1.216*(lam[valid]**3) +
1.647*(lam[valid]**4)
    out[valid] = term1 * term2 * term3
    return out

def K_c(r_nm, a_nm):
    lam = _lambda(a_nm, r_nm)
    kt = K_t(r_nm, a_nm)
    ks = K_s(r_nm, a_nm)
    phi = np.where(lam < 1.0, (1.0 - lam)**2, 0.0)
    valid = (lam < 1.0) & np.isfinite(kt) & (kt != 0.0)
    out = np.zeros_like(r_nm, dtype=float)
    out[valid] = ((2.0 - phi[valid]) * ks[valid]) / (2.0 * kt[valid])
    return out

def K_d(r_nm, a_nm):
    kt = K_t(r_nm, a_nm)
    out = np.full_like(r_nm, np.inf, dtype=float)
    mask = np.isfinite(kt) & (kt != 0.0)
    out[mask] = 6.0 * np.pi / kt[mask]
    return out

def S_inf(r_nm, a_nm):
    """Infinite-Pe sieving for a single pore radius (fraction 0..1)."""
    lam = _lambda(a_nm, r_nm)
    return np.clip((1.0 - lam)**2, 0.0, 1.0) * K_c(r_nm, a_nm)

def Pe_m(r_nm, a_nm, D_inf_m2s):
    """Wall Péclet number for a single pore radius; D∞ is bulk diffusion [m^2/s]."""
    ratio = K_c(r_nm, a_nm) / K_d(r_nm, a_nm)
    return (J_v * delta_m / epsilon) * np.where(np.isfinite(ratio), ratio, 0.0) / D_inf_m2s

def S_a_single(r_nm, a_nm, D_inf_m2s):
    """
    Apparent sieving at finite Pe for a single pore radius.
    Numerically guarded form: S∞ * e^Pe / (S∞ + e^Pe - 1).
    """
    Sinf = S_inf(r_nm, a_nm)
    Pe  = np.clip(Pe_m(r_nm, a_nm, D_inf_m2s), -700.0, 700.0)  # avoid overflow
    ePe = np.exp(Pe)
    denom = Sinf + ePe - 1.0
    denom = np.where(denom != 0.0, denom, 1e-12)
    return Sinf * ePe / denom

# --------------------- 5) Log-normal PSD (phenomenological) --------------------
def lognorm_pdf(r_nm, rbar_nm, sigma_abs_nm):
    """
    Log-normal over radius with mean r̂ and absolute width σ (phenomenological).
    For stability at the grid ends, we clamp r_nm to positive values.
    """
    r = np.clip(r_nm, 1e-9, None)
    # Convert (r̂, σ) to log-space variance via σ_l^2 = ln(1 + (σ/r̂)^2)
    sigma_l2 = np.log(1.0 + (sigma_abs_nm / rbar_nm)**2)
    denom = r * np.sqrt(2.0 * np.pi * sigma_l2)
    z = np.log(r / rbar_nm)
    return np.exp(-(z**2) / (2.0 * sigma_l2)) / denom

# --------------------- 6) Heterogeneous average (throughput-weighted) --------------------
def S_het(a_nm_scalar, D_inf_m2s_scalar, rbar_nm, y_rel):
    """
    Throughput-weighted sieving for one dextran size.
    p(r) = lognormal(r̂, σ=y*r̂); weighting r^4 corresponds to Poiseuille flow.
    """
    sigma_nm = y_rel * rbar_nm
    p = lognorm_pdf(r_grid_nm, rbar_nm, sigma_nm)                 # number-PSD shape
```

```python
    w = p * (r_grid_nm**4)                              # throughput weighting
    S = S_a_single(r_grid_nm, a_nm_scalar, D_inf_m2s_scalar)  # single-pore sieving
    num = np.trapz(S * w, r_grid_nm)                    # ∫ S p r^4 dr
    den = np.trapz(w, r_grid_nm)                        # ∫ p r^4 dr
    return num / den if den > 0 else np.inf

def S_het_vector(a_nm_vec, D_inf_vec, rbar_nm, y_rel):
    return np.array([S_het(a_i, D_i, rbar_nm, y_rel) for a_i, D_i in zip(a_nm_vec, D_inf_vec)],
dtype=float)

# -------------------- 7) Objective and optimization --------------------
def objective(params):
    """
    Minimize mean absolute error between experimental S_exp and S_het.
    params = [r̂ (nm), y (σ/r̂)] ; y in [0.1, 0.5] keeps shapes reasonable.
    """
    rbar_nm, y = params
    if (rbar_nm <= 0) or (y < 0.1) or (y > 0.5):
        return np.inf
    S_pred = S_het_vector(a_nm, D_inf, rbar_nm, y)
    # Penalize any NaNs/Infs to steer the optimizer away from unstable regions
    if not np.all(np.isfinite(S_pred)):
        return 1e6
    return np.mean(np.abs(S_exp - S_pred))

bounds = [(4.0, 10.0), (0.2, 0.5)]        # r̂ in nm, y = σ/r̂ (dimensionless)
x0     = [6.0, 0.5]                        # initial guess
res = minimize(objective, x0, bounds=bounds, method='L-BFGS-B')

# -------------------- 8) Plot retention fits (R = 1 - S) --------------------
if res.success:
    rbar_opt, y_opt = res.x
    sigma_opt = y_opt * rbar_opt
    S_fit = S_het_vector(a_nm, D_inf, rbar_opt, y_opt)
    R_exp = 1.0 - S_exp
    R_fit = 1.0 - S_fit

    plt.figure(figsize=(9, 5.5))
    plt.scatter(a_nm*2.0, R_exp, s=18, color='k', label='Experimental R', zorder=3)
    plt.plot(a_nm*2.0, R_fit, '--', lw=2.0, color='tab:red', label='Model fit')
    plt.xlabel('Dextran Diameter, D = 2a (nm)', fontsize=13)
    plt.ylabel('Retention, R (fraction)', fontsize=13)  # use 0..1 to match curves
    plt.title('Dextran Retention: Bungay–Brenner Fit (constrained by ε, δ_m)', fontsize=14)
    plt.grid(True, alpha=0.3); plt.legend(frameon=False)
    plt.tight_layout(); plt.show()

    print(f"[Fit] mean pore DIAMETER: {2.0*rbar_opt:.3f} nm   (r̂ = {rbar_opt:.3f} nm)")
    print(f"[Fit] relative width     : y = σ/r̂ = {y_opt:.3f}   (σ = {sigma_opt:.3f} nm)")
else:
    print(f"[Optimization failed] {res.message}")
    # Exit early if you prefer; we continue so the rest of the SI cell still runs.

# -------------------- 9) Optional: plot fitted PSD (for overlay with SEM histogram) --------------------
# This is cosmetic and *not used* in the fit; it just visualizes the inferred p(r).
try:
    r_plot = np.linspace(r_min_nm, r_max_nm, 1000)
    p_plot = lognorm_pdf(r_plot, rbar_opt, sigma_opt)
    # Convert to diameter for publication plotting
    d_plot = 2.0 * r_plot
    plt.figure(figsize=(9, 5.2))
    plt.plot(d_plot, p_plot, color='tab:green', lw=2.0, label='Inferred PSD (radius→diameter)')
    plt.xlabel('Pore Maximum Inscribed Circle Diameter (nm)', fontsize=13)
    plt.ylabel('Frequency (arb.)', fontsize=13)
    plt.title('Fitted Log-normal PSD (from retention)', fontsize=14)
    plt.grid(True, alpha=0.3); plt.legend(frameon=False)
    plt.tight_layout(); plt.show()
except Exception as e:
    print(f"[PSD plot skipped] {e}")
```

```
# -------------------- 10) Save R vs D (for plotting in Illustrator/Origin) --------------------
save_path = r"C:\Users\szein\Desktop\Re-visit\review\Bungay and Brenner .xlsx"
x_nm = (a_nm * 2.0).astype(float)
R_exp = (1.0 - S_exp).astype(float)
R_fit = (1.0 - S_het_vector(a_nm, D_inf, rbar_opt, y_opt)).astype(float) if res.success else
np.full_like(R_exp, np.nan)

out = pd.DataFrame({"Dextran D (nm)": x_nm, "R (exp)": R_exp, "R (calc)": R_fit})

# Create workbook or replace the 'R_fit' sheet if it exists
try:
    with pd.ExcelWriter(save_path, engine="openpyxl", mode="a", if_sheet_exists="replace") as w:
        out.to_excel(w, sheet_name="R_fit", index=False)
except FileNotFoundError:
    with pd.ExcelWriter(save_path, engine="openpyxl", mode="w") as w:
        out.to_excel(w, sheet_name="R_fit", index=False)

print(f"[Saved] {save_path}  (sheet: R_fit)")
```

## References

(1) Nasir, A. M.; Adam, M. R.; Mohamad Kamal, S. N. E. A.; Jaafar, J.; Othman, M. H. D.; Ismail, A. F.; Aziz, F.; Yusof, N.; Bilad, M. R.; Mohamud, R.; A. Rahman, M.; Wan Salleh, W. N. A Review of the Potential of Conventional and Advanced Membrane Technology in the Removal of Pathogens from Wastewater. *Sep. Purif. Technol.* **2022**, *286*, 120454. https://doi.org/10.1016/j.seppur.2022.120454.

(2) Zhang, Q.; Zhou, R.; Peng, X.; Li, N.; Dai, Z. Development of Support Layers and Their Impact on the Performance of Thin Film Composite Membranes (TFC) for Water Treatment. *Polymers* **2023**, *15* (15). https://doi.org/10.3390/polym15153290.

(3) Hampu, N.; Werber, J. R.; Chan, W. Y.; Feinberg, E. C.; Hillmyer, M. A. Next-Generation Ultrafiltration Membranes Enabled by Block Polymers. *ACS Nano* **2020**, *14* (12), 16446–16471. https://doi.org/10.1021/acsnano.0c07883.

(4) Deen, W. M. Hindered Transport of Large Molecules in Liquid-filled Pores. *AIChE J.* **1987**, *33* (9), 1409–1425. https://doi.org/10.1002/aic.690330902.

(5) Ferry, J. D. STATISTICAL EVALUATION OF SIEVE CONSTANTS IN ULTRAFILTRATION. *J. Gen. Physiol.* **1936**, *20* (1), 95–104. https://doi.org/10.1085/jgp.20.1.95.

(6) Zeman, L.; Wales, M. Steric Rejection of Polymeric Solutes by Membranes with Uniform Pore Size Distribution. *Sep. Sci. Technol.* **1981**, *16* (3), 275–290. https://doi.org/10.1080/01496398108068519.

(7) Mehta, A.; Zydney, A. L. Permeability and Selectivity Analysis for Ultrafiltration Membranes. *J. Membr. Sci.* **2005**, *249* (1–2), 245–249. https://doi.org/10.1016/j.memsci.2004.09.040.

(8) Werber, J.; Elimelech, M. Materials for Next-Generation Desalination and Water Purification Membranes. *Nat. Rev. Mater.* **2016**, *1*, 16018. https://doi.org/10.1038/natrevmats.2016.18.

(9) Ramon, G. Z.; Wong, M. C. Y.; Hoek, E. M. V. Transport through Composite Membrane, Part 1: Is There an Optimal Support Membrane? *J. Membr. Sci.* **2012**, *415–416*, 298–305. https://doi.org/10.1016/j.memsci.2012.05.013.

(10) Jiang, Z.; Karan, S.; Livingston, A. G. Water Transport through Ultrathin Polyamide Nanofilms Used for Reverse Osmosis. *Adv. Mater.* **2018**, *30* (15), 1705973. https://doi.org/10.1002/adma.201705973.

(11) Enninful, H. R. N. B.; Schneider, D.; Enke, D.; Valiullin, R. Impact of Geometrical Disorder on Phase Equilibria of Fluids and Solids Confined in Mesoporous Materials. *Langmuir* **2021**, *37* (12), 3521–3537. https://doi.org/10.1021/acs.langmuir.0c03047.

(12) Sanz, J. M.; Jardines, D.; Bottino, A.; Capannelli, G.; Hernández, A.; Calvo, J. I. Liquid–Liquid Porometry for an Accurate Membrane Characterization. *Desalination* **2006**, *200* (1–3), 195–197. https://doi.org/10.1016/j.desal.2006.03.293.

(13) Maalal, O.; Prat, M.; Peinador, R.; Lasseux, D. Evaluation of Pore Size Distribution via Fluid-Fluid Displacement Porosimetry: The Viscous Bias. *Int. J. Multiph. Flow* **2022**, *149*, 103983. https://doi.org/10.1016/j.ijmultiphaseflow.2022.103983.

(14) Tovbin, Yu. K. Universality of an Estimate of the Minimum Size of an Equilibrium Phase. *Russ. J. Phys. Chem. A* **2010**, *84* (9), 1640–1643. https://doi.org/10.1134/S0036024410090359.

(15) Pidstryhach Institute for Applied Problems of Mechanics and Mathematics of National Academy of Sciences of Ukraine; Holubets, T. Investigation of the Structural Properties of Porous Material According to the Sorption Isotherms and Drainage Curves. *Math. Model. Comput.* **2016**, *3* (1), 23–32. https://doi.org/10.23939/mmc2016.01.023.

(16) Denoyel, R.; Llewellyn, P.; Beurroies, I.; Rouquerol, J.; Rouquerol, F.; Luciani, L. Comparing the Basic Phenomena Involved in Three Methods of Pore-size Characterization: Gas Adsorption, Liquid Intrusion and Thermoporometry. *Part. Part. Syst. Charact.* **2004**, *21* (2), 128–137. https://doi.org/10.1002/ppsc.200400929.

(17) Yang, J.-W.; Cui, Y.-J.; Mokni, N.; Ormea, E. Investigation into the Mercury Intrusion Porosimetry (MIP) and Micro-Computed Tomography ($\mu$CT) Methods for Determining the Pore Size Distribution of MX80 Bentonite Pellet. *Acta Geotech.* **2023**, *19*, 1–13. https://doi.org/10.1007/s11440-023-01863-y.

(18) Linden, S.; Cheng, L.; Wiegmann, A. *Specialized Methods for Direct Numerical Simulations in Porous Media*; 2018. https://doi.org/10.30423/report.m2m-2018-01.

(19) Tanis-Kanbur, M. B.; Peinador, R. I.; Calvo, J. I.; Hernández, A.; Chew, J. W. Porosimetric Membrane Characterization Techniques: A Review. *J. Membr. Sci.* **2021**, *619*, 118750. https://doi.org/10.1016/j.memsci.2020.118750.

(20) Zheng, L.; Li, H.; Yu, H.; Kang, G.; Xu, T.; Yu, J.; Li, X.; Xu, H. "Modified" Liquid–Liquid Displacement Porometry and Its Applications in Pd-Based Composite Membranes. *Membranes* **2018**, *8* (2), 29. https://doi.org/10.3390/membranes8020029.

(21) Lopez Marquez, A.; Gareis, I. E.; Dias, F. J.; Gerhard, C.; Lezcano, M. F. Methods to Characterize Electrospun Scaffold Morphology: A Critical Review. *Polymers* **2022**, *14* (3), 467. https://doi.org/10.3390/polym14030467.

(22) Singh, R. Chapter 1 - Introduction to Membrane Technology. In *Membrane Technology and Engineering for Water Purification (Second Edition)*; Singh, R., Ed.; Butterworth-Heinemann: Oxford, 2015; pp 1–80. https://doi.org/10.1016/B978-0-444-63362-0.00001-X.

(23) Yehl, C. J.; Zydney, A. L. Characterization of Dextran Transport and Molecular Weight Cutoff (MWCO) of Large Pore Size Hollow Fiber Ultrafiltration Membranes. *J. Membr. Sci.* **2021**, *622*, 119025. https://doi.org/10.1016/j.memsci.2020.119025.

(24) Wang, K. The Effects of Flow Angle and Shear Rate within the Spinneret on the Separation Performance of Poly(Ethersulfone) (PES) Ultrafiltration Hollow Fiber Membranes. *J. Membr. Sci.* **2004**, *240* (1–2), 67–79. https://doi.org/10.1016/j.memsci.2004.04.012.

(25) Wickramasinghe, S. R.; Bower, S. E.; Chen, Z.; Mukherjee, A.; Husson, S. M. Relating the Pore Size Distribution of Ultrafiltration Membranes to Dextran Rejection. *J. Membr. Sci.* **2009**, *340* (1–2), 1–8. https://doi.org/10.1016/j.memsci.2009.04.056.

(26) Sun, P. Z.; Yagmurcukardes, M.; Zhang, R.; Kuang, W. J.; Lozada-Hidalgo, M.; Liu, B. L.; Cheng, H.-M.; Wang, F. C.; Peeters, F. M.; Grigorieva, I. V.; Geim, A. K. Exponentially Selective Molecular Sieving through Angstrom Pores. *Nat. Commun.* **2021**, *12* (1), 7170. https://doi.org/10.1038/s41467-021-27347-9.

(27) Abdullah, S. Z.; Bérubé, P. R.; Horne, D. J. SEM Imaging of Membranes: Importance of Sample Preparation and Imaging Parameters. *J. Membr. Sci.* **2014**, *463*, 113–125. https://doi.org/10.1016/j.memsci.2014.03.048.

(28) Alqaheem, Y.; Alomair, A. A. Microscopy and Spectroscopy Techniques for Characterization of Polymeric Membranes. *Membranes* **2020**, *10* (2), 33. https://doi.org/10.3390/membranes10020033.

(29)  Ziel, R.; Haus, A.; Tulke, A. Quantification of the Pore Size Distribution (Porosity Profiles) in Microfiltration Membranes by SEM, TEM and Computer Image Analysis. *J. Membr. Sci.* **2008**, *323* (2), 241–246. https://doi.org/10.1016/j.memsci.2008.05.057.

(30)  Wang, L.; Wang, X. Study of Membrane Morphology by Microscopic Image Analysis and Membrane Structure Parameter Model. *J. Membr. Sci.* **2006**, *283* (1–2), 109–115. https://doi.org/10.1016/j.memsci.2006.06.017.

(31)  Shi, M.; Wang, Z.; Zhao, S.; Wang, J.; Wang, S. A Support Surface Pore Structure Re-Construction Method to Enhance the Flux of TFC RO Membrane. *J. Membr. Sci.* **2017**, *541*, 39–52. https://doi.org/10.1016/j.memsci.2017.06.087.

(32)  Liu, B.; Chen, C.; Zhao, P.; Li, T.; Liu, C.; Wang, Q.; Chen, Y.; Crittenden, J. Thin-Film Composite Forward Osmosis Membranes with Substrate Layer Composed of Polysulfone Blended with PEG or Polysulfone Grafted PEG Methyl Ether Methacrylate. *Front. Chem. Sci. Eng.* **2016**, *10* (4), 562–574. https://doi.org/10.1007/s11705-016-1588-9.

(33)  Shi, M.; Dong, C.; Wang, Z.; Tian, X.; Zhao, S.; Wang, J. Support Surface Pore Structures Matter: Effects of Support Surface Pore Structures on the TFC Gas Separation Membrane Performance over a Wide Pressure Range. *Chin. J. Chem. Eng.* **2019**, *27* (8), 1807–1816. https://doi.org/10.1016/j.cjche.2018.12.009.

(34)  Yan, W.; Wang, Z.; Wu, J.; Zhao, S.; Wang, J.; Wang, S. Enhancing the Flux of Brackish Water TFC RO Membrane by Improving Support Surface Porosity via a Secondary Pore-Forming Method. *J. Membr. Sci.* **2016**, *498*, 227–241. https://doi.org/10.1016/j.memsci.2015.10.029.

(35)  Wang, S.; Li, Q.; He, B.; Gao, M.; Ji, Y.; Cui, Z.; Yan, F.; Ma, X.; Younas, M.; Li, J. Preparation of Small-Pore Ultrafiltration Membranes with High Surface Porosity by In Situ $CO_2$ Nanobubble-Assisted NIPS. *ACS Appl. Mater. Interfaces* **2022**, *14* (6), 8633–8643. https://doi.org/10.1021/acsami.1c23760.

(36)  Joy, D. C.; Joy, C. S. Low Voltage Scanning Electron Microscopy. *Micron* **1996**, *27* (3), 247–263. https://doi.org/10.1016/0968-4328(96)00023-6.

(37)  Wu, J.; Xiao, M.; Quezada-Renteria, J. A.; Hou, Z.; Hoek, E. M. V. Sample Preparation Matters: Scanning Electron Microscopic Characterization of Polymeric Membranes. *J. Membr. Sci. Lett.* **2024**, *4* (1), 100073. https://doi.org/10.1016/j.memlet.2024.100073.

(38)  Molina, S.; Landaburu-Aguirre, J.; Rodríguez-Sáez, L.; García-Pacheco, R.; De La Campa, J. G.; García-Calvo, E. Effect of Sodium Hypochlorite Exposure on Polysulfone Recycled UF Membranes and Their Surface Characterization. *Polym. Degrad. Stab.* **2018**, *150*, 46–56. https://doi.org/10.1016/j.polymdegradstab.2018.02.012.

(39)  Darvishmanesh, S.; Degrève, J.; Van Der Bruggen, B. Performance of Solvent-Pretreated Polyimide Nanofiltration Membranes for Separation of Dissolved Dyes from Toluene. *Ind. Eng. Chem. Res.* **2010**, *49* (19), 9330–9338. https://doi.org/10.1021/ie101050k.

(40)  *Scanning Electron Microscopy and X-Ray Microanalysis*, 3rd ed.; Goldstein, J., Ed.; Kluwer Academic/Plenum Publishers: New York, 2003.

(41)  Wang, A. Y.; Sharma, V.; Saini, H.; Tingen, J. N.; Flores, A.; Liu, D.; Safain, M. G.; Kryzanski, J.; McPhail, E. D.; Arkun, K.; Riesenburger, R. I. Machine Learning Quantification of Amyloid Deposits in Histological Images of Ligamentum Flavum. *J. Pathol. Inform.* **2022**, *13*, 100013. https://doi.org/10.1016/j.jpi.2022.100013.

(42)  Ortega-Ramírez, P.; Pot, V.; Laville, P.; Schlüter, S.; Amor-Quiroz, D. A.; Hadjar, D.; Mazurier, A.; Lacoste, M.; Caurel, C.; Pouteau, V.; Chenu, C.; Basile-Doelsch, I.; Henault, C.; Garnier, P. Pore Distances of Particulate Organic Matter Predict N 2 O Emissions from Intact Soil at Moist Conditions. *Geoderma* **2023**, *429*, 116224. https://doi.org/10.1016/j.geoderma.2022.116224.

(43)  Zeman, L. J.; Zydney, A. *Microfiltration and Ultrafiltration: Principles and Applications*; 2017; p 618. https://doi.org/10.1201/9780203747223.

(44)  Dechadilok, P.; Deen, W. M. Hindrance Factors for Diffusion and Convection in Pores. *Ind. Eng. Chem. Res.* **2006**, *45* (21), 6953–6959. https://doi.org/10.1021/ie051387n.

(45) Kasemset, S.; Wang, L.; He, Z.; Miller, D. J.; Kirschner, A.; Freeman, B. D.; Sharma, M. M. Influence of Polydopamine Deposition Conditions on Hydraulic Permeability, Sieving Coefficients, Pore Size and Pore Size Distribution for a Polysulfone Ultrafiltration Membrane. *J. Membr. Sci.* **2017**, *522*, 100–115. https://doi.org/10.1016/j.memsci.2016.07.016.

(46) Vrijenhoek, E. M.; Hong, S.; Elimelech, M. Influence of Membrane Surface Properties on Initial Rate of Colloidal Fouling of Reverse Osmosis and Nanofiltration Membranes. *J. Membr. Sci.* **2001**, *188* (1), 115–128. https://doi.org/10.1016/S0376-7388(01)00376-3.

(47) Tan, K. S.; Lam, C. K.; Tan, W. C.; Ooi, H. S.; Lim, Z. H. A Review of Image Processing and Quantification Analysis for Solid Oxide Fuel Cell. *Energy AI* **2024**, *16*, 100354. https://doi.org/10.1016/j.egyai.2024.100354.

(48) Shi, B.; Patel, M.; Yu, D.; Yan, J.; Li, Z.; Petriw, D.; Pruyn, T.; Smyth, K.; Passeport, E.; Miller, R. J. D.; Howe, J. Y. Automatic Quantification and Classification of Microplastics in Scanning Electron Micrographs via Deep Learning. *Sci. Total Environ.* **2022**, *825*, 153903. https://doi.org/10.1016/j.scitotenv.2022.153903.

(49) Lin, Y.; Joy, D. C. A New Examination of Secondary Electron Yield Data. *Surf. Interface Anal.* **2005**, *37* (11), 895–900. https://doi.org/10.1002/sia.2107.

(50) Schatten, H. Low Voltage High-Resolution SEM (LVHRSEM) for Biological Structural and Molecular Analysis. *Biol. Specim. Prep. Preserv. High Resolut. Microsc.* **2011**, *42* (2), 175–185. https://doi.org/10.1016/j.micron.2010.08.008.

(51) *Polymer Microscopy*; Springer New York: New York, NY, 2008. https://doi.org/10.1007/978-0-387-72628-1.

(52) Khan, S. B.; Wu, H.; Fei, Z.; Ning, S.; Zhang, Z. Antireflective Coatings with Enhanced Adhesion Strength. *Nanoscale* **2017**, *9* (31), 11047–11054. https://doi.org/10.1039/C7NR02334K.

(53) Rauschenbach, B. Ion Beam Sputtering Induced Glancing Angle Deposition; 2022; pp 613–662. https://doi.org/10.1007/978-3-030-97277-6_11.

(54) Liu, M.-J.; Zhang, M.; Zhang, X.-F.; Li, G.-R.; Zhang, Q.; Li, C.-X.; Li, C.-J.; Yang, G.-J. Transport and Deposition Behaviors of Vapor Coating Materials in Plasma Spray-Physical Vapor Deposition. *Appl. Surf. Sci.* **2019**, *486*, 80–92. https://doi.org/10.1016/j.apsusc.2019.04.224.

(55) Xiao, M.; Wang, X.; Hou, Z.; Alan Quezada Renteria, J.; Dlamini, D. S.; Jassby, D.; Hoek, E. M. V. Comparison of Classical Hydrodynamic Models of Transport through Porous Membranes. *Sep. Purif. Technol.* **2025**, *360*, 131189. https://doi.org/10.1016/j.seppur.2024.131189.

(56) Zeman, L. J.; Wales, M. Steric Rejection of Polymeric Solutes by Membranes with Uniform Pore Size Distribution. *Sep. Sci. Technol.* **1981**, *16*, 275–290.

(57) Kasemset, S.; Wang, L.; He, Z.; Miller, D. J.; Kirschner, A.; Freeman, B. D.; Sharma, M. M. Influence of Polydopamine Deposition Conditions on Hydraulic Permeability, Sieving Coefficients, Pore Size and Pore Size Distribution for a Polysulfone Ultrafiltration Membrane. *J. Membr. Sci.* **2017**, *522*, 100–115. https://doi.org/10.1016/j.memsci.2016.07.016.

(58) Brickey, K. P.; Zydney, A. L.; Gomez, E. D. FIB-SEM Tomography Reveals the Nanoscale 3D Morphology of Virus Removal Filters. *J. Membr. Sci.* **2021**, *640*, 119766. https://doi.org/10.1016/j.memsci.2021.119766.

(59) Sundaramoorthi, G.; Hadwiger, M.; Ben-Romdhane, M.; Behzad, A. R.; Madhavan, P.; Nunes, S. P. 3D Membrane Imaging and Porosity Visualization. *Ind. Eng. Chem. Res.* **2016**, *55* (12), 3689–3695. https://doi.org/10.1021/acs.iecr.6b00387.

(60) Roberge, H.; Moreau, P.; Couallier, E.; Abellan, P. Determination of the Key Structural Factors Affecting Permeability and Selectivity of PAN and PES Polymeric Filtration Membranes Using 3D FIB/SEM. *J. Membr. Sci.* **2022**, *653*, 120530. https://doi.org/10.1016/j.memsci.2022.120530.

(61) Chamani, H.; Rabbani, A.; Russell, K. P.; Zydney, A. L.; Gomez, E. D.; Hattrick-Simpers, J.; Werber, J. R. Data-Science-Based Reconstruction of 3-D Membrane Pore Structure Using a Single 2-D Micrograph. *J. Membr. Sci.* **2023**, *678*, 121673. https://doi.org/10.1016/j.memsci.2023.121673.

(62) Ritt, C. L.; Werber, J. R.; Wang, M.; Yang, Z.; Zhao, Y.; Kulik, H. J.; Elimelech, M. Ionization Behavior of Nanoporous Polyamide Membranes. *Proc. Natl. Acad. Sci. U. S. A.* **2020**, *117* (48), 30191–30200. https://doi.org/10.1073/pnas.2008421117.

(63) Karan, S.; Jiang, Z.; Livingston, A. G. Sub-10 Nm Polyamide Nanofilms with Ultrafast Solvent Transport for Molecular Separation. *Science* **2015**, *348* (6241), 1347–1351. https://doi.org/10.1126/science.aaa5058.

(64) Gan, Q.; Hu, Y.; Wu, C.; Yang, Z.; Peng, L. E.; Tang, C. Y. Nanofoamed Polyamide Membranes: Mechanisms, Developments, and Environmental Implications. *Environ. Sci. Technol.* **2024**, *58* (47), 20812–20829. https://doi.org/10.1021/acs.est.4c06434.

(65) Culp, T. E.; Khara, B.; Brickey, K. P.; Geitner, M.; Zimudzi, T. J.; Wilbur, J. D.; Jons, S. D.; Roy, A.; Paul, M.; Ganapathysubramanian, B.; Zydney, A. L.; Kumar, M.; Gomez, E. D. Nanoscale Control of Internal Inhomogeneity Enhances Water Transport in Desalination Membranes. *Science* **2021**, *75* (January), 72–75.

(66) Culp, T. E.; Shen, Y.; Geitner, M.; Paul, M.; Roy, A.; Behr, M. J.; Rosenberg, S.; Gu, J.; Kumar, M.; Gomez, E. D. Electron Tomography Reveals Details of the Internal Microstructure of Desalination Membranes. *Proc. Natl. Acad. Sci.* **2018**, *115* (35), 201804708. https://doi.org/10.1073/pnas.1804708115.

(67) Pacheco, F.; Sougrat, R.; Reinhard, M.; Leckie, J. O.; Pinnau, I. 3D Visualization of the Internal Nanostructure of Polyamide Thin Films in RO Membranes. *J. Membr. Sci.* **2016**, *501*, 33–44. https://doi.org/10.1016/j.memsci.2015.10.061.

(68) Kłosowski, M. M.; McGilvery, C. M.; Li, Y.; Abellan, P.; Ramasse, Q.; Cabral, J. T.; Livingston, A. G.; Porter, A. E. Micro-to Nano-Scale Characterisation of Polyamide Structures of the SW30HR RO Membrane Using Advanced Electron Microscopy and Stain Tracers. *J. Membr. Sci.* **2016**, *520*, 465–476. https://doi.org/10.1016/j.memsci.2016.07.063.

(69) Palacio, L.; Prádanos, P.; Calvo, J. I.; Hernández, A. Porosity Measurements by a Gas Penetration Method and Other Techniques Applied to Membrane Characterization. *Thin Solid Films* **1999**, *348* (1), 22–29. https://doi.org/10.1016/S0040-6090(99)00197-2.

(70) Xiao, M.; Yang, F.; Im, S.; Dlamini, D. S.; Jassby, D.; Mahendra, S.; Honda, R.; Hoek, E. M. V. Characterizing Surface Porosity of Porous Membranes via Contact Angle Measurements. *J. Membr. Sci. Lett.* **2022**, *2* (1), 100022. https://doi.org/10.1016/j.memlet.2022.100022.

(71) Lenaerts, N.; Verbeke, R.; Davenport, D. M.; Caspers, S.; Eyley, S.; Kantre, K.-A.; Volodine, A.; Helm, R.; Butterling, M.; Liedke, M. O.; Wagner, A.; Thielemans, W.; Meersschaut, J.; Dickmann, M.; Vankelecom, I. F. J. Influence of Support Pore Size and Porosity on Epoxide-Based TFC Membranes. *J. Membr. Sci.* **2025**, *722*, 123900. https://doi.org/10.1016/j.memsci.2025.123900.

(72) Xie, W.; Tiraferri, A.; Liu, B.; Tang, P.; Wang, F.; Chen, S.; Figoli, A.; Chu, L.-Y. First Exploration on a Poly(Vinyl Chloride) Ultrafiltration Membrane Prepared by Using the Sustainable Green Solvent PolarClean. *ACS Sustain. Chem. Eng.* **2020**, *8* (1), 91–101. https://doi.org/10.1021/acssuschemeng.9b04287.

(73) Baker, R. W. *MEMBRANE TECHNOLOGY AND APPLICATIONS*; 2004. https://doi.org/10.1002/0470020393.

(74) Porter, M. C. Concentration Polarization with Membrane Ultrafiltration. *Prod. RD* **1972**, *11* (3), 234–248. https://doi.org/10.1021/i360043a002.